\documentclass[prd,onecolumn,notitlepage,eqsecnum,nofootinbib,floatfix]{revtex4-1}
\usepackage{amssymb}
\usepackage{amsmath}
\usepackage{amsfonts} 
\usepackage{bm}
\usepackage{graphicx}
\usepackage{dcolumn}
\DeclareSymbolFont{EulerScript}{U}{eus}{m}{n}
\SetSymbolFont{EulerScript}{bold}{U}{eus}{b}{n}
\DeclareSymbolFontAlphabet\scrpt{EulerScript}
\allowdisplaybreaks
\newcommand{\W}{{\scrpt W}} 
\newcommand{\stf}[1]{{\langle #1 \rangle}}
\newcommand{\uu}{\mathfrak{u}} 
\begin{document}
\title{Particle hanging on a string near a Schwarzschild black hole}  
\author{Michael LaHaye and Eric Poisson}  
\affiliation{Department of Physics, University of Guelph, Guelph, Ontario, N1G 2W1, Canada} 
\date{July 29, 2021} 
\begin{abstract} 
The literature features many instances of spacetimes containing two black holes held apart by a thin distribution of matter (strut or strings) on the axis joining the holes. For all such spacetimes, the Einstein field equations are integrated with an energy-momentum tensor that does not include a contribution from the axial matter; the presence of this matter is inferred instead from the existence of a conical singularity in the spacetime. And for all such spacetimes, the axial matter is characterized by a pressure (or tension) equal to its linear energy density, which are both constant along the length of the strut (or strings); the matter is therefore revealed to have a very specific equation of state. Our purpose with this paper is to show that the axial matter can be introduced at the very start of the exercise, through the specification of a distributional energy-momentum tensor, and that one can choose for it any equation of state. To evade no-go theorems regarding line sources in general relativity, which are too singular to be accommodated by the theory's nonlinearities, we retreat to a perturbative expansion of the gravitational field, using the Schwarzschild metric as a description of the background spacetime. Instead of a second black hole, our prototypical system features a point particle at a fixed position outside the Schwarzschild black hole, attached to a string extending to infinity. While this string prevents the particle from falling toward the black hole, a second string is attached to the black hole to prevent it from falling toward the particle. All this matter is described in terms of a distributional energy-momentum tensor, and we examine different equations of state for the strings. To integrate the field equations we introduce a new ``Weyl'' gauge for the metric perturbation, which allows us to find closed-form expressions for the gravitational potentials. Our solutions are linearized versions of multi-hole spacetimes, and some of them feature strings with a varying tension, unequal to the energy density. We describe the properties of these spacetimes, and begin an exploration of their extended thermodynamics.  
\end{abstract} 
\maketitle

\section{Introduction and summary} 
\label{sec:intro} 

\subsection{Multi-hole spacetimes and conical singularities}

In 1922, in the earliest days of relativistic gravitation, Bach and Weyl \cite{bach-weyl:22a} produced an exact solution to the Einstein field equations describing two Schwarzschild black holes held apart by a thin strut; the presence of the strut was revealed by a conical singularity in the spacetime (see Ref.~\cite{bach-weyl:22b} for an English translation of their article). In 1964, in the earliest days of the golden age of black-hole research, Israel and Khan \cite{israel-khan:64} generalized the Bach-Weyl solution to any number of black holes, provided that these are assembled in a collinear sequence; again struts ensure that the holes are kept at a fixed distance from one another. Further generalizations kept coming: the black holes were allowed to rotate in Refs.~\cite{kramer-neugenauer:80, breton-manko:95, manko-etal:08, manko-ruiz:17, cabrera-mungia-etal:17, cabrera-munguia:18, manko-ruiz:19}, and endowed with an electric charge in Refs.~\cite{manko:07, manko-ruiz-sanchezmondragon:09, cabrera-munguia-etal:20}.

We pause and point out that a conical singularity can be associated either with a strut or a string. A strut has a positive longitudinal stress (a positive pressure), while a string has a negative stress (a positive tension). Our focus in these introductory remarks shall be on struts; it will eventually shift to strings. 

The presence of conical singularities in these spacetimes can be viewed as an unwanted feature: to be physically meaningful, the spacetime should be free of such singularities. Indeed, a thread of the literature has adopted this sensible point of view, and sought singularity-free solutions in which the black holes are kept apart either by an electrostatic repulsion or a spin-spin interaction \cite{wald:72}. Investigations have concluded that charged black holes can sometimes be held apart without a strut \cite{alekseev-belinkski:07}, but that uncharged, rotating black holes cannot \cite{neugebauer-henning:09, henning-neugebauer:11, crusciel-etal:11}. 

Another thread of the literature has embraced the struts as idealized physical objects, and welcomed the opportunities they provide in the construction of static (or stationary) multi-hole spacetimes. A motivation to examine such spacetimes comes from the study of the thermodynamic properties of multi-hole systems \cite{costa-perry:00, krtous-zelnikov:19a, ramirez-garcia-manko:20, garcia-manko-ramirez:21, gregory-lim-scoins:21}. These studies were complemented with descriptions of the thermodynamics of black holes with conical defects \cite{aryal-ford-vilenkin:86, martinez-york:90, herdeiro-etal:10, appels-gregory-kubiznak:17}, and black holes accelerated by means of a cosmic string or a cosmological constant \cite{appels-gregory-kubiznak:16, anabalon-etal:18, anabalon-etal:19};
the spacetime of a black hole accelerated by a massive string was presented in Ref.~\cite{camps-emparan:10}. Intriguing aspects of these thermodynamics are that the total mass of the spacetime appears as an enthalpy variable in the first law \cite{kastor-ray-traschen:09, dolan:11}, and that conical singularities necessitate the introduction of new state functions, including a thermodynamic length for each strut \cite{appels-gregory-kubiznak:16}. Another motivation is provided by Ref.~\cite{lahaye-poisson:20}: a string is adopted as the physical agent responsible for keeping a charged particle stationary in a black-hole spacetime, and the particle's self-force \cite{smith-will:80} is measured by the string's tension. We follow this tradition in this paper, and take a conical singularity seriously as the manifestation of an idealized physical object.   

Much of the literature reviewed in the preceding paragraphs takes its foundation in Weyl's canonical metric (for a review, see Chapter 10 of Ref.~\cite{griffiths-podolsky:09})
\begin{equation} 
ds^2 = -e^{-2U}\, dt^2 + e^{2(U+\gamma)} (d\rho^2 + dz^2) + e^{2U}\rho^2\, d\phi^2, 
\label{weyl_original} 
\end{equation} 
in which the potentials $U$ and $\gamma$ depend on $\rho$ and $z$ only; the metric is static and axially symmetric, and with an additional potential it can be generalized to describe a stationary gravitational field. The ability to generate multi-hole solutions from this metric originates from the fact that the field equation for $U$ is linear; in vacuum it takes the form of Laplace's equation. Solutions can therefore be formed by superposition, and once $U$ is known, $\gamma$ can in principle be obtained by quadratures. For these solutions, it is found that in general, $\gamma$ fails to vanish on part of the $z$-axis, a condition that signals the presence of a conical singularity. Indeed, for a spacetime with the metric of Eq.~(\ref{weyl_original}), the ratio of proper circumference to proper radius for a small circle $\rho = \mbox{constant}$ around the $z$-axis is given by $2\pi \exp(-\gamma^{\rm axis})$, where $\gamma^{\rm axis} := \gamma(\rho=0,z)$. Elementary flatness demands that this ratio be precisely $2\pi$, and for this we must have $\gamma^{\rm axis} = 0$. Failure to achieve this implies that an angular deficit measured by $2\pi[1 - \exp(-\gamma^{\rm axis})]$ has been introduced in the geometry; the spacetime contains a conical singularity. As was shown by Israel \cite{israel:77a}, the singularity signals the presence of a thin distribution of matter on the axis --- a strut or string. It is this axial matter that is physically responsible for holding the black holes apart, and keeping the spacetime static (or stationary).   

\subsection{Objectives of this work}

A review of this literature reveals some remarkable aspects that seemed to us worthy of further reflection. First, it is interesting to observe that the axial matter (strut or string) is revealed only after the fact, that is, after the task of integrating the Einstein field equations is completed. Indeed, as we have seen, the presence of matter on the axis is inferred at the end, from the nonzero value of $\gamma^{\rm axis}$, instead of being incorporated at the start of the exercise, in the form of an energy-momentum tensor. Second, in all the cases examined thus far, the strut or string turns out to have a longitudinal stress that is precisely equal to its linear energy density; moreover, the stress is constant along the object. The absence of a gradient implies that the axial matter is weightless, a property confirmed by the fact that it makes no contribution to the gravitational potential $U$. In view of all this, one wonders whether more control could be placed into the design of a multi-hole spacetime. Shouldn't it be possible, for example, to specify an equation of state for the axial matter, write down an energy-momentum tensor for it, and find a solution to the field equations that takes explicit account of this matter source? In particular, shouldn't it be possible to dictate, at the start of the exercise, that the axial matter is to be a massive strut (or string), with a varying pressure (or tension)? One of our objectives with this paper is to restore such control.    

A second objective is to provide a resolution to the following puzzle. The form of Eq.~(\ref{weyl_original}) for the Weyl metric necessarily implies the restriction $G^\rho_{\ \rho} + G^z_{\ z} = 0$ on the Einstein tensor, and therefore a similar restriction on the energy-momentum tensor of any matter source. Axial matter, however, is expected to come with a nonzero $T^z_{\ z}$ and a vanishing $T^\rho_{\ \rho}$, and it should therefore produce a violation of the stated condition. How is it that the matter revealed by a conical singularity ends up violating a restriction on the energy-momentum tensor imposed by the assumed form of the metric?

To restore control to spacetime design, and to address the puzzle, we shall introduce, instead of Eq.~(\ref{weyl_original}), a general metric for a static and axially symmetric spacetime, one that does not feature any other constraint on the energy-momentum tensor. And we shall incorporate distributional terms in this tensor, so that axial matter can explicitly be placed in the spacetime at the very start of the construction. We shall then endeavor to solve the Einstein field equations for various models of this matter.   

It is known that by virtue of the nonlinearities of general relativity, a line source cannot, in general, be defined in terms of a distribution-valued energy-momentum tensor \cite{geroch-traschen:87}. Because the restored control and elucidation of the puzzle require these distributional sources, we shall have to shy away from a fully nonlinear description of the gravitational field. We will, instead, construct spacetimes containing axial matter by assuming that all energy densities and stresses are sufficiently small that they produce small perturbations of a given spacetime. These will be obtained by linearizing the field equations around the selected background spacetime; there is no obstacle to the introduction of distributional sources at such a linearized level. Because we wish our spacetimes to contain a black hole, we will take the background spacetime to be described by the Schwarzschild metric. And to respect the perturbative nature of the construction, we shall replace the additional hole with a small-mass particle, also described by a distributional energy-momentum tensor. (These is no obstacle to a generalization to multiple particles.) 

Our mission in this paper is therefore to construct perturbations of the Schwarzschild spacetime that are produced by a point particle attached to various types of axial matter. Our prototypical system shall be this: a particle of mass $m$ is held in place at position $r = r_0$ outside a Schwarzschild black hole of mass $M$. We prevent the particle from falling toward the black hole by attaching it to a string, which extends all the way to infinity on the upper axis of the spacetime. And we prevent the black hole from falling toward the particle by attaching it to a second string, which is placed on the lower axis. (We do not consider struts in this paper. The methods to be developed, however, apply to any type of axial matter.) The spacetime is static and axisymmetric with respect to the axis defined by the strings. Each string may be massless, with a tension $T$ equal to its energy density $\mu$. Or it may be massive, with $T \neq \mu$. In general we do not expect the perturbed spacetime to have a metric that can be put in the form of Eq.~(\ref{weyl_original}); after all, the distributional energy-momentum tensor violates the condition $T^\rho_{\ \rho} + T^z_{\ z} = 0$. We shall find, however, that in most circumstances, the metric does in fact take a form equivalent to Eq.~(\ref{weyl_original}). 

In spite of the perturbative nature of the constructions, we shall find that our gravitational potentials are not small everywhere. Near the particle, $U$ will be found to diverge as $r^{-1}$ when $r \to 0$, where $r$ is the distance to the particle. This divergence is expected, and is an artefact of the assumed pointlike nature of the particle. In a more elaborate construction, the particle would be replaced by a finite-sized body, and the $r^{-1}$ divergence would be regulated. A similar situation occurs near a massive string, where $U$ diverges as $\ln r$ when $r \to 0$; a more sophisticated construction featuring a finite-sized string would regularize this behavior. These divergences are localized and essentially harmless. They do not call to question the perturbative nature of our calculations; one is simply reminded to take the idealizations of a point particle and infinitely thin string with a grain of salt. Another type of divergence, however, is more serious. In the case of massive string, we shall find that $U$ diverges as $\ln r$ when $r \to \infty$. This is an artefact of the assumed infinite length of the string, and such a logarithmic divergence is present even in the Newtonian field of an infinite line mass. This divergence implies that our linearized calculations cannot be trusted beyond a given distance from the hole-particle-string system. A way to avoid this pathology would be to truncate the massive string to a finite segment, and to attach it to a massless string that extends the remaining way to infinity. (There is no divergence issue with a massless string, which makes no contribution to $U$.) The techniques introduced below could easily be exploited to calculate the gravitational potentials for this more satisfactory construction, but we shall not pursue this here.  

\subsection{Overview of our results}

We begin in Sec.~\ref{sec:perturbation} with a brief review of metric perturbations of the Schwarzschild spacetime, specialized to static and axially symmetric situations. A {\it Weyl class} of perturbations is then introduced in Sec.~\ref{sec:Weyl-class}. These are perturbations of the Schwarzschild metric that can be expressed in a form directly related to Eq.~(\ref{weyl_original}); the relation involves a transformation from cylindrical to spherical coordinates. Perturbations in the Weyl class are restricted by the same condition on the energy-momentum tensor; in spherical coordinates the restriction reads $T^r_{\ r} + T^\theta_{\ \theta} = 0$. Because of this condition, a perturbation of the Schwarzschild spacetime will not, in general, belong to the Weyl class. In Sec.~\ref{sec:weyl-gauge} we introduce a {\it Weyl gauge} for static and axisymmetric, but otherwise generic, perturbations of the Schwarzschild metric. The gauge is designed to keep the metric as close as possible to the spherical version of Eq.~(\ref{weyl_original}), but without the restriction on the energy-momentum tensor. We explore the properties of this gauge, in particular the fact that it is not unique, and explain under which special circumstances a perturbation in Weyl gauge can belong to the Weyl class. The energy-momentum tensor of our system of particle and strings is constructed in Sec.~\ref{sec:EMtensor}. The Einstein field equations are written down and formally integrated in Sec.~\ref{sec:EFE}. 

The remainder of the paper is devoted to various applications of this formalism, starting with simple warmup problems. In Sec.~\ref{sec:tidal} we examine the simplest type of perturbation, one describing a tidal deformation of the Schwarzschild spacetime (no particle, no string). The perturbation in Weyl gauge is compared to its better known expression in Regge-Wheeler gauge. In Sec.~\ref{sec:strings} we construct the spacetime of a black hole attached to massless strings of different tensions, one on the upper axis, the other on the lower axis (strings, no particle). We recover a linearized form of the $C$-metric (see Sec.~14.1 of Ref.~\cite{griffiths-podolsky:09} for a review of the exact solution). In Sec.~\ref{sec:massive-noBH} we introduce our favorite model of a massive string, one with 
\begin{equation} 
\sigma := \mu - T = \mbox{constant}. 
\label{massive_model} 
\end{equation} 
We recall that a massless string has $\sigma = 0$, and here we choose the constant to be nonvanishing. We calculate the gravitational field of this string as a perturbation of flat spacetime (massive string, no particle, no black hole). We recover a linearized form of the Levi-Civita metric (see Sec.~10.2 of Ref.~\cite{griffiths-podolsky:09}).  

Next we turn to more ambitious applications of the formalism. In Sec.~\ref{sec:particle-massless} we integrate the perturbation equations for a system of particle, massless string on the upper axis, and massless string on the lower axis. We show that the perturbation belongs to the Weyl class, and obtain explicit expressions for the potentials $U$ and $\gamma$ --- see Eqs.~(\ref{U_particle}), (\ref{D_def1}), and (\ref{gamma_particle}). We explore the thermodynamics of this spacetime by computing its total mass $M_{\rm tot}$, as well as the surface area $A$ and surface gravity $\kappa$ of the black hole. We obtain a first law of the form [Eq.~(\ref{firstlaw_massless})] 
\begin{equation} 
d M_{\rm tot} = \frac{\kappa}{8\pi}\, dA - \lambda\, dT + z\, dm, 
\label{1law_massless} 
\end{equation} 
in which $T$ is the (equal) tension in the upper and lower strings, and $m$ is the particle's mass. Because the strings are massless, they do not participate in the accounting of the total mass, which is therefore the sum of black-hole mass, particle mass, and gravitational binding energy. As expected, the law features the string's thermodynamic length $\lambda$, and the variable conjugate to $m$ is $z$, a redshift factor relating the energy of photons emitted at the particle's position and received at infinity. Most of the results presented in this section were obtained previously in Ref.~\cite{lahaye-poisson:20} using different techniques. We reproduce them here because they follow from a natural application of perturbation theory in the Weyl gauge.  

In Sec.~\ref{sec:particle-massive} we exploit the formalism to generate new results. We integrate the perturbation equations for a system of particle, massive string on the upper axis, and massless string on the lower axis; the massive string is again taken to satisfy Eq.~(\ref{massive_model}). In this case we find again that the perturbation belongs to the Weyl class, and obtain explicit expressions for the potentials $U$ and $\gamma$ --- see Eqs.~(\ref{U_massive_comp}) and (\ref{gamma_massive_comp}). We examine the impact of the massive string on the thermodynamic properties of the spacetime. While there is no obstacle to the computation of $A$ and $\kappa$, the fact that the geometry is no longer asymptotically flat --- the potential $U$ diverges logarithmically --- creates a difficulty in defining a notion of total mass $M_{\rm tot}$. In addition, the massive nature of the string implies that its tension and density vary along its length, and that the definition of state variables is no longer straightforward. In spite of these conceptual difficulties, we show that a formulation of the first law can nevertheless be given. It takes the form [Eq.~(\ref{firstlaw_massive})]  
\begin{equation} 
d M_{\rm tot} = \frac{\kappa}{8\pi}\, dA - \lambda\, dT_\infty + \omega\, d\sigma + z\, dm, 
\label{1law_massive}
\end{equation} 
where $T_\infty$ is the string tension measured at infinity (either on the upper or lower axis), and $\sigma$ is the parameter introduced in Eq.~(\ref{massive_model}), which, for this specific model of a massive string, makes a plausible candidate of state variable. To this we join $\omega$ as a conjugate variable, which can be interpreted as a new kind of thermodynamic length. The least compelling ingredient appearing in the first law is $M_{\rm tot}$, which is formally {\it the same} as the one appearing in Eq.~(\ref{1law_massless}). In other words, our ``total mass'' accounts for the black hole and particle and binding energy, but it ignores entirely the (infinite) contribution from the massive string. So while Eq.~(\ref{1law_massive}) is a valid relation among changes of various quantities that appear in the solution, its physical interpretation as a first law remains lacking. Nevertheless, this successful attempt at extending the thermodynamics of black holes and massless strings to massive strings should motivate further work on this topic.   

In Sec.~\ref{sec:generic}, the final section of the paper, we go away from the specific model of Eq.~(\ref{massive_model}) and examine massive strings with a well-motivated equation of state. The enhanced realism of the string model, however, comes at the price of a lost ability to integrate the perturbation equations exactly. We therefore let the particle and string lie in the weak-field region of the Schwarzschild spacetime, and construct approximate solutions to the equations. This retreat, fortunately, comes with its own measure of success: we are able to find a weak-field solution for a generic string that satisfies a broad class of plausible equations of state. Once again we find that the perturbation belongs to the Weyl class, and we obtain explicit expressions for the potentials $U$ and $\gamma$ --- see Eqs.~(\ref{potentials_generic}).  

In all the cases reviewed in the preceding paragraphs, we observe that $\gamma^{\rm axis}$ is in an intimate relationship with the string tension, whether the string is massless or massive, and whether the tension is constant or not. We find that 
\begin{equation} 
\gamma^{\rm axis}(z) = 4 T(z), 
\label{gamma_vs_T} 
\end{equation} 
where the dependence on $z$ indicates that these quantities depend on position along the axis. A special case of this result was encountered previously in the literature \cite{israel:77a}: the statement that $\gamma^{\rm axis} = 4T$, with the dependence on $z$ removed. By this we mean the following. The field equations for $\gamma$ imply that $\partial_z \gamma = 0$ when $\rho = 0$ (on axis), provided that $U$ is nonsingular there. In a typical situation featuring a black hole, a particle, and massless strings, $U$ is singular at the black hole and particle, and it is well-behaved everywhere else. In this situation, $\gamma^{\rm axis}$ is piecewise constant, but it jumps from one constant to another across the black hole, and at the particle. With $\gamma^{\rm axis} = 4T$ we mean to capture the piecewise constant behavior of these quantities; the equation does not capture the jumps. Equation (\ref{gamma_vs_T}) means something else: it states that $\gamma^{\rm axis}$ and $T$ can both vary along the axis, which they do when the string is massive, but that they are always proportional to each other. The evasion of the condition $\partial_z \gamma = 0$ when $\rho = 0$ for a massive string comes from the fact that $U$ is then singular on the axis --- it diverges logarithmically.  

Some technical developments are relegated to appendices. A sum over tensorial spherical harmonics is evaluated in Appendix~\ref{sec:summation1}, ready to be used in Sec.~\ref{sec:Weyl-class}. In Appendix~\ref{sec:regularity} we identify the conditions under which a metric perturbation in Weyl gauge is regular at the black-hole horizon. Relations between hypergeometric and Legendre functions are established in Appendix~\ref{sec:FvsL}, to be exploited in Sec.~\ref{sec:tidal}; this material is duplicated from Appendix B of Ref.~\cite{poisson:21a} for ease of reference. In Appendix~\ref{sec:integrals} we compute integrals featuring a product of Legendre functions, to aid integration of the perturbation equations in Sec.~\ref{sec:particle-massless}. Finally, the calculation of $U$ and $\gamma$ in Secs.~\ref{sec:particle-massless}, \ref{sec:particle-massive}, and \ref{sec:generic} requires the evaluation of a large number of infinite sums over multipole order $\ell$, and this is carried out in Appendix~\ref{sec:summation}. 

\subsection{Design control restored; puzzle resolved} 

We have restored control to spacetime design. We have also resolved the puzzle. The resolution is simply that the restriction $T^\rho_{\ \rho} + T^z_{\ z} = 0$ on the energy-momentum tensor is not meant to apply to the axis, at which the cylindrical coordinates of Eq.~(\ref{weyl_original}) are singular. As we show throughout the paper, using more suitable spherical coordinates, it is entirely permissible to introduce a distributional energy-momentum tensor that violates the restriction on axis, and nevertheless obtain a metric that can be cast in the form of Eq.~(\ref{weyl_original}).

\section{Metric perturbation} 
\label{sec:perturbation} 

We begin with a review of the formalism of metric perturbations of the Schwarzschild spacetime, specialized to static and axisymmetric situations. The material is drawn entirely from Ref.~\cite{martel-poisson:05}.  

The background Schwarzschild metric $g_{\alpha\beta}$ is expressed in the usual $(t,r,\theta,\phi)$ coordinates, and is given by
\begin{equation}
ds^2 = -f\, dt^2 + f^{-1}\, dr^2 + r^2 (d\theta^2 + \sin^2\theta\, d\phi^2),
\end{equation}
where $f := 1-2M/r$. It is useful to group the $(t,r)$ coordinates into $x^a$, and the $(\theta,\phi)$ coordinates into $\theta^A$. We let $\Omega_{AB} := \mbox{diag}[1, \sin^2\theta]$ be the metric on the unit 2-sphere, and $\Omega^{AB}$ be its matrix inverse. We denote by $D_A$ the covariant derivative operator compatible with $\Omega_{AB}$.  

The metric perturbation $p_{\alpha\beta}$ is taken to be static and axially symmetric, and we consider its even-parity sector only; for the sources examined in this work, the odd-parity sector plays no role and can be ignored. We expand the perturbation in scalar, vector, and tensor harmonics based on Legendre polynomials. We write
\begin{subequations}
\label{perturbation} 
\begin{align} 
p_{ab} &= \sum_{\ell=0}^\infty h^\ell_{ab}(r) P^\ell, \\
p_{aB} &= \sum_{\ell=1}^\infty j^\ell_a(r) P^\ell_A, \\
p_{AB} &= r^2 \Omega_{AB} \sum_{\ell=0}^\infty K^\ell(r) P^\ell
+ r^2 \sum_{\ell=2}^\infty G^\ell(r) P^\ell_{AB},
\end{align}
\end{subequations}
where $P^\ell := P_\ell(\cos\theta)$ are Legendre polynomials, and
\begin{equation}
P^\ell_A := D_A P^\ell, \qquad
P^\ell_{AB} := \Bigl[ D_A D_B + \frac{1}{2} \ell(\ell+1) \Omega_{AB} \Bigr] P^\ell. 
\label{tensor_harmonics} 
\end{equation}
Legendre's equation implies that $\Omega^{AB} P^\ell_{AB} = 0$: the tensorial harmonics are tracefree. Explicitly, the nonvanishing components of the vector and tensor harmonics are 
\begin{equation} 
P_\theta^\ell =\frac{d P^\ell}{d\theta}, \qquad
P_{\theta\theta}^\ell = -\frac{\cos\theta}{\sin\theta} \frac{d P^\ell}{d\theta}
- \frac{1}{2} \ell(\ell+1) P^\ell, \qquad
P_{\phi\phi}^\ell = \sin\theta \cos\theta \frac{d P^\ell}{d\theta}
+ \frac{1}{2} \ell(\ell+1) \sin^2\theta\, P^\ell.
\end{equation}
They satisfy the orthogonality relations
\begin{subequations} 
\label{ortho_P} 
\begin{align} 
\int P^\ell\, P^{\ell'}\, \sin\theta\, d\theta &= \frac{2}{2\ell+1}\delta_{\ell \ell'}, \\
\int \Omega^{AB} P^\ell_A\, P^{\ell'}_B\, \sin\theta\, d\theta
&= \frac{2\ell(\ell+1)}{2\ell+1} \delta_{\ell \ell'}, \\
\int \Omega^{AC} \Omega^{BD} P^\ell_{AB}\, P^{\ell'}_{CD}\, \sin\theta\, d\theta
&= \frac{(\ell-1)\ell(\ell+1)(\ell+2)}{2\ell+1} \delta_{\ell \ell'}. 
\end{align}
\end{subequations}

For a static perturbation we have that $h^\ell_{tr} = 0 = j^\ell_t$. These conditions restrict the gauge freedom to a vector $\Xi_\alpha$ with nonvanishing components 
\begin{equation}
\Xi_r = \sum_{\ell=0}^\infty \xi^\ell_r(r) P^\ell, \qquad
\Xi_A = \sum_{\ell=1}^\infty \xi^\ell(r) P^\ell_A.
\end{equation}
It produces the changes
\begin{subequations}
\label{gauge_transf} 
\begin{align}
\Delta h^\ell_{tt} &= \frac{2Mf}{r^2}\, \xi^\ell_r, \\ 
\Delta h^\ell_{rr} &= -2 \frac{d \xi^\ell_r}{dr} - \frac{2M}{r^2 f} \xi^\ell_r, \\
\Delta j^\ell_r &= -\frac{d \xi^\ell}{dr} + \frac{2}{r} \xi^\ell - \xi^\ell_r, \\
\Delta K^\ell &= -\frac{2f}{r} \xi^\ell_r + \frac{\ell(\ell+1)}{r^2} \xi^\ell, \\
\Delta G^\ell &= -\frac{2}{r^2} \xi^\ell
\end{align}
\end{subequations}
in the metric perturbations. The equations for $\Delta h^\ell_{tt}$, $\Delta h^\ell_{rr}$, and $\Delta K^\ell$ are valid for $\ell \geq 0$, the equation for $\Delta j^\ell_r$ is valid for $\ell \geq 1$, and the equation for $\Delta G^\ell$ is valid for $\ell \geq 2$. An additional gauge freedom exists for $\ell = 0$. It consists of a rescaling of the time coordinate described by $\xi^t = \alpha t$, or $\xi_t = -\alpha f t$, where $\alpha$ is a dimensionless constant; this produces $\Delta h_{tt} =2\alpha f$.

The Regge-Wheeler gauge \cite{regge-wheeler:57} sets $j^\ell_r = 0$ (for $\ell \geq 1$) and $G^\ell = 0$ (for $\ell \geq 2$). It is easy to check that this determines the gauge vector completely. The Regge-Wheeler is unique, and metric perturbations in this gauge are completely gauge-fixed (and therefore gauge invariant). 

\section{Weyl class}
\label{sec:Weyl-class}

A static and axially symmetric perturbation of the Schwarzschild spacetime belongs to the {\it Weyl class} if it can be presented in the form 
\begin{equation}
ds^2 = -e^{-2U}f\, dt^2 + e^{2(U+\gamma)} \bigl( f^{-1}\, dr^2 + r^2\, d\theta^2 \bigr)
+ e^{2U} r^2\sin^2\theta\, d\phi^2,
\label{weylclass_metric} 
\end{equation}
where $U(r,\theta)$ and $\gamma(r,\theta)$ are the metric perturbations. Because these are small, it is understood that $e^{-2U} = 1 - 2U$ and $e^{2\gamma} = 1 + 2\gamma$.

Equation (\ref{weylclass_metric}) is related to Eq.~(\ref{weyl_original}) in the following way. Starting from the cylindrical form of the Weyl metric, we write $U = U_0 + \bar{U}$ and $\gamma = \gamma_0 + \bar{\gamma}$, where $U_0$, $\gamma_0$ are the potentials associated with the Schwarzschild solution, and $\bar{U}$, $\bar{\gamma}$ are perturbations. The background potentials are given by (see Sec.~10.3 of Ref.~\cite{griffiths-podolsky:09}) 
\begin{equation} 
U_0 = -\frac{1}{2} \ln \frac{R_+ + R_- - 2M}{R_+ + R_- + 2M}, \qquad 
\gamma_0 = \frac{1}{2} \ln \frac{(R_+ + R_-)^2 - 4M^2}{4 R_+ R_-}, 
\end{equation} 
where 
\begin{equation} 
R_\pm := \sqrt{\rho^2 + (z \mp M)^2}. 
\end{equation} 
In a Newtonian interpretation, $U_0$ is the potential of a thin rod of mass $M$ and length $2M$ placed on the symmetry axis; the rod is centered at $z=0$, and $R_+$ is the Euclidean distance to the positive end of the rod (at $z = M$), while $R_-$ is the distance to its negative end (at $z = -M$). Next we perform a transformation from cylindrical to spherical coordinates, given by 
\begin{equation} 
\rho = r \sqrt{f} \sin\theta, \qquad 
z = (r-M)\cos\theta.
\end{equation} 
With $R_\pm = r - M \mp M \cos\theta$, $e^{-2U_0} = f$, $e^{2\gamma_0} = r^2 f/(R_+ R_-)$,  
$d\rho^2 + dz^2 = (R_+ R_-/r^2)( f^{-1}\, dr^2 + r^2\, d\phi^2)$, we find that the metric takes the form of Eq.~(\ref{weylclass_metric}), with $\bar{U}$ now denoted $U$, and $\bar{\gamma}$ now denoted $\gamma$. 

The linearized field equations for the metric of Eq.~(\ref{weylclass_metric}) are
\begin{subequations}
\label{Weylclass_EFE} 
\begin{align}
4\pi r^2 \bigl( T^t_{\ t} - T^\phi_{\ \phi} \bigr) &=
\partial_r (r^2 f\, \partial_r U)
+ \frac{1}{\sin\theta} \partial_\theta(\sin\theta\, \partial_\theta U), \\
8\pi r^2 T^r_{\ r} = -8\pi r^2 T^\theta_{\ \theta}
&= 2M\, \partial_r U + (r-M)\, \partial_r \gamma - \frac{\cos\theta}{\sin\theta}\, \partial_\theta \gamma, \\
8\pi r^2 T^r_{\ \theta} &= 2M\, \partial_\theta U + r^2 f \frac{\cos\theta}{\sin\theta}\, \partial_r \gamma
+ (r-M)\, \partial_\theta \gamma
\label{Weylclass_EFEc}; 
\end{align}
\end{subequations}
they come with the important restriction $T^r_{\ r} + T^\theta_{\ \theta} = 0$ on the perturbing energy-momentum tensor. Because of this restriction, the metric of Eq.~(\ref{weylclass_metric}) is not merely the result of a choice of gauge for the perturbation.

The metric perturbation associated with Eq.~(\ref{weylclass_metric}) is given by
\begin{equation}
p_{tt} = 2U f, \qquad
p_{rr} = 2(U + \gamma) f^{-1},  \qquad
p_{AB} = r^2 (2U\, \Omega_{AB} + 2\gamma\, e_A e_B),
\end{equation} 
with all other components vanishing; here $e_A := \partial_A \theta$ is normal to surfaces of constant $\theta$. The perturbation can be cast in the language of Sec.~\ref{sec:perturbation}. We decompose $U$ and $\gamma$ as
\begin{equation}
U = \sum_{\ell=0}^\infty u_\ell(r) P_\ell(\cos\theta), \qquad
\gamma = \sum_{\ell=0}^\infty g_\ell(r) P_\ell(\cos\theta),
\label{Ug_decomp}
\end{equation}
and see immediately that
\begin{equation}
h^\ell_{tt} = 2 u_\ell f, \qquad
h^\ell_{rr} = 2 (u_\ell + g_\ell) f^{-1}.
\end{equation}
We also have that $j^\ell_r = 0$. Taking the two-dimensional trace of $p_{AB}$ reveals that
\begin{equation}
K^\ell = 2u_\ell + g_\ell.
\end{equation}
The remaining, tracefree part of $p_{AB}$ then gives rise to the equality
\begin{equation}
\sum_{\ell=2}^\infty G^\ell P^\ell_{AB} = 2 e_\stf{AB} \sum_{\ell=0}^\infty g_\ell P^\ell,
\label{weylclass_Gcond} 
\end{equation}
where $e_\stf{AB} := e_A e_B - \frac{1}{2} \Omega_{AB}$.

Equation (\ref{weylclass_Gcond}) is solved for $G^\ell$ in Appendix~\ref{sec:summation1}. We obtain
\begin{equation} 
\frac{1}{4} (\ell-1)\ell(\ell+1)(\ell+2) G^\ell = (2\ell+1) S_\ell
- \frac{1}{2} (\ell-1) \ell\, g_\ell, \qquad \ell \geq 2, 
\label{weylclass_G}
\end{equation} 
where 
\begin{equation}
S_\ell := \left\{
\begin{array}{ll}
  g_0 + g_2 + \cdots + g_{\ell-2} & \qquad \mbox{$\ell$ even} \\
  g_1 + g_3 + \cdots + g_{\ell-2} & \qquad \mbox{$\ell$ odd}
\end{array} \right. . 
\label{Sdef}
\end{equation}
This quantity satisfies the recursion relation
\begin{equation}
S_{\ell + 2} = S_\ell + g_\ell
\label{S_recursion}
\end{equation}
together with the initial conditions
\begin{equation}
S_2 = g_0, \qquad S_3 = g_1.
\label{S_init}
\end{equation}
Equation (\ref{weylclass_G}) states that each $G^\ell$ is constructed algebraically from $g_{\ell'}$s with $\ell' = \{ \ell, \ell-2, \cdots \}$, all the way down to $\ell'=0$ when $\ell$ is even, or $\ell'=1$ when $\ell$ is odd.

\section{Weyl gauge}
\label{sec:weyl-gauge} 

We define a {\it Weyl gauge} for the perturbation by making the assignments
\begin{equation}
h^\ell_{tt} = 2 u_\ell f, \qquad
h^\ell_{rr} = 2 (u_\ell + g_\ell) f^{-1}, \qquad
K^\ell = 2u_\ell + g_\ell, 
\label{Wgauge}
\end{equation}
together with $j^\ell_r = 0$. These are the same relations as in the Weyl class of Sec.~\ref{sec:Weyl-class}, but we do not impose Eq.~(\ref{weylclass_G}). {\it A perturbation in Weyl gauge shall also belong to the Weyl class when this additional condition results from the field equations.} In general, however, a perturbation in Weyl gauge will not belong to the Weyl class, because of the restriction on the perturbing energy-momentum tensor encountered previously.

The Weyl gauge produces a metric perturbation that is regular at $r = 2M$ provided that $u_\ell$, $g_\ell$, and $G_\ell$ are bounded there, and provided also that 
\begin{equation}
2 u_\ell(r=2M) + g_\ell(r=2M) = 0  \qquad (\ell \geq 1). 
\label{regularity_2M} 
\end{equation}
This statement is established in Appendix~\ref{sec:regularity}. 

The gauge conditions are $f^{-1} h_{tt}^\ell + f h_{rr}^\ell - 2 K^\ell = 0$ for $\ell \geq 0$, and $j_r^\ell = 0$ for $\ell \geq 1$. These can always be imposed by a suitable choice of gauge vector. The gauge, however, is not unique: Equations (\ref{gauge_transf}) imply that
\begin{subequations}
\begin{align}
\Delta \bigl( f^{-1} h_{tt}^\ell + f h_{rr}^\ell - 2 K^\ell \bigr) &=
-2 f \biggl( \frac{d}{dr} - \frac{2}{r} \biggr) \xi^\ell_r - \frac{2\ell(\ell+1)}{r^2} \xi^\ell, \\
\Delta j^\ell_r &= -\biggl( \frac{d}{dr} - \frac{2}{r} \biggr)  \xi^\ell - \xi^\ell_r,
\end{align}
\end{subequations}
and these changes vanish for any pair $(\xi_r^\ell, \xi^\ell)$ that satisfies
\begin{equation}
r^2 f \frac{d^2 \xi^\ell}{dr^2} - 4rf \frac{d \xi^\ell}{dr} - \bigl[ \ell(\ell+1) - 6f \bigr] \xi^\ell = 0
\label{gauge_residual_deq} 
\end{equation}
and $\xi^\ell_r = -d\xi^\ell/dr + 2\xi^\ell/r$. The corresponding changes in the perturbation variables are
\begin{subequations}
\label{gauge_residual_changes1} 
\begin{align}
\Delta u_\ell &= -\frac{M}{r^3} \biggl( r \frac{d\xi^\ell}{dr} - 2 \xi^\ell \biggr), \\
\Delta g_\ell &= \frac{2}{r} \biggl(1 - \frac{M}{r} \biggr) \frac{d\xi^\ell}{dr}
+ \frac{1}{r^2} \biggl[ \ell(\ell+1) - 4 + \frac{4M}{r} \biggr] \xi^\ell, \\
\Delta G_\ell &= -\frac{2}{r^2} \xi^\ell. 
\label{DeltaG_residual} 
\end{align}
\end{subequations}
These results apply when $\ell \geq 1$, and Eq.~(\ref{DeltaG_residual}) is restricted to $\ell \geq 2$. For $\ell = 0$ we have that $\xi$ is not defined, and we find that $\xi_r$ must satisfy $d\xi_r/dr - 2\xi_r/r = 0$. The solution is $\xi_r = (\beta/M) r^2$, where $\beta$ is a dimensionless constant. This produces the changes $\Delta u = \beta$, $\Delta g = -2\beta(r/M-1)$ in the $\ell=0$ perturbation variables.

It is useful to rewrite Eq.~(\ref{gauge_residual_deq}) in terms of the new independent variable
\begin{equation}
x := r/M - 1.
\label{xdef}
\end{equation}
This gives
\begin{equation}
(x^2-1) \frac{d^2 \xi^\ell}{dx^2} - 4(x-1) \frac{d \xi^\ell}{dx}
- \biggl[ \ell(\ell+1) - 6 \frac{x-1}{x+1} \biggr] \xi^\ell = 0.
\end{equation}
The linearly independent solutions to this equation are
\begin{equation}
\xi^\ell = M^2 \Bigl\{ (x-1)(x+1)^3 P'_\ell(x),\  (x-1)(x+1)^3 Q'_\ell(x) \Bigr\},
\label{gauge_residual_sols} 
\end{equation}
with a prime indicating differentiation with respect to $x$. Making the substitution in Eq.~(\ref{gauge_residual_changes1}), we obtain
\begin{subequations}
\label{gauge_residual_changes2} 
\begin{align}
\Delta u_\ell &= -\ell(\ell+1) \Bigl\{ P_\ell, \ Q_\ell \Bigr\}, \\
\Delta g_\ell &= \ell(\ell+1) \Bigl\{ (x^2-1) P'_\ell + 2x P_\ell, \
  (x^2-1) Q'_\ell + 2x Q_\ell \Bigr\}, \\
\Delta G_\ell &= -2 \Bigl\{ (x^2-1) P'_\ell, \ (x^2-1) Q'_\ell \Bigr\}.
\end{align}
\end{subequations}
We recall that these results apply when $\ell \geq 1$, or $\ell \geq 2$ in the case of $\Delta G_\ell$. 

The gauge vector
\begin{equation}
\xi^{\rm W \to RW} = \frac{1}{2} r^2 G^{\rm W}, \qquad
\xi_r^{\rm W \to RW} = -\biggl( \frac{d}{dr} - \frac{2}{r} \biggr) \xi^{\rm W \to RW}
\end{equation}
sends a perturbation from the Weyl gauge (W) to the Regge-Wheeler gauge (RW). The new perturbation variables are
\begin{subequations}
\begin{align}
h^{\rm RW}_{tt} &= 2f \biggl( u^{\rm W} + \frac{M}{r^2} \xi_r^{\rm W \to RW} \biggr)\, \\
h^{\rm RW}_{rr} &= 2 f^{-1} \biggl( u^{\rm W} + g^{\rm W}
- f \frac{d}{dr} \xi_r^{\rm W \to RW}  - \frac{M}{r^2} \xi_r^{\rm W \to RW} \biggr), \\
K^{\rm RW} &= 2 u^{\rm W} + g^{\rm W} - \frac{2f}{r} \xi_r^{\rm W \to RW}
+ \frac{\ell(\ell+1)}{r^2} \xi^{\rm W \to RW}.
\end{align} 
\end{subequations}
Here we omit the label $\ell$ on the variables, to avoid a clutter of notation.

\section{Energy-momentum tensor} 
\label{sec:EMtensor} 

In later portions of this paper we shall consider a Schwarzschild black hole perturbed by a system of particle and strings. In this section we construct the energy-momentum tensor for the matter sources. 

The particle has a mass $m$ and it moves on a world line $\gamma$ described parametrically by $x^\alpha = Z^\alpha(\tau)$, where $\tau$ is proper time. Its velocity vector is $v^\alpha = dZ^\alpha/d\tau$, and its energy-momentum tensor is
\begin{equation}
T_{\rm p}^{\alpha\beta} = m \int_\gamma v^\alpha v^\beta \delta(x,Z)\, d\tau,
\end{equation}
where $\delta(x,Z) = \delta(x-Z)/\sqrt{-g}$ is a scalarized Dirac distribution; $\delta(x-Z)$ is the usual four-dimensional delta function, and $g$ is the metric determinant. We place the particle at $r = r_0$, $\theta = 0$, and assign to it the arbitrary azimuthal coordinate $\phi_0$. The only nonvanishing component of the velocity vector is then $v^t = f_0^{-1/2}$, where $f_0 := 1-2M/r_0$. The only nonvanishing component of the energy-momentum tensor is
\begin{equation}
T^{\ t}_{{\rm p}\ t} = -\frac{m f_0^{1/2}}{r_0^2} \delta(r-r_0) \delta(\cos\theta-1) \delta(\phi - \phi_0).
\label{Tp}
\end{equation}
The quantity $m f_0^{1/2}$ is recognized as the particle's Killing energy in the Schwarzschild spacetime. 

The particle is held in place with a thin string, which extends from the particle to infinity along $\theta = 0$. Because it is placed on the upper $z$-axis, we assign the label ``up'' to this string, and we give it a linear energy density $\mu_{\rm up}$ and a tension $T_{\rm up}$. The string moves on a world sheet $\W$ with intrinsic coordinates $\xi^a$. The embedding relations are $x^\alpha = X^\alpha(\xi^a)$, and $e^\alpha_a := \partial X^\alpha/\partial \xi^a$ are tangent vectors on $\W$. The world sheet's intrinsic metric is $\gamma_{ab} :=
g_{\alpha\beta} e^\alpha_a e^\beta_b$, and $u^a$ is a velocity field on $\W$. The string's energy-momentum tensor is
\begin{equation}
T^{\alpha\beta}_{\rm up} = \int_\W t_{\rm up}^{ab} e^\alpha_a e^\beta_b\, \delta(x,X)\, \sqrt{-\gamma}\, d^2\xi,
\end{equation}
where
\begin{equation}
t^{ab}_{\rm up} = \mu_{\rm up} u^a u^b - T_{\rm up} (\gamma^{ab} + u^a u^b)
\label{tab_up} 
\end{equation}
is the intrinsic energy-momentum tensor, and where $\sqrt{-\gamma}\, d^2\xi$ is the element of surface area on the world sheet. In our case the intrinsic coordinates are $\xi^a = (t,r)$, the embedding relations are $t = t$, $r = r$, $\theta = 0$, and $\phi = \phi_0$ (the particle's azimuthal coordinate). The string is static, and the only nonvanishing component of its velocity vector is $u^t = f^{-1/2}$. With all this, we have that the nonvanishing components of the energy-momentum tensor are
\begin{equation}
T^{\ \ t}_{{\rm up}\ t} = -\frac{\mu_{\rm up}}{r^2}\, \Theta(r-r_0) \delta(\cos\theta - 1) \delta(\phi-\phi_0), \qquad
T^{\ \ r}_{{\rm up}\ r} = -\frac{T_{\rm up}}{r^2}\, \Theta(r-r_0) \delta(\cos\theta - 1) \delta(\phi-\phi_0), 
\label{Tup}
\end{equation}
where $\Theta(r-r_0)$ is the Heaviside step function, equal to 1 when $r > r_0$ and 0 otherwise. 

The black hole also is held in place with a thin string, which extends from the event horizon to infinity along $\theta = \pi$. Because this string is placed on the lower $z$-axis, we assign the label ``dn'' to it (short for ``down''), and we give it a linear energy density $\mu_{\rm dn}$ and a tension $T_{\rm dn}$. The nonvanishing components of its energy-momentum tensor are
\begin{equation}
T^{\ \ t}_{{\rm dn}\ t} = -\frac{\mu_{\rm dn}}{r^2}\, \delta(\cos\theta + 1) \delta(\phi-\phi_0), \qquad
T^{\ \ r}_{{\rm dn}\ r} = -\frac{T_{\rm dn}}{r^2}\, \delta(\cos\theta + 1) \delta(\phi-\phi_0). 
\label{Tdn}
\end{equation}
In this case there is no need to incorporate a step function. 

The total energy-momentum tensor
\begin{equation}
T^{\alpha\beta} = T^{\alpha\beta}_{\rm p} + T^{\alpha\beta}_{\rm up} + T^{\alpha\beta}_{\rm dn}
\end{equation}
is conserved. Because the particle is attached to the upper string, but not interacting with the lower string, we have that
\begin{equation}
\nabla_\beta \bigl( T^{\alpha\beta}_{\rm p} + T^{\alpha\beta}_{\rm up} \bigr) = 0. 
\end{equation}
This equation produces
\begin{equation} 
\frac{d T_{\rm up}}{dr} = \frac{M}{r^2 f} (\mu_{\rm up} - T_{\rm up})
\label{conservation-up} 
\end{equation}
together with the boundary condition
\begin{equation}
T_{\rm up}(r=r_0) = \frac{mM}{r_0^2 f_0^{1/2}}.
\label{T0}
\end{equation}
The quantity on the right-hand side is $m$ times the acceleration of a test mass at position $r=r_0$ in the Schwarzschild spacetime; the conservation equation therefore implies that $T = m a$, a statement of Newton's second law.

The energy-momentum tensor of the lower string is conserved separately, and this statement gives rise to
\begin{equation}
\frac{d T_{\rm dn}}{dr} = \frac{M}{r^2 f} (\mu_{\rm dn} - T_{\rm dn}).
\label{conservation-dn} 
\end{equation}
The upper string is massless when $\mu_{\rm up} = T_{\rm up}$, and similarly, the lower string is massless when $\mu_{\rm dn} = T_{\rm dn}$; the conservation equations then imply that the tension is constant along each string. When the lower string is massive, Eq.~(\ref{conservation-dn}) implies that $\mu_{\rm dn} - T_{\rm dn}$ must approach zero when $r \to 2M$. This condition is likely impossible to fulfill for most equations of state, and below we shall always take the lower string to be massless.  

The only remaining task for this section is to decompose the energy-momentum tensor in Legendre polynomials. This is accomplished with
\begin{subequations}
\label{delta_Legendre}
\begin{align}
\delta(\cos\theta-1) \delta(\phi-\phi_0)
&= \sum_{\ell=0}^\infty \frac{2\ell+1}{4\pi} P_\ell(\cos\theta), \\
\delta(\cos\theta+1) \delta(\phi-\phi_0)
&= \sum_{\ell=0}^\infty \frac{2\ell+1}{4\pi} (-1)^\ell P_\ell(\cos\theta).
\end{align}
\end{subequations}
These equations follow as direct consequences of the completeness relation for spherical harmonics, together with the properties $P_\ell(1) = 1$ and $P_\ell(-1) = (-1)^\ell$ of Legendre polynomials.

\section{Field equations in the Weyl gauge}
\label{sec:EFE}

The Einstein field equations $G^{\alpha}_{\ \beta} = 8\pi T^\alpha_{\ \beta}$ are linearized with respect to the metric perturbation $p_{\alpha\beta}$, which is presented in the Weyl gauge of Sec.~\ref{sec:weyl-gauge}. We incorporate the energy-momentum tensor of Sec.~\ref{sec:EMtensor}, and obtain an equation for $u_\ell$,
\begin{equation}
r^2 f \frac{d^2 u_\ell}{dr^2} + 2(r-M) \frac{d u_\ell}{dr} - \ell(\ell+1) u_\ell
= -(2\ell+1) \Bigl[ m f_0^{1/2}\, \delta(r-r_0) +( \mu_{\rm up} - T_{\rm up} ) \Theta(r-r_0)
+ (-1)^\ell ( \mu_{\rm dn} - T_{\rm dn} ) \Bigr],
\label{eq-u}
\end{equation}
an equation for $g_\ell$,
\begin{equation}
r^2 f \frac{d^2 g_\ell}{dr^2} - \ell(\ell+1) g_\ell = 4 M \frac{d u_\ell}{dr},
\label{eq-g} 
\end{equation}
and an equation for $G_\ell$,
\begin{equation}
\frac{1}{4}(\ell-1)\ell(\ell+1)(\ell+2) G_\ell = 2 M \frac{d u_\ell}{dr} + (r-M) \frac{d g_\ell}{dr}
-\frac{1}{2} \bigl[ \ell(\ell+1) + 2 \bigr] g_\ell
+ 2(2\ell+1) \bigl[ T_{\rm up} \Theta(r-r_0) + (-1)^\ell T_{\rm dn} \bigr].  
\label{eq-G}
\end{equation}
All three equations are valid for $\ell \geq 0$. For $\ell = 0$ and $\ell = 1$, Eq.~(\ref{eq-G}) plays the role of a constraint equation implicating $du_\ell/dr$, $dg_\ell/dr$, and $g_\ell$. 

It is useful to re-express the field equations in terms of $x := r/M - 1$, previously introduced in Eq.~(\ref{xdef}). We have
\begin{subequations}
\label{EFEx}
\begin{align}
(x^2-1) \frac{d^2 u_\ell}{dx^2} + 2 x \frac{d u_\ell}{dx} - \ell(\ell+1) u_\ell
&= -(2\ell+1) \bigl[ k \delta(x-x_0) + ( \mu_{\rm up} - T_{\rm up} ) \Theta(x - x_0)
+ (-1)^\ell ( \mu_{\rm dn} - T_{\rm dn} ) \bigr],
\label{EFEu} \\
(x^2 - 1) \frac{d^2 g_\ell}{dx^2} - \ell(\ell+1) g_\ell &= 4 \frac{d u_\ell}{dx},
\label{EFEg} \\
\frac{1}{4}(\ell-1)\ell(\ell+1)(\ell+2) G_\ell &= (2\ell+1) \hat{S}_\ell - \frac{1}{2} (\ell-1) \ell\, g_\ell, 
\label{EFEG}
\end{align} 
\end{subequations}
where 
\begin{equation} 
(2\ell+1) \hat{S}_\ell := 2 \frac{d u_\ell}{dx} + x \frac{d g_\ell}{dx}
- (\ell+1) g_\ell + 2(2\ell+1) \bigl[ T_{\rm up} \Theta(x-x_0) + (-1)^\ell T_{\rm dn} \bigr]. 
\label{Shat_def}
\end{equation}  
We have also introduced 
\begin{equation}
k := \frac{m}{M} f_0^{1/2} = \frac{m}{M} \sqrt{\frac{x_0-1}{x_0+1}}  
\label{kdef}
\end{equation}
and $x_0 := r_0/M - 1$. 

The homogeneous version of Eq.~(\ref{EFEu}) has the linearly independent solutions
\begin{equation}
u_\ell^{\rm homo} = \Bigl\{ P_\ell(x), \ Q_\ell(x) \Bigr\}.
\label{u_homo}
\end{equation}
The first solution is finite at $x = 1$ (the event horizon of the Schwarzschild spacetime) and diverges at $x = \infty$; the second solution vanishes at infinity, but diverges logarithmically at the horizon. These solutions can be exploited to construct a Green's function for Eq.~(\ref{EFEu}), and we find that a particular solution is
\begin{equation}
u^{\rm part}_\ell = -Q_\ell(x) \int^x P_\ell(x') W_\ell(x')\, dx' - P_\ell(x) \int_x Q_\ell(x') W_\ell(x')\, dx',
\label{u_part} 
\end{equation}
with $W_\ell$ denoting the right-hand side of Eq.~(\ref{EFEu}). The general solution is obtained by adding to this a linear superposition of the functions listed in Eq.~(\ref{u_homo}). The physical solution to Eq.~(\ref{EFEu}) is identified by imposing regularity at $x=\infty$ and smoothness at $x=1$.

The homogeneous version of Eq.~(\ref{EFEg}) has the linearly independent solutions
\begin{equation}
g_\ell^{\rm homo} = \Bigl\{  (x^2-1) P_\ell'(x), \ (x^2-1) Q_\ell'(x) \Bigr\}
\label{g_homo}
\end{equation}
when $\ell \geq 1$; a prime on a Legendre function indicates differentiation with respect to $x$. Using these to construct a Green's function, we find that a particular solution to Eq.~(\ref{EFEg}) is
\begin{equation}
g^{\rm part}_\ell = \frac{4(x^2-1)}{\ell(\ell+1)} \biggl[
Q'_\ell(x) \int^x P'_\ell(x') u'_\ell(x')\, dx' + P'_\ell(x) \int_x Q'_\ell(x') u'_\ell(x')\, dx' \biggr].
\label{g_part}
\end{equation}
The case $\ell = 0$ requires a separate treatment. The homogeneous solutions are then $x-1$ and $1$, and the particular solution is
\begin{equation} 
g^{\rm part}_0 = -4 \int^x \frac{u'_0(x')}{x'+1}\, dx'
- 4(x-1) \int_x \frac{u'_0(x')}{x^{\prime 2}-1}\, dx'.
\label{g0_part}
\end{equation} 
The general solution is obtained by combining particular and homogeneous solutions, and again, the physical solution to Eq.~(\ref{EFEg}) is identified by imposing regularity at $x=\infty$ and smoothness at $x=1$.

We observed back in Sec.~\ref{sec:weyl-gauge} that the Weyl gauge is not unique. In principle, the solutions to Eqs.~(\ref{EFEx}) can be modified at will by effecting the changes described by Eqs.~(\ref{gauge_residual_changes2}). These changes, however, are likely to produce a perturbation that is no longer smooth at $x=1$ or regular at $x=\infty$. In most situations, therefore, the residual gauge freedom cannot be exercised and the Weyl gauge is effectively unique. It may be noted that $\Delta u_\ell$, $\Delta g_\ell$, and $\Delta G_\ell$ satisfy the vacuum field equations ---- Eqs.~(\ref{EFEx}) with all material sources set to zero. It is also interesting to note that $\Delta u_\ell$ and $u_\ell^{\rm homo}$ are given by the same set of functions.  

It was pointed out at the beginning of Sec.~\ref{sec:weyl-gauge} that a solution to Eqs.~(\ref{EFEx}) will not, in general, produce a perturbation that belongs the Weyl class of Sec.~\ref{sec:Weyl-class}. The reason is that the $G_\ell$ obtained from Eq.~(\ref{EFEG}) will fail, in general, to satisfy the condition of Eq.~(\ref{weylclass_G}). In exceptional circumstances, however, this condition will be satisfied, and the perturbation will belong to the Weyl class. To test whether this is the case, we examine the quantity $\hat{S}_\ell$ defined by Eq.~(\ref{Shat_def}). According to Eqs.~(\ref{weylclass_G}) and (\ref{EFEG}), the perturbation belongs to the Weyl class provided that $\hat{S}_\ell = S_\ell$, where $S_\ell$ is defined by Eq.~(\ref{Sdef}). Alternatively, following the discussion at the end of Sec.~\ref{sec:Weyl-class}, the perturbation is in the Weyl class when $\hat{S}_\ell$ satisfies the recursion relation
\begin{equation}
\hat{S}_{\ell+2} = \hat{S}_\ell + g_\ell
\label{Shat_recursion} 
\end{equation} 
together with the initial conditions
\begin{equation}
\hat{S}_2 = g_0, \qquad \hat{S}_3 = g_1.
\label{Shat_initial}
\end{equation}
To see whether a perturbation belongs to the Weyl class, one therefore computes $\hat{S}_\ell$ with the help of Eq.~(\ref{Shat_def}), and verifies if it satisfies the recursion relation and initial conditions.  

\section{Application: Tidal perturbation}
\label{sec:tidal}

We may now consider some applications of the formalism put in place in the preceding section. We begin with a simple case, a tidal perturbation of the Schwarzschild black hole (no particle, no strings). We take the metric  perturbation to be a pure multipole of order $\ell$, with $\ell \geq 2$. Everywhere in this section we omit the label $\ell$ on the perturbation variables. 

\subsection{Regge-Wheeler gauge}

A tidal perturbation of a Schwarzschild black hole was previously worked out in the Regge-Wheeler gauge by Binnington and Poisson \cite{binnington-poisson:09}. They obtain
\begin{equation}
h^{\rm RW}_{tt} = 2 u^{\rm RW} f, \qquad
h^{\rm RW}_{tt} = 2 u^{\rm RW} f^{-1}
\end{equation} 
with
\begin{equation}
u^{\rm RW} = z^{-\ell} (1-z) F(-\ell+2,-\ell;-2\ell;z),  
\end{equation}
where $z := 2M/r$ and $F(a,b;c;z)$ is the hypergeometric function. They also get 
\begin{equation}
K^{\rm RW} = \frac{2}{\ell-1} z^{-\ell} \bigl[ (\ell+1) F(-\ell,-\ell;-2\ell;z)
- 2 F(-\ell-1,-\ell;-2\ell; z) \bigr]
\end{equation}
for the remaining perturbation variable. The tidal field is normalized so that $u^{\rm RW} \sim (r/2M)^\ell$ and $K^{\rm RW} \sim 2(r/2M)^\ell$ when $r \gg 2M$.

An alternative representation of the tidal field in terms of Legendre polynomials is
\begin{subequations}
\label{tidal_RW} 
\begin{align} 
u^{\rm RW} &= \mu_\ell \bigl[ -2x P'_\ell(x) + \ell(\ell+1) P_\ell(x) \bigr], \\
K^{\rm RW} &= 2 \mu_\ell \bigl[ -2(x-1) P'_\ell(x) + \ell(\ell+1) P_\ell(x) \bigr],
\end{align}
\end{subequations}
where a prime indicates differentiation with respect to $x$, and where 
\begin{equation}
\mu_\ell := \frac{(\ell-2)!\, (\ell-1)!}{2(2\ell-1)!}. 
\label{mu_def} 
\end{equation}
The equivalence between these representations is established on the basis of results collected in Appendix~\ref{sec:FvsL}.

\subsection{Weyl gauge}

A tidal perturbation must satisfy the vacuum field equations, and $u$ must therefore be a solution to Eq.~(\ref{EFEu}) with a zero right-hand side. The solution must be smooth at $r = 2M$, and this requirement selects
\begin{equation}
u^{\rm W} = \lambda_\ell P_\ell(x),
\label{u_tidalW} 
\end{equation}
where $\lambda_\ell$ is a normalization factor. We normalize the potential so that it behaves as $(r/2M)^\ell$ when $r \gg 2M$, and therefore set (refer to Appendix~\ref{sec:FvsL})
\begin{equation}
\lambda_\ell = \frac{(\ell!)^2}{(2\ell)!}. 
\label{lambda_def} 
\end{equation}

With $u$ thus identified, $g$ must be a solution to Eq.~(\ref{EFEg}). The particular solution of Eq.~(\ref{g_part}) is
\begin{equation}
g^{\rm part} = \frac{4\lambda_\ell (x^2-1)}{\ell(\ell+1)} \biggl[
Q'_\ell(x) \int^x P'_\ell(x') P'_\ell(x')\, dx'
+ P'_\ell(x) \int_x P'_\ell(x') Q'_\ell(x')\, dx' \biggr],
\end{equation}
and the integrals are evaluated in Appendix~\ref{sec:integrals}. We obtain
\begin{equation}
g^{\rm part} = \lambda_\ell \bigl[ c_1(x^2-1) P'_\ell(x) + c_2(x^2-1) Q'_\ell(x) - 2x P_\ell(x) \bigr], 
\end{equation}
where $c_1$ and $c_2$ are arbitrary constants. The freedom to add solutions to the homogeneous equation, described by Eq.~(\ref{g_homo}), allows us to eliminate the term implicating $Q'_\ell(x)$, which is not smooth at $x=1$. It also allows us to shift arbitrarily the value of $c_1$. The physical solution to the field equations is therefore
\begin{equation}
g^{\rm W} = \lambda_\ell \bigl[ c(x^2-1) P'_\ell(x) - 2x P_\ell(x) \bigr], 
\label{g_tidalW} 
\end{equation}
where $c$ remains as an arbitrary constant. We verify that $u$ and $g$ satisfy the regularity condition of Eq.~(\ref{regularity_2M}) for any value of $c$.  

With $u$ and $g$ determined, we finally obtain $G$ from Eq.~(\ref{EFEG}). The result is
\begin{equation}
G^{\rm W} = \frac{4\lambda_\ell}{(\ell-1)\ell(\ell+1)(\ell+2)} \Bigl\{
- \bigl[ 2 + \tfrac{1}{2}(\ell^2+\ell+2)c \bigr] (x^2-1) P_\ell'(x)
+ (c+1) \ell(\ell+1) x P_\ell(x) \Bigr\}.
\label{G_tidalW}
\end{equation} 
Equations (\ref{u_tidalW}), (\ref{g_tidalW}), and (\ref{G_tidalW}) give a complete description of a tidal perturbation in the Weyl gauge. Because Eq.~(\ref{weylclass_G}) is violated, the perturbation does not belong to the Weyl class. 

\subsection{Transformation to Regge-Wheeler gauge}

The meaning of the constant $c$ can be elucidated by subjecting the perturbation to a change of gauge, from Weyl gauge to Regge-Wheeler gauge. The operation should reproduce the results of Eqs.~(\ref{tidal_RW}).

The gauge transformation was described at the end of Sec.~\ref{sec:weyl-gauge}. The gauge vector is given by
\begin{equation}
\xi^{\rm W \to RW} = \frac{1}{2} M^2 (x+1)^2 G^{\rm W}, \qquad
\xi_r^{\rm W \to RW} = -\frac{1}{M} \biggl( \frac{d}{dx} - \frac{2}{x+1} \biggr) \xi^{\rm W \to RW},
\end{equation}
and the transformation is given by
\begin{subequations}
\begin{align}
u^{\rm W \to RW} &= u^{\rm W} + \frac{1}{M} \frac{1}{(x+1)^2} \xi_r^{\rm W \to RW}, \\
u^{\rm W \to RW} &= u^{\rm W} + g^{\rm W} - \frac{1}{M} \biggl[ \frac{x-1}{x+1} \frac{d}{dx}
+ \frac{1}{(x+1)^2} \biggr] \xi_r^{\rm W \to RW}, \\
K^{\rm W \to RW} &= 2u^{\rm W} + g^{\rm W} - \frac{2}{M} \frac{x-1}{(x+1)^2} \xi_r^{\rm W \to RW}
+ \frac{1}{M^2} \frac{\ell(\ell+1)}{(x+1)^2} \xi^{\rm W \to RW}.
\end{align}
\end{subequations}
The calculation yields
\begin{subequations}
\begin{align}
u^{\rm W \to RW} &= \frac{(c+1)\lambda_\ell}{(\ell-1)(\ell+2)}
  \bigl[ -2x P'_\ell(x) + \ell(\ell+1) P_\ell(x) \bigr], \\
K^{\rm RW} &= \frac{2(c+1)\lambda_\ell}{(\ell-1)(\ell+2)}
  \bigl[ -2(x-1) P'_\ell(x) + \ell(\ell+1) P_\ell(x) \bigr]. 
\end{align}
\end{subequations}
Comparing with Eqs.~(\ref{tidal_RW}) and accounting for Eq.~(\ref{mu_def}), we see that we have a match provided that $c + 1 = (\ell+2)/\ell$, or $c = 2/\ell$.

This calculation informs us that a choice of $c$ corresponds to a choice of normalization for the tidal perturbation. A specific normalization was imposed on $u^{\rm RW}$, and it seemed as if the same normalization was adopted for $u^{\rm W}$. The normalization of $u^{\rm W}$, however, can be altered at will by the residual gauge freedom contained in the Weyl gauge; as we saw back in Eq.~(\ref{gauge_residual_changes2}), $\Delta u$ is of the same functional form as the tidal potential of Eq.~(\ref{u_tidalW}). The choice of normalization for $u^{\rm W}$ is therefore immaterial, and it is the constant $c$ that assumes responsibility for the physical normalization of the tidal perturbation.  

\subsection{Calibrated tidal potentials} 

As we saw, the choice $c = 2/\ell$ returns a tidal perturbation that matches the normalization adopted in the Regge-Wheeler gauge. Making this choice, the Weyl-gauge potentials become
\begin{subequations}
\begin{align}
u^{\rm Wcal} &= \lambda_\ell P_\ell(x), \\
g^{\rm Wcal} &= \frac{2\lambda_\ell}{\ell} \bigl[ (x^2-1) P'_\ell(x) - \ell x P_\ell(x) \bigr], \\
G^{\rm Wcal} &= -\frac{4\lambda_\ell}{(\ell-1)\ell^2} \bigl[ (x^2-1) P'_\ell(x) - \ell x P_\ell(x) \bigr].
\end{align}
\end{subequations}
The superscript ``Wcal'' indicates that the potentials are calibrated by their Regge-Wheeler counterparts, so that a gauge transformation returns a precise match for Eqs.~(\ref{tidal_RW}). A standard identity involving Legendre polynomials brings the potentials to the simpler form
\begin{subequations}
\begin{align}
u^{\rm Wcal} &= \lambda_\ell P_\ell(x), \\
g^{\rm Wcal} &= -2\lambda_\ell P_{\ell-1}(x), \\ 
G^{\rm Wcal} &= \frac{4\lambda_\ell}{(\ell-1)\ell} P_{\ell-1}(x). 
\end{align}
\end{subequations}

\subsection{Minimal gauge}

We observed that $u^{\rm W}$ can be normalized arbitrarily, and that the choice does not affect the physical description of the tidal perturbation. The residual gauge transformation
\begin{subequations}
\begin{align}
\Delta u &= -\lambda_\ell P_\ell(x), \\
\Delta g &= \lambda_\ell \bigl[ (x^2-1) P'_\ell(x) + 2x P_\ell(x) \bigr], \\
\Delta G &= -\frac{2\lambda_\ell}{\ell(\ell+1)} (x^2 - 1) P'_\ell(x),
\end{align} 
\end{subequations} 
drawn from Eqs.~(\ref{gauge_residual_changes2}), allows us to eliminate $u^{\rm W}$ altogether. In this refinement of the Weyl gauge, called here the ``minimal gauge'', the tidal perturbation is described by the potentials
\begin{subequations}
\begin{align}
u^{\rm min} &= 0, \\
g^{\rm min} &= (c+1) \lambda_\ell (x^2-1) P'_\ell(x), \\
G^{\rm min} &= \frac{2(c+1)\lambda_\ell}{(\ell-1)\ell(\ell+1)(\ell+2)} \bigl[
-(\ell^2+\ell+2) (x^2-1) P'_\ell(x) + 2\ell(\ell+1) x P_\ell(x) \bigr].
\end{align}
\end{subequations}
This expression makes the point rather clearly that the normalization of the tidal perturbation is governed by $c + 1$; the choice $c + 1 = (\ell+2)/\ell$ continues to provide a calibration against a description in Regge-Wheeler gauge.

\section{Application: Massless strings} 
\label{sec:strings} 

In this section we integrate the perturbation equations for a system of two massless strings attached to the black hole; there is no particle in the system. The first string extends along the upper $z$-axis, and has an energy density $\mu_{\rm up}$ equal to its tension $T_{\rm up}$; both are constant along the string. The second string is placed on the lower $z$-axis, and has a density $\mu_{\rm dn}$ equal to its tension $T_{\rm dn}$; these also are constant. We do not assume that $T_{\rm up} = T_{\rm dn}$, and shall see that unbalanced tensions produce an acceleration of the black hole in the perturbed spacetime.

The variables $u_\ell$ satisfy Eq.~(\ref{EFEu}) with a zero right-hand side, and $g_\ell$ must be a solution to Eq.~(\ref{EFEg}). For $\ell = 0$ we set $u_0 = c_1$ and $g_0 = c_2 + c_3 x$, where the $c_n$s are constants. The constraint equation (\ref{EFEG}) yields $c_2 = 2(T_{\rm up} + T_{\rm dn})$, and we choose $c_1 = c_3 = 0$. We therefore have
\begin{equation}
u_0 = 0, \qquad
g_0 = 2(T_{\rm up} + T_{\rm dn}). 
\end{equation}
For $\ell = 1$ we get that $u_1 = c_1 x$ and $g_1 = -2c_1 + c_2(x^2-1)$. The constraint implies that $c_1 = -\frac{1}{3} c_2 - (T_{\rm up} - T_{\rm dn})$. We choose $c_2 = 0$ and therefore obtain
\begin{equation}
u_1 = -(T_{\rm up} - T_{\rm dn}) x, \qquad
g_1 = 2 (T_{\rm up} - T_{\rm dn}). 
\end{equation}
We observe that the regularity condition of Eq.~(\ref{regularity_2M}) is satisfied. For $\ell \geq 2$ we simply set $u_\ell = g_\ell = 0$; nonzero values would describe a tidal deformation of the black hole, as described in Sec.~\ref{sec:tidal}. Equation (\ref{EFEG}) then returns
\begin{equation}
G_\ell = \frac{8(2\ell+1)}{(\ell-1)\ell(\ell+1)(\ell+2)} \bigl[ T_{\rm up} + (-1)^\ell T_{\rm dn} \bigr].
\end{equation}
It is easy to verify that this satisfies the condition of Eq.~(\ref{weylclass_Gcond}); the perturbation belongs to the Weyl class. 

The gravitational perturbation created by the massless strings is therefore described by the Weyl-class potentials
\begin{equation}
U = -(T_{\rm up} - T_{\rm dn})  (r/M - 1) \cos\theta, \qquad
\gamma = 2 (T_{\rm up} + T_{\rm dn}) + 2 (T_{\rm up} - T_{\rm dn}) \cos\theta.
\end{equation}
The line element is given by Eq.~(\ref{weylclass_metric}). This is a form of the $C$-metric (see Sec.~14.1 of Ref.~\cite{griffiths-podolsky:09}), linearized with respect to the tension parameters. We note that $\gamma(\theta = 0) = 4 T_{\rm up}$ while $\gamma(\theta = \pi) = 4 T_{\rm dn}$, in agreement with Eq.~(\ref{gamma_vs_T}). 

When $r \gg 2M$, the temporal component of the metric is given asymptotically by
\begin{equation}
g_{tt} = -2 \frac{T_{\rm up} - T_{\rm dn}}{M}\, r \cos\theta + O(1).
\end{equation}
The linear growth in $r\cos\theta$ indicates that the black hole is accelerated in the $z$-direction. The acceleration $a$ is given by $M a = T_{\rm up} - T_{\rm dn}$. As claimed, unbalanced tensions across the black hole give rise to an acceleration in the perturbed spacetime. 

\section{Application: Massive string; no black hole} 
\label{sec:massive-noBH} 

In this section we construct the linearized gravitational field of a single massive string; there is no black hole in the spacetime, and no particle. Our treatment is based on the field equations of Sec.~\ref{sec:EFE}, in which we set $M = 0$ and $m = 0$. Because there is no black hole and no particle, there is no distinction between the ``up'' and ``down'' strings; we write $\mu := \mu_{\rm up} = \mu_{\rm dn}$ and $T := T_{\rm up} = T_{\rm dn}$. 

\subsection{String model and field equations} 

Our model of a massive string is a simple one. We let 
\begin{equation} 
\sigma := \mu - T = \mbox{constant}.  
\end{equation} 
The conservation equations (\ref{conservation-up}) and (\ref{conservation-dn}), specialized to the case $M = 0$, imply that $T$ is a constant. In this section, therefore, the energy density and tension are taken to be unequal constants. 

Equation  (\ref{eq-u}) becomes 
\begin{equation}
r^2 \frac{d^2 u_\ell}{dr^2} + 2r \frac{d u_\ell}{dr} - \ell(\ell+1) u_\ell
= -(2\ell+1) \bigl[ 1 + (-1)^\ell \bigr] \sigma, 
\end{equation}
and the solution is 
\begin{equation} 
u_\ell = \frac{2\ell+1}{\ell(\ell+1)} \bigl[ 1 + (-1)^\ell \bigr] \sigma \qquad (\ell \neq 0). 
\end{equation} 
The freedom to add solutions to the homogeneous equation cannot be exercised, because these behave as $r^\ell$ and $r^{-(\ell+1)}$, which fail to be regular at either $r=0$ or $r=\infty$.  For $\ell = 0$ we have 
$u_0 = c_1 -2\sigma\, \ln r$. In this case we have the freedom to add a constant, which is denoted $c_1$. Its value will be chosen below. 

Equation (\ref{eq-g}) becomes 
\begin{equation}
r^2 \frac{d^2 g_\ell}{dr^2} - \ell(\ell+1) g_\ell = 0. 
\end{equation} 
The nontrivial solutions behave as $r^{-\ell}$ and $r^{\ell+1}$, and they must both be discarded when $\ell \neq 0$. We therefore have 
\begin{equation} 
g_\ell = 0 \qquad (\ell \neq 0). 
\end{equation} 
For $\ell = 0$ we eliminate the term proportional to $r$, and retain $g_0 = c_2$, where $c_2$ is a constant that will be determined presently. 

Equation (\ref{eq-G}) becomes 
\begin{equation}
\frac{1}{4}(\ell-1)\ell(\ell+1)(\ell+2) G_\ell = r \frac{d g_\ell}{dr}
-\frac{1}{2} \bigl[ \ell(\ell+1) + 2 \bigr] g_\ell
+ 2(2\ell+1) \bigl[ 1 + (-1)^\ell \bigr] T.  
\end{equation}
For $\ell = 0$ the equation returns $c_2 = 4 T$. For $\ell = 1$ it delivers $0 = 0$, and for $\ell \geq 2$ we obtain 
\begin{equation} 
G_\ell = \frac{8 (2\ell+1)}{(\ell-1)\ell(\ell+1)(\ell+2)} \bigl[ 1 + (-1)^\ell \bigr] T. 
\end{equation} 
We then find that Eq.~(\ref{weylclass_G}) is satisfied, because $S_\ell$ is equal to $g_0 = 4T$ when $\ell$ is even, and vanishes when $\ell$ is odd. The perturbation therefore belongs to the Weyl class. 

\subsection{Potentials and metric} 

The sum of Eq.~(\ref{Ug_decomp}) is carried out with the help of Eqs.~(\ref{SF0}) and (\ref{SF00}). We obtain 
\begin{equation} 
U = U_0 - \sigma \ln (r^2\sin^2\theta), 
\end{equation} 
where $U_0 := c_1 - 2\sigma(1-\ln 2)$. The sum for $\gamma$ is immediate, and we get 
\begin{equation} 
\gamma = 4 T. 
\end{equation} 
This assignment gives us a special case of Eq.~(\ref{gamma_vs_T}). 

We insert the potentials within the metric of Eq.~(\ref{weylclass_metric}), in which we set $f = 1$. The factors of $e^{\pm 2U_0}$ can be eliminated with a rescaling of the coordinates, $t \to e^{-U_0} t$ and $r \to e^{U_0} r$. This gives us the freedom to set $U_0 = 0$, which provides a choice of constant $c_1$. The metric becomes 
\begin{equation} 
ds^2 = -\bigl[ 1 + 2\sigma \ln(r^2\sin^2\theta) \bigr]\, dt^2 
+ \bigl[ 1 - 2\sigma \ln(r^2\sin^2\theta) + 8T \bigr]\, (dr^2 + r^2\, d\theta^2) 
+ \bigl[ 1 - 2\sigma \ln(r^2\sin^2\theta) \bigr] r^2\sin^2\theta\, d\phi^2. 
\end{equation} 
A transformation to cylindrical coordinates $\rho = r\sin\theta$, $z = r\cos\theta$ produces 
\begin{equation} 
ds^2 = -\bigl( 1 + 2\sigma \ln\rho^2 \bigr)\, dt^2 
+ \bigl( 1 - 2\sigma \ln\rho^2 + 8T \bigr)\, (d\rho^2 + dz^2)
+ \bigl( 1 - 2\sigma \ln\rho^2 \bigr)\rho^2\, d\phi^2. 
\end{equation} 
If we restore the original exponential notation, this is 
\begin{equation} 
ds^2 = -\rho^{4\sigma}\, dt^2 + e^{8T} \rho^{-4\sigma} (d\rho^2 + dz^2)
+ \rho^{2(1-2\sigma)}\, d\phi^2, 
\label{LC_lin}
\end{equation} 
with $\sigma$ and $T$ both considered to be small. As expected for an infinitely long, massive string, the metric is singular on the axis $(\rho = 0)$ and at infinity $(\rho = \infty)$.  

\subsection{Levi-Civita metric} 

It is easy to promote the metric of Eq.~(\ref{LC_lin}) to an exact solution to the vacuum field equations,  away from $\rho = 0$ and $\rho = \infty$. The result is the Levi-Civita metric, reviewed in Sec.~10.2 of Ref.~\cite{griffiths-podolsky:09}. 

The Levi-Civita solution is obtained by imposing a cylindrical symmetry on the Weyl metric of Eq.~(\ref{weyl_original}). Setting $U = U(\rho)$, we find that the field equations return $U = -\sigma \ln\rho^2$ up to the addition of an irrelevant constant; this agrees with our previous expression. Setting $\gamma = \gamma(\rho)$ we also get that $\gamma = \gamma_0 + 2\sigma^2 \ln\rho^2$, where $\gamma_0$ is a constant; this agrees with our previous result when we neglect the second term and make the association $\gamma_0 = 4T$. 

Inserting these results within the metric, we arrive at 
\begin{equation} 
ds^2 = -\rho^{4\sigma}\, dt^2 + e^{2\gamma_0} \rho^{-4\sigma(1-2\sigma)} (d\rho^2 + dz^2)
+ \rho^{2(1-2\sigma)}\, d\phi^2,  
\label{LC_exact}
\end{equation} 
the exact version of Eq.~(\ref{LC_lin}). While the Levi-Civita metric does not come with an immediate interpretation for the parameters $\sigma$ and $\gamma_0$, its linearized version makes explicit contact with our model of a massive string. 

\section{Application: Particle and massless string} 
\label{sec:particle-massless}

In this section we consider a particle of mass $m$ held in place at $r = r_0$ outside a Schwarzschild black hole. The particle is tied to a massless string with tension $T_{\rm up}$, and the black hole is attached to another string with tension $T_{\rm dn}$. The tension in each string is constant, and we assume from the outset that $T_{\rm up} = T_{\rm dn}$, so that the black hole is not accelerated in the perturbed spacetime; the tensions are henceforth denoted $T$. 

\subsection{Field equations and solutions} 

The field equations of Sec.~\ref{sec:EFE} for $u_\ell$ and $g_\ell$ become 
\begin{subequations}
\label{EFEparticle}
\begin{align}
(x^2-1) u''_\ell + 2 x u'_\ell - \ell(\ell+1) u_\ell
&= -(2\ell+1) k\, \delta(x-x_0),
\label{EFEu_particle} \\
(x^2 - 1) g''_\ell - \ell(\ell+1) g_\ell &= 4u'_\ell,
\label{EFEg_particle}
\end{align} 
\end{subequations}
where $x := r/M - 1$, $x_0 := r_0/M - 1$, and $k := (m/M) \sqrt{f_0}$ with $f_0 = (x_0-1)/(x_0+1)$. These equations are accompanied by Eqs.~(\ref{EFEG}) and (\ref{Shat_def}), which provide constraints when $\ell = 0, 1$, and which determine $G_\ell$ when $\ell \geq 2$. We recall that the string tension $T$ is given by Eq.~(\ref{T0}); this becomes 
\begin{equation} 
T = \frac{k}{x_0^2-1} 
\label{T_vs_k} 
\end{equation} 
when expressed in terms of $k$ and $x_0$. 

The solution for $u_\ell$ is obtained by inserting $W_\ell = -(2\ell+1) k\, \delta(x-x_0)$ within Eq.~(\ref{u_part}). This gives 
\begin{equation} 
u_\ell = (2\ell+1) k \left\{ 
\begin{array}{ll} 
Q_\ell(x_0) P_\ell(x)  & \quad x < x_0 \\ 
P_\ell(x_0) Q_\ell(x)  & \quad x > x_0 
\end{array}
\right. . 
\label{u_particle} 
\end{equation} 
The solution for $g_\ell$ is found by substituting Eq.~(\ref{u_particle}) into Eq.~(\ref{g_part}). A particular solution to Eq.~(\ref{EFEg_particle}) is
\begin{equation} 
g^{\rm part}_\ell = \frac{4(2\ell - 1)}{\ell(\ell+1)} k (x^2-1) \biggl[ 
Q'_\ell(x) \int_1^x P'_\ell(x') V_\ell(x')\, dx' 
+ P'_\ell(x) \int_x^\infty Q'_\ell(x') V_\ell(x')\, dx \biggr], 
\end{equation} 
where 
\begin{equation} 
V_\ell(x') := \left\{ 
\begin{array}{ll} 
Q_\ell(x_0) P'_\ell(x')  & \quad x' < x_0 \\ 
P_\ell(x_0) Q'_\ell(x')  & \quad x' > x_0 
\end{array}
\right. . 
\end{equation} 
When $x < x_0$ the first integral involves the $x' < x_0$ member of $V_\ell$; the second integral is broken up into two domains, the first from $x$ to $x_0$ involving also the $x' < x_0$ member of $V_\ell$, and the second from $x_0$ to $\infty$ involving the $x' > x_0$ member of $V_\ell$. When $x > x_0$, the first integral is broken up into one from $1$ to $x_0$ involving the $x' < x_0$ member of $V_\ell$, and another from $x_0$ to $x$ involving the $x' > x_0$ member of $V_\ell$; the second integral implicates the $x' > x_0$ member of $V_\ell$. Each integral is of the form given in Appendix~\ref{sec:integrals}, and after a fairly long computation we obtain 
\begin{equation} 
g^{\rm part}_\ell = -\frac{2(2\ell+1)}{\ell(\ell+1)} k \left\{ 
\begin{array}{ll} 
\ell(\ell+1) Q_\ell(x_0) x P_\ell(x) + \ell(\ell+1) Q_\ell(x_0) (x^2-1) Q'_\ell(x) 
+ x_0 Q'_\ell(x_0) (x^2-1) P'_\ell(x) & \quad x < x_0 \\ 
\ell(\ell+1) P_\ell(x_0) x Q_\ell(x) + \ell(\ell+1) Q_\ell(x_0) (x^2-1) Q'_\ell(x) 
+ x_0 P'_\ell(x_0) (x^2-1) Q'_\ell(x) & \quad x > x_0 
\end{array} 
\right. . 
\end{equation} 
The result was simplified by making repeated use of the Wronskian identity $P_\ell(x) Q'_\ell(x) - P'_\ell(x) Q_\ell(x) = -(x^2-1)^{-1}$. 

We observe that the term proportional to $(x^2-1) Q'_\ell(x)$ is common to both members of $g^{\rm part}_\ell$, and that it fails to be smooth at $x = 1$. We eliminate this term by subtracting a corresponding solution to the homogeneous equation for $g_\ell$, as described by Eq.~(\ref{g_homo}). The physical solution is therefore 
\begin{equation} 
g_\ell = -\frac{2(2\ell+1)}{\ell(\ell+1)} k \left\{ 
\begin{array}{ll} 
\ell(\ell+1) Q_\ell(x_0) x P_\ell(x) + x_0 Q'_\ell(x_0) (x^2-1) P'_\ell(x) & \quad x < x_0 \\ 
\ell(\ell+1) P_\ell(x_0) x Q_\ell(x) + x_0 P'_\ell(x_0) (x^2-1) Q'_\ell(x) & \quad x > x_0 
\end{array} 
\right. . 
\label{g_particle} 
\end{equation} 
It can be verified that $g_\ell(x)$ is continuous and differentiable at $x = x_0$; its second derivative, however, is discontinuous, in view of the discontinuity in $u'_\ell$. It can also be verified that the regularity condition of Eq.~(\ref{regularity_2M}) is satisfied. Furthermore, the constraint of Eq.~(\ref{EFEG}) is enforced when $\ell = 1$.  

The solution of Eq.~(\ref{g_particle}) does not apply when $\ell = 0$. In this case we have that $u_0 = k Q_0(x_0)$ when $x < x_0$ and $u_0 = k Q_0(x)$ when $x > x_0$, where $Q_0(x) = -\frac{1}{2} \ln[(x-1)/(x+1)]$. When $x < x_0$ the solution to Eq.~(\ref{EFEg_particle}) is $g = c_1 + c_2 x$, where $c_1$ and $c_2$ are constants. When $x > x_0$ the solution is instead $g = c_3 + c_4 x + kx \ln[(x-1)/(x+1)]$. Continuity and differentiability at $x=x_0$ determines two of the four constants, which we pick to be $c_2$ and $c_3$. The constraint of Eq.~(\ref{EFEG}) then allows us to determine $c_1$, which is given by $c_1 = 2T = 2k/(x_0^2-1)$. The fourth constant, $c_4$, remains arbitrary, and we set it to zero to avoid a linear growth of $g$ when $x > x_0$. With all this, we find that 
\begin{equation} 
g_0 = \frac{2k}{x_0^2-1} + k \biggl( \ln\frac{x_0-1}{x_0+1} + \frac{2x_0}{x_0^2-1} \biggr) x 
\label{g0_small} 
\end{equation} 
when $x < x_0$, and 
\begin{equation} 
g_0 = \frac{2k(x_0^2+1)}{x_0^2-1} + kx \ln\frac{x-1}{x+1} 
\label{g0_large} 
\end{equation} 
when $x > x_0$.   

With $u_\ell$ and $g_\ell$ thus determined, Eq.~(\ref{EFEG}) provides expressions for $G_\ell$ when $\ell \geq 2$. We follow the strategy described at the end of Sec.~\ref{sec:EFE} to establish that the perturbation belongs to the Weyl class. First, we compute $\hat{S}_\ell$ according to Eq.~(\ref{Shat_def}), and get 
\begin{equation} 
\hat{S}_\ell 
= 2k \left\{ 
\begin{array}{ll} 
-Q'_{\ell-1}(x_0)\, P_{\ell-1}(x) + (-1)^\ell (x_0^2 - 1)^{-1} & \quad x < x_0 \\ 
-P'_{\ell-1}(x_0)\, Q_{\ell-1}(x) + [1 + (-1)^\ell] (x_0^2 - 1)^{-1} & \quad x > x_0 
\end{array} 
\right.. 
\end{equation} 
Second, we verify that $\hat{S}_2 = g_0$ and $\hat{S}_3 = g_1$, so that $\hat{S}_\ell$ satisfies the initial conditions of Eq.~(\ref{Shat_initial}). And third, we compute 
\begin{equation} 
\hat{S}_{\ell+2} - \hat{S}_\ell = -\frac{2(2\ell+1)}{\ell(\ell+1)} k \left\{ 
\begin{array}{ll} 
\ell(\ell+1) Q_\ell(x_0) x P_\ell(x) + x_0 Q'_\ell(x_0) (x^2-1) P'_\ell(x) & \quad x < x_0 \\ 
\ell(\ell+1) P_\ell(x_0) x Q_\ell(x) + x_0 P'_\ell(x_0) (x^2-1) Q'_\ell(x) & \quad x > x_0 
\end{array} 
\right., 
\end{equation} 
and thereby prove that $\hat{S}_\ell$ satisfies the recursion relation of Eq.~(\ref{Shat_recursion}). The perturbation does indeed belong to the Weyl class. 

\subsection{Summed potentials} 

The potentials $U(r,\theta)$ and $\gamma(r,\theta)$ are obtained from Eq.~(\ref{Ug_decomp}), with $u_\ell$ given by Eq.~(\ref{u_particle}) and $g_\ell$ by Eqs.~(\ref{g_particle}), (\ref{g0_small}), and (\ref{g0_large}). The sums can be evaluated with formulae developed in Appendix~\ref{sec:summation}.  

For $U$ we get 
\begin{equation} 
U = \frac{k}{D} 
\label{U_particle} 
\end{equation} 
with the help of Eq.~(\ref{SF1}), where $k := (m/M) \sqrt{f_0}$ and 
\begin{equation} 
D := \bigl( x^2 - 2x_0 x \cos\theta + x_0^2 - \sin^2\theta \bigr)^{1/2}
\label{D_def1} 
\end{equation} 
is the spatial distance between $x$ and $x_0$ in the Schwarzschild spacetime. This is immediately recognized as the potential of a point particle of mass $m$ and Killing energy $m \sqrt{f_0}$.  

For $\gamma$, the sum over $\ell$ must separate out the contribution from $\ell = 0$; we therefore write 
\begin{equation} 
\gamma = g_0 + \sum_{\ell=1}^\infty g_\ell(x)\, P_\ell(\cos\theta). 
\end{equation} 
The sum involves two sets of terms, one involving the product $Q_\ell(x_0) P_\ell(x)$ or $P_\ell(x_0) Q_\ell(x)$, the other involving $Q'_\ell(x_0) P'_\ell(x)$ or $P'_\ell(x_0) Q'_\ell(x)$. The sum over the first set of terms is handled with Eq.~(\ref{SF1}), properly written so that the sum begins at $\ell = 1$. The sum over the second set is handled with Eq.~(\ref{SF2}). After some simplifying algebra, we arrive at 
\begin{equation} 
\gamma =\frac{2k}{x_0^2 - 1} \biggl( \frac{x - x_0\cos\theta}{D} + 1 \biggr). 
\label{gamma_particle} 
\end{equation} 
We recall that the string tension is $T = k/(x_0^2-1)$. 

From Eq.~(\ref{gamma_particle}) we infer that 
\begin{subequations}
\begin{align} 
\gamma(x,\theta=0) &= \left\{ 
\begin{array}{ll} 
0 & \quad x < x_0 \\ 
4T & \quad x > x_0 
\end{array} 
\right., \\
\gamma(x,\theta=\pi) &= 4 T. 
\end{align}
\end{subequations} 
This reveals the existence of a conical singularity on the upper axis when $x > x_0$ (above the particle), and everywhere on the lower axis. The field equations imply that $\gamma^{\rm axis}$ is either zero (between black hole and particle) or equal to $4 T$ (everywhere else). This is a special case of Eq.~(\ref{gamma_vs_T}). 

\subsection{Black-hole properties and first law} 
\label{subsec:firstlaw} 

The metric of a particle of mass $m$ held in place at position $r = r_0$ outside a Schwarzschild black hole of mass $M$ is given by Eq.~(\ref{weylclass_metric}), with the potentials of Eqs.~(\ref{U_particle}) and (\ref{gamma_particle}); both particle and black hole are supported with a massless string of tension $T = k/(x_0^2-1)$, where $k := (m/M) [(x_0-1)/(x_0+1)]^{1/2}$. We recall that $x := r/M - 1$ and $x_0 := r_0/M - 1$. 

The Komar mass associated with a closed 2-surface $S$ is (see, for example, Sec.~4.3.3 of Ref.~\cite{poisson:b04})
\begin{equation} 
M_{\rm K}(S) = \frac{1}{4\pi} \oint_S \nabla^\alpha t^\beta\, n_\alpha r_\beta\, dS, 
\end{equation} 
where $t^\alpha$ is the timelike Killing vector, $n_\alpha$ the surface's unit timelike normal, $r_\alpha$ its unit spacelike normal, and $dS$ the element of surface area. We take $S$ to be a surface of constant $t$ and $r$. In the $(t,r,\theta,\phi)$ coordinates, we have that $t^\alpha = (1,0,0,0)$, $n_\alpha = e^{-U} f^{1/2}(-1,0,0,0)$, $r_\alpha = e^{U+\gamma} f^{-1/2}(0,1,0,0)$, and $dS = e^{2U+\gamma} r^2\sin\theta\, d\theta d\phi$. Evaluation of the integral gives 
\begin{equation} 
M_{\rm K}(r) = M - r^2 f \langle \partial_r U \rangle, 
\label{komar1} 
\end{equation} 
where  
\begin{equation} 
\langle \partial_r U \rangle := \frac{1}{2} \int_0^\pi \partial_r U\, \sin\theta\, d\theta 
\label{komar2}
\end{equation} 
is the average of $\partial_r U$ over the 2-surface. For large $r$ we obtain 
\begin{equation} 
M_{\rm K}(r) = M + k M + O(r^{-1}), 
\end{equation} 
in which we recognize $k M = m[(x_0-1)/(x_0+1)]^{1/2} = m(1-2M/r_0)^{1/2}$ as the particle's Killing energy in the Schwarzschild spacetime. The $r \to \infty$ limit of the Komar mass defines the total mass of the spacetime: 
\begin{equation} 
M_{\rm tot} = M + k M. 
\label{Mtot_massless} 
\end{equation} 
This coincides with its Arnowitt-Deser-Misner mass. 

With the metric of Eq.~(\ref{weylclass_metric}), the event horizon of the perturbed black hole is still situated at $r = 2M$, or $x = 1$, where $g_{tt} = 0$. The field equation (\ref{Weylclass_EFEc}), evaluated on the horizon, produces $\partial_\theta(2U + \gamma) = 0$. It follows that 
\begin{equation} 
\beta := e^{2U+\gamma} \Bigr|_{r = 2M} = \mbox{constant, independent of $\theta$}. 
\label{beta_def}
\end{equation} 
As we shall see presently, this observation is behind the validity of the zeroth law of black-hole mechanics in this class of spacetimes. 

The element of surface area on the event horizon is $dS = (2M)^2 \beta\sin\theta\, d\theta d\phi$. Integration yields 
\begin{equation} 
A = 4\pi (2M)^2 \beta 
\label{area} 
\end{equation} 
for the black-hole area. The surface gravity $\kappa$ is obtained from $\kappa^2 = -\frac{1}{2} (\nabla_\alpha t_\beta) (\nabla^\alpha t^\beta)$, with the right-hand side evaluated at $r = 2M$. This gives 
\begin{equation} 
\kappa = \frac{1}{4M\beta}. 
\label{kappa} 
\end{equation} 
As was previously stated, Eq.~(\ref{beta_def}) ensures that $\kappa$ is uniform on the horizon, as required by the zeroth law. The Smarr mass $M_{\rm Smarr} := \kappa A/(4\pi)$ evaluates to 
\begin{equation} 
M_{\rm Smarr} = M.  
\label{Smarr} 
\end{equation} 
The results of Eqs.~(\ref{beta_def}), (\ref{area}), (\ref{kappa}), and (\ref{Smarr}) hold for any metric of the form of Eq.~(\ref{weylclass_metric}). In fact, they are {\it exact consequences} of this metric, which remain true even when $U$ and $\gamma$ are no longer assumed to be small. 

For the application at hand, and the potentials of Eqs.~(\ref{U_particle}) and (\ref{gamma_particle}), we find that 
\begin{equation} 
\beta = 1 + \frac{2k}{x_0-1}. 
\end{equation} 
This result is perturbative in $m$, which hides within the constant $k$. 

The quantities $M_{\rm tot}$, $A$, and $T$ are functions of the parameters $M$, $m$, and $x_0$. As a matter of mathematical identity we have that 
\begin{subequations} 
\begin{align} 
dM_{\rm tot} &= \frac{\partial M_{\rm tot}}{\partial M}\, dM 
+ \frac{\partial M_{\rm tot}}{\partial m}\, dm
+ \frac{\partial M_{\rm tot}}{\partial x_0}\, dx_0, \\ 
dA &= \frac{\partial A}{\partial M}\, dM 
+ \frac{\partial A}{\partial m}\, dm
+ \frac{\partial A}{\partial x_0}\, dx_0, \\ 
dT &= \frac{\partial T}{\partial M}\, dM 
+ \frac{\partial T}{\partial m}\, dm
+ \frac{\partial T}{\partial x_0}\, dx_0,
\end{align} 
\end{subequations} 
and an examination of these relations reveals that 
\begin{equation} 
d M_{\rm tot} = \frac{\kappa}{8\pi}\, dA  - \lambda\, dT + z\, dm, 
\label{firstlaw_massless}
\end{equation} 
where 
\begin{equation} 
\lambda := M(x_0 + 1) = r_0 
\label{lambda} 
\end{equation} 
is the string's ``thermodynamic length'', and 
\begin{equation} 
z := \sqrt{\frac{x_0-1}{x_0+1}} = \sqrt{1 - 2M/r_0} 
\label{redshift}
\end{equation} 
is the redshift factor for a photon emitted at $r = r_0$ and received at infinity in the Schwarzschild spacetime. Equation (\ref{firstlaw_massless}) is the first law of black-hole mechanics for the perturbed spacetime. It expresses the fact that $M_{\rm tot}$ can be viewed as a function of $A$, $T$, and $m$ instead of as a function of $M$, $m$, and $x_0$.  

It might be noted that in Eq.~(\ref{firstlaw_massless}), the quantities $dT$ and $dm$ are perturbations, while $dA$ and $dM_{\rm tot}$ contain background and perturbative terms. It follows from this observation that $\kappa$ also contains background and perturbative terms, but that $\lambda$ and $z$ are purely background quantities, defined in the Schwarzschild spacetime.  

\section{Application: Particle and massive string} 
\label{sec:particle-massive}

In this section we continue to place a particle of mass $m$ at position $r = r_0$ outside a Schwarzschild black hole, but we now replace the massless upper string of Sec.~\ref{sec:particle-massless} with a massive string.  This shall have an energy density $\mu_{\rm up}$ and tension $T_{\rm up}$ that are not equal to each other. We recall that the string tension is subjected to the conservation equation (\ref{conservation-up}) and the boundary condition of Eq.~(\ref{T0}). We continue to hold the black hole with a massless string, so that $\mu_{\rm dn} = T_{\rm dn} = \mbox{constant}$.  

\subsection{Source superposition} 
\label{subsec:difference} 

In Sec.~\ref{sec:particle-massless} we found the gravitational field of a particle of mass $m$ at $r = r_0$, held in place by a massless string with 
\begin{equation} 
\mu^{\rm massless}_{\rm up} = T^{\rm massless}_{\rm up} =\frac{k}{x_0^2 - 1}, \qquad 
k = \frac{m}{M} \sqrt{\frac{x_0-1}{x_0+1}},
\end{equation} 
where $x_0 := r_0/M - 1$. The black hole also was held with a massless string, with 
\begin{equation} 
\mu^{\rm massless}_{\rm dn} = T^{\rm massless}_{\rm dn} = \frac{k}{x_0^2 - 1}. 
\end{equation} 
The balanced tensions ensured that the black hole was unaccelerated in the perturbed spacetime. 

We make good use of this solution when we replace the massless upper string with a massive one. The idea is to write 
\begin{equation} 
\mbox{particle} + \mbox{massive string} = \bigl( \mbox{particle} + \mbox{massless string} \bigr) 
+ \bigl( \mbox{massive string} - \mbox{massless string} \bigr) 
\end{equation} 
and to exploit the superposition principle afforded by the linearized field equations. Because we already have the solution to the first problem on the right-hand side of the equation, we can place our attention entirely on the second one, and add the solutions in the final step. For this second problem we have no particle, but we have a ``massive difference string'' on the upper axis, extending from $r = r_0$ to infinity, and a ``massless difference string'' everywhere on the lower axis.  

The energy density and tension of the actual massive string are denoted $\mu^{\rm massive}_{\rm up}$ and $T^{\rm massive}_{\rm up}$, respectively. The density and tension of the upper difference string are then 
\begin{equation} 
\mu^{\rm diff}_{\rm up} = \mu^{\rm massive}_{\rm up} - \frac{k}{x_0^2-1}, \qquad
T^{\rm diff}_{\rm up} = T^{\rm massive}_{\rm up} - \frac{k}{x_0^2-1}. 
\end{equation} 
By virtue of Eqs.~(\ref{conservation-up}) and (\ref{T0}), we have that 
\begin{equation} 
\frac{d T^{\rm diff}_{\rm up}}{dx} = \frac{\mu^{\rm diff}_{\rm up} - T^{\rm diff}_{\rm up}}{x^2 -1}, \qquad 
T^{\rm diff}_{\rm up}(x=x_0) = 0, 
\label{cons-Tdiff} 
\end{equation} 
where $x := r/M - 1$. On the other hand, the actual string that holds the black hole is still massless, with an energy density $\mu^{\rm massless}_{\rm dn}$ and a tension $T^{\rm massless}_{\rm dn}$ that are equal to each other. The density and tension of the lower difference string are then 
\begin{equation} 
\mu^{\rm diff}_{\rm dn} = T^{\rm diff}_{\rm dn} = T^{\rm massless}_{\rm dn} - \frac{k}{x_0^2-1}. 
\end{equation} 
The conservation equation implies that $T^{\rm diff}_{\rm dn}$ is constant along the string. 

Our focus from this point on shall be on the ``difference strings'', and to unclutter the notation we shall write 
\begin{equation} 
\mu_{\rm up} := \mu^{\rm diff}_{\rm up}, \qquad 
T_{\rm up} := T^{\rm diff}_{\rm up}, \qquad 
\mu_{\rm dn} := \mu^{\rm diff}_{\rm dn}, \qquad 
T_{\rm dn} := T^{\rm diff}_{\rm dn}. 
\end{equation} 
We adopt the model of a massive string introduced in Sec.~\ref{sec:massive-noBH}, and set 
\begin{equation} 
\sigma := \mu_{\rm up} - T_{\rm up} = \mbox{constant}. 
\end{equation} 
For this model, Eq.~(\ref{cons-Tdiff}) produces 
\begin{equation} 
T_{\rm up} = \frac{1}{2} \sigma \biggl( \ln\frac{x-1}{x+1} 
+ \ln\frac{x_0+1}{x_0-1} \biggr).
\label{Tup_massive} 
\end{equation} 
We expect that the black hole will remain unaccelerated when the string tensions are balanced at infinity. Based on this expectation, we anticipate that the field equations will demand that 
\begin{equation} 
T_{\rm dn} = \frac{1}{2} \sigma \ln\frac{x_0+1}{x_0-1}. 
\label{Tdn_massive} 
\end{equation} 
We shall see that this is indeed the correct expression for the tension in the lower difference string. 

\subsection{Field equations and solutions} 

With the choices made in the preceding subsection, we find that the perturbation equations of Sec.~\ref{sec:EFE} for $u_\ell$ and $g_\ell$ become 
\begin{subequations}
\label{EFE_diff}
\begin{align}
(x^2-1) u''_\ell + 2 x u'_\ell - \ell(\ell+1) u_\ell
&= -(2\ell+1) \sigma \Theta(x-x_0), 
\label{EFE_diff_u} \\
(x^2 - 1) g''_\ell - \ell(\ell+1) g_\ell &= 4u'_\ell. 
\label{EFE_diff_g} 
\end{align} 
\end{subequations}
These equations are accompanied by Eqs.~(\ref{EFEG}) and (\ref{Shat_def}), which provide constraints when $\ell = 0, 1$, and which determine $G_\ell$ when $\ell \geq 2$. It is useful to note that the equations are very similar to Eqs.~(\ref{EFEparticle}), with the step function associated with the massive string replacing the delta function associated with the particle. This observation implies that a solution to the ($\mbox{particle} + \mbox{massless string}$) problem can be recovered from a solution to the massive string problem, by differentiating with respect to $x_0$, inserting a minus sign, and replacing $\sigma$ with $k$. This correspondence will be used below as a check on our solutions.  

The techniques developed in Sec.~\ref{sec:EFE} and exploited in Sec.~\ref{sec:particle-massless} can be put to the task of integrating Eqs.~(\ref{EFE_diff}). Our experience with these techniques, however, allows us to identify some shortcuts. 

Equation (\ref{EFE_diff_u}) is homogeneous when $x < x_0$, and according to Eq.~(\ref{u_homo}), its solution must be a superposition of Legendre functions; regularity at the horizon requires the elimination of $Q_\ell(x)$. When $x > x_0$, a particular solution to Eq.~(\ref{EFE_diff_u}) is the constant $(2\ell+1) \sigma/[\ell(\ell+1)]$, and to this we may add any solution to the homogeneous equation; regularity at infinity requires the elimination of $P_\ell(x)$. The global solution to Eq.~(\ref{EFE_diff_u}) must be continuous and differentiable at $x = x_0$. Combining all these requirements, we obtain 
\begin{subequations} 
\label{u_massive} 
\begin{align} 
u_\ell^< &= -\sigma \frac{2\ell+1}{\ell(\ell+1)} (x_0^2-1) Q_\ell'(x_0)\, P_\ell(x), \\ 
u_\ell^> &= -\sigma \frac{2\ell+1}{\ell(\ell+1)} \Bigl[ (x_0^2-1) P_\ell'(x_0)\, Q_\ell(x) - 1 \Bigr]. 
\end{align} 
\end{subequations} 
Here $u_\ell^<$ is the solution for $x < x_0$, while $u_\ell^>$ is the solution for $x > x_0$. To arrive at these results we made use of the Wronskian identity $P_\ell(x) Q'_\ell(x) - P'_\ell(x) Q_\ell(x) = -(x^2-1)^{-1}$. It is easy to verify that differentiation with respect to $x_0$ returns Eq.~(\ref{u_particle}), as was explained previously. 

The preceding results do not apply when $\ell = 0$, and this case requires a separate treatment. The only change is in the particular solution when $x > x_0$, which becomes $-\frac{1}{2} \sigma \ln(x^2-1)$. Following the same steps as before, we find that the global solution is given by
\begin{subequations} 
\label{u_massive_L0} 
\begin{align} 
u_0^< &= 0, \\ 
u_0^> &= \frac{1}{2} \sigma \biggl[ (x_0-1) \ln\frac{x-1}{x_0-1} - (x_0+1) \ln\frac{x+1}{x_0+1} \biggr].  
\end{align} 
\end{subequations} 
The solution is defined up to the addition of an overall constant. Here we chose the constant so that $u_0$ vanishes when $x < x_0$. Other choices are of course possible, and this issue will be re-examined at a later stage. 

We next turn to Eq.~(\ref{EFE_diff_g}). With $u_\ell(x)$ given by Eq.~(\ref{u_massive}), it is easy to check that 
\begin{equation} 
g^{\rm part}_\ell = 2\sigma \frac{2\ell+1}{\ell(\ell+1)} \left\{ 
\begin{array}{ll} 
(x_0^2-1) Q_\ell'(x_0)\, x P_\ell(x) & \quad x < x_0 \\ 
(x_0^2-1) P_\ell'(x_0)\, x Q_\ell(x) & \quad x > x_0
\end{array} \right. 
\end{equation} 
is a particular solution to the equation. To this we may add solutions to the homogeneous equation, which were identified in Eq.~(\ref{g_homo}). The global solution is identified by imposing smoothness at the horizon and regularity at infinity, as well as continuity and differentiability at $x = x_0$. We arrive at 
\begin{subequations} 
\label{g_massive} 
\begin{align} 
g_\ell^< &= 2\sigma \frac{2\ell+1}{\ell(\ell+1)} \biggl\{ (x_0^2-1) Q'_\ell(x_0)\, x P_\ell(x) 
+ \Bigl[ x_0 Q_\ell(x_0) - \frac{1}{\ell(\ell+1)} (x_0^2-1) Q'_\ell(x_0) \Bigr] (x^2-1) P'_\ell(x) \biggr\}, \\ 
g_\ell^> &= 2\sigma \frac{2\ell+1}{\ell(\ell+1)} \biggl\{ (x_0^2-1) P'_\ell(x_0)\, x Q_\ell(x) 
+ \Bigl[ x_0 P_\ell(x_0) - \frac{1}{\ell(\ell+1)} (x_0^2-1) P'_\ell(x_0) \Bigr] (x^2-1) Q'_\ell(x) \biggr\}. 
\end{align}
\end{subequations} 
It can be verified that Eq.~(\ref{g_particle}) is recovered after differentiation with respect to $x_0$. A little more work reveals that the functions can be simplified to 
\begin{subequations} 
\label{g_massive_alt} 
\begin{align} 
g_\ell^< &= 2\sigma \bigl[ Q_{\ell+1}(x_0)\, P_{\ell+1}(x) - Q_{\ell-1}(x_0)\, P_{\ell-1}(x) \bigr], \\ 
g_\ell^> &= 2\sigma \bigl[P_{\ell+1}(x_0)\, Q_{\ell+1}(x) - P_{\ell-1}(x_0)\, Q_{\ell-1}(x) \bigr]. 
\end{align}
\end{subequations} 
The translation involves extensive use of identities satisfied by Legendre functions. 

Again the case $\ell = 0$ requires a separate treatment. We obtain 
\begin{subequations} 
\label{g_massive_L0_tmp} 
\begin{align} 
g_0^< &= -\sigma \Bigl[ (x_0 x - 1) \ln\frac{x_0-1}{x_0+1} + 2x + c \Bigr], \\ 
g_0^> &= -\sigma \Bigl[ (x_0 x - 1) \ln\frac{x-1}{x+1} + 2x_0 + c \Bigr],
\end{align} 
\end{subequations} 
where $c$ is a constant that will be determined presently. 

To complete the solution we must impose the constraints contained in Eq.~(\ref{EFEG}). For $\ell = 0$ the equation returns 
\begin{equation} 
\sigma \biggl( c - \ln\frac{x_0-1}{x_0+1} \biggr) + 2 T_{\rm dn} = 0, 
\end{equation} 
and for $\ell = 1$ we get 
\begin{equation}  
\sigma \ln\frac{x_0-1}{x_0+1} + 2 T_{\rm dn} = 0. 
\end{equation} 
The second equation determines $T_{\rm dn}$, and we recover the statement of Eq.~(\ref{Tdn_massive}). The first equation then implies that $c = 2\ln[(x_0-1)/(x_0+1)]$. Making the substitution in Eq.~(\ref{g_massive_L0_tmp}), we find that $g_0$ becomes 
\begin{subequations} 
\label{g_massive_L0} 
\begin{align} 
g_0^< &= -\sigma \biggl[ (x_0 x + 1) \ln\frac{x_0-1}{x_0+1} + 2x \biggr], \\ 
g_0^> &= -\sigma \biggl[ (x_0 x - 1) \ln\frac{x-1}{x+1} + 2 \ln\frac{x_0-1}{x_0+1} + 2x_0 \biggr]. 
\end{align} 
\end{subequations} 
Differentiation with respect to $x_0$ produces Eqs.~(\ref{g0_small}) and (\ref{g0_large}). For $\ell \geq 2$, Eq.~(\ref{EFEG}) produces an explicit expression for $G_\ell$. There is no need to display this here. 

The next order of business is to show that the perturbation belongs to the Weyl class. We compute $\hat{S}_\ell$ with the help of Eq.~(\ref{Shat_def}), and obtain 
\begin{subequations} 
\begin{align} 
\hat{S}_\ell^< &= \sigma \biggl\{ 2 Q_{\ell-1}(x_0) P_{\ell-1}(x) + (-1)^\ell \ln\frac{x_0+1}{x_0-1} \biggr\}, \\ 
\hat{S}_\ell^> &= \sigma \biggl\{ 2 P_{\ell-1}(x_0) Q_{\ell-1}(x) + \ln\frac{x-1}{x+1} 
+ \bigl[1 + (-1)^\ell \bigr] \ln\frac{x_0+1}{x_0-1} \biggr\}
\end{align} 
\end{subequations} 
after simplification. From this it follows that $\hat{S}_2 = g_0$ and $\hat{S}_3 = g_1$, and as can be seen from Eqs.~(\ref{g_massive_alt}) and (\ref{g_massive_L0}), we also have that $\hat{S}_{\ell+2} - \hat{S}_\ell = g_\ell$. The conditions of Eqs.~(\ref{Shat_recursion}) and (\ref{Shat_initial}) are satisfied, and this implies that the perturbation is indeed in the Weyl class. 

\subsection{Summed potentials} 

The potential $U(x,\theta)$ is obtained by evaluating the sum of Eq.~(\ref{Ug_decomp}) with the $u_\ell$s calculated in the preceding subsection. For $x < x_0$ we make use of the summation formula (\ref{SF3a}), and we arrive at 
\begin{equation} 
U = -\sigma \ln(D + x_0 - x\cos\theta) + U_0, 
\label{U_massive} 
\end{equation} 
where 
\begin{equation} 
D := (x^2 - 2x_0 x\cos\theta + x_0^2 - \sin^2\theta)^{1/2} 
\end{equation} 
was first introduced in Eq.~(\ref{D_def1}), and $U_0 := \sigma[\frac{1}{2} (x_0+1)\ln(x_0+1) - \frac{1}{2}(x_0-1)\ln(x_0-1) - 1 + \ln 2]$ is a constant. For $x > x_0$ we invoke Eqs.~(\ref{SF0}) and (\ref{SF3b}) instead, and again land on Eq.~(\ref{U_massive}). 

As was pointed out in Sec.~\ref{sec:massive-noBH}, a constant term $U_0$ in the potential can always be eliminated with a rescaling of the $t$ and $r$ coordinates. We exercise this freedom to set $U_0 = 0$ in Eq.~(\ref{U_massive}). Another way to achieve this result would have been to shift $u_0$, as displayed in Eq.~(\ref{u_massive_L0}), by the constant $U_0$.  

The potential $\gamma(x,\theta)$ is also obtained from Eq.~(\ref{Ug_decomp}), in which we insert the $g_\ell$s 
computed previously. For $x < x_0$ we invoke Eq.~(\ref{SF3a}) multiplied by $x$, Eq.~(\ref{SF3b}) with $x$ and $y \equiv x_0$ interchanged, and Eq.~(\ref{SF4}). Many terms cancel out in these sums, and we obtain 
\begin{equation} 
\gamma =\sigma \ln \frac{(x_0+1) \Phi_-}{(x_0-1) \Phi_+}, 
\label{gamma_massive1} 
\end{equation} 
where 
\begin{equation} 
\Phi_\pm := (x_0 \pm \cos\theta)(D + x_0 \pm \cos\theta) - (x_0\cos\theta \pm 1)(x \pm 1). 
\end{equation} 
For $x > x_0$ we use Eq.~(\ref{SF3b}) multiplied by $x$, Eq.~(\ref{SF3a}) with $x$ and $y$ interchanged, and Eq.~(\ref{SF4}), also with $x$ and $y$ interchanged. This time we arrive at 
\begin{equation} 
\gamma = \sigma \ln \frac{(x_0+1)^2(x-1) \Psi_-}{(x_0-1)^2(x+1)\Psi_+}, 
\label{gamma_massive2} 
\end{equation} 
where 
\begin{equation} 
\Psi_\pm := (x \pm \cos\theta)(D + x \pm \cos\theta) - (x_0 \pm 1)(x\cos\theta \pm 1).  
\end{equation} 
The identities 
\begin{equation} 
\Phi_+ \Psi_- = (x_0-1)(x+1)(1-\cos\theta)\, \Upsilon, \qquad 
\Phi_- \Psi_+ = (x_0+1)(x-1)(1-\cos\theta)\, \Upsilon, 
\end{equation} 
with 
\begin{equation} 
\Upsilon := D^2 + (x+y)D + (xy-\cos\theta)(1+\cos\theta),  
\end{equation} 
guarantee that Eqs.~(\ref{gamma_massive1}) and (\ref{gamma_massive2}) are equivalent. The expression of Eq.~(\ref{gamma_massive1}) is simpler, and unlike Eq.~(\ref{gamma_massive2}), it does not feature an apparent singularity at the event horizon, situated at $x=1$. 

The potentials behave as 
\begin{equation} 
U(x \gg 1) \sim -\sigma \biggl\{ \ln\bigl[ x(1-\cos\theta) \bigr] + \frac{x_0}{x} + O(x^{-2}) \biggr\}, \qquad 
\gamma(x \gg 1) \sim 2\sigma \biggl\{ \ln\frac{x_0+1}{x_0-1} - \frac{1+\cos\theta}{x} + O(x^{-2}) \biggr\} 
\end{equation} 
when $x := r/M - 1$ is large. As expected for an infinite line source, $U$ diverges logarithmically at large distances; the spacetime is not asymptotically flat. At the event horizon the potentials become 
\begin{equation} 
U(x=1) = -\sigma \ln\bigl[ 2(x_0 - \cos\theta) \bigr], \qquad 
\gamma(x=1) = 2\sigma \ln\frac{x_0 - \cos\theta}{x_0 - 1}. 
\end{equation} 

We also find that $U$ diverges logarithmically when evaluated on the massive string (on the upper axis, with $x > x_0$), but that it is bounded everywhere else on the axis. The calculation of $\gamma$ on the upper axis is complicated by the fact that when $x > x_0$, both $\Phi_-$ and $\Phi_+$ go to zero when $\theta = 0$; it is therefore necessary to take a limit $\theta \to 0$. We have that 
\begin{equation} 
\Phi_- \sim \frac{(x_0+1)(x-1)^2}{2(x-x_0)} \theta^2,\qquad 
\Phi_+ \sim \frac{(x_0-1)(x+1)^2}{2(x-x_0)} \theta^2, 
\end{equation} 
and we arrive at 
\begin{equation} 
\gamma(x > x_0, \theta \to 0) = 2\sigma \ln \frac{(x_0+1)(x-1)}{(x_0+1)(x+1)}. 
\end{equation} 
By virtue of Eq.~(\ref{Tup_massive}), this implies that $\gamma = 4 T_{\rm up}$ on this portion of the upper axis. Below $x = x_0$ the calculation is straightforward, and we obtain 
\begin{equation} 
\gamma(x < x_0, \theta = 0) = 0. 
\end{equation} 
On the lower axis the calculation is equally straightforward, and we obtain 
\begin{equation} 
\gamma(\theta = \pi) = 2 \sigma \ln\frac{x_0+1}{x_0-1}. 
\end{equation} 
According to Eq.~(\ref{Tdn_massive}), we have that $\gamma = 4 T_{\rm dn}$ everywhere on the lower axis. It is a remarkable fact that the value of $\gamma$ on the upper and lower axis can be linked to the string tension, even when the string is massive and the tension is not constant. This result was previously announced in Eq.~(\ref{gamma_vs_T}). 

\subsection{Complete potentials} 

To conclude, we recall that the potentials $U$ and $\gamma$ obtained here are those of the ``difference strings'' introduced in Sec.~\ref{subsec:difference}. The solution to the ($\mbox{particle} + \mbox{massive string}$) problem  is then given by the sum of these potentials and those computed in Sec.~\ref{sec:particle-massless}. The complete potentials are 
\begin{equation} 
U_{\rm complete} = \frac{k}{D} -\sigma \ln(D + x_0 - x\cos\theta) 
\label{U_massive_comp}
\end{equation} 
and 
\begin{equation} 
\gamma_{\rm complete} = \frac{2k}{x_0^2 - 1} \biggl( \frac{x - x_0\cos\theta}{D} + 1 \biggr)
+ \sigma \ln \frac{(x_0+1) \Phi_-}{(x_0-1) \Phi_+}. 
\label{gamma_massive_comp} 
\end{equation} 
We recall that $x := r/M - 1$, $x_0 := r_0/M - 1$, 
\begin{equation} 
k := \frac{m}{M} \sqrt{\frac{x_0-1}{x_0+1}}, \qquad 
\sigma := \mu^{\rm massive}_{\rm up} - T^{\rm massive}_{\rm up} = \mbox{constant}, 
\end{equation} 
and that 
\begin{equation} 
D := (x^2 - 2x_0 x\cos\theta + x_0^2 - \sin^2\theta)^{1/2}. 
\end{equation} 
According to our previous results, we have that 
\begin{equation} 
\gamma^{\rm axis}_{\rm complete} = \left\{ 
\begin{array}{ll}
4T^{\rm massive}_{\rm up} & \quad \mbox{upper axis},\ x > x_0 \\ 
0 & \quad \mbox{upper axis},\ x < x_0 \\ 
4T^{\rm massless}_{\rm dn} & \quad \mbox{lower axis} 
\end{array} \right., 
\end{equation} 
where 
\begin{equation} 
T^{\rm massive}_{\rm up} = \frac{k}{x_0^2-1}
+ \frac{1}{2} \sigma \biggl( \ln\frac{x-1}{x+1} 
+ \ln\frac{x_0+1}{x_0-1} \biggr) 
\end{equation} 
is the tension in the massive string holding the particle, while  
\begin{equation} 
T^{\rm massless}_{\rm dn} = \frac{k}{x_0^2-1}
+ \frac{1}{2} \sigma \ln\frac{x_0+1}{x_0-1} 
\end{equation} 
is the tension in the massless string holding the black hole. 

\subsection{Black-hole properties and first law} 

Our discussion here parallels the one of Sec.~\ref{subsec:firstlaw}. The starting points are the same: the metric of Eq.~(\ref{weylclass_metric}), the Komar mass of Eqs.~(\ref{komar1}) and (\ref{komar2}), and the black-hole quantities of Eqs.~(\ref{beta_def}), (\ref{area}), (\ref{kappa}), and (\ref{Smarr}). The difference is that we now work with the potentials of Eqs.~(\ref{U_massive_comp}) and (\ref{gamma_massive_comp}). 

The Komar mass for a large 2-surface of constant $t$ and $r$ is now given by 
\begin{equation} 
M_{\rm K}(r) = \sigma r + M + k M - \sigma M(x_0 + 1) + O(r^{-1}). 
\end{equation} 
This diverges in the limit $r \to \infty$, as should be expected for an infinite massive string. The finite piece of the Komar mass, $M + kM - \sigma M(x_0 + 1)$, provides a plausible candidate for a total mass $M_{\rm tot}$ that could be implicated in the first law. As we shall see, however, this candidate will eventually be rejected. 

The horizon quantity defined in Eq.~(\ref{beta_def}) is now given by 
\begin{equation} 
\beta = 1 + \frac{2k}{x_0-1} - 2\sigma \ln\bigl[ 2(x_0-1) \bigr]; 
\end{equation} 
it is perturbative in both $m$ (which is hidden in $k$) and $\sigma$. This is inserted within Eqs.~(\ref{area}) and (\ref{kappa}) to obtain the area and surface gravity of the perturbed black hole. 

We wish to generalize the first law of Eq.~(\ref{firstlaw_massless}) to the case of a massive string. To the extent that a  ``total mass'' can be defined, we expect that it should be generalized to 
\begin{equation} 
M_{\rm tot} = M + k M + \sigma E(M,x_0) 
\end{equation} 
to account for $\sigma$, the second perturbation parameter. The quantity $E(M, x_0)$ is an unknown function of $M$ and $x_0$; it cannot depend on $m$ because the total mass should be of the first order in both $m$ and $\sigma$. For thermodynamic variables external to the black hole we choose $\sigma$ and 
\begin{equation} 
T_\infty := \frac{k}{x_0^2-1} + \frac{1}{2} \sigma \ln\frac{x_0+1}{x_0-1}, 
\end{equation} 
the tension measured in either string at infinity. 

We expect the first law to take the new form 
\begin{equation} 
d M_{\rm tot} = \frac{\kappa}{8\pi}\, dA  - \lambda\, dT_\infty + \omega\, d\sigma + z\, dm, 
\label{firstlaw_massive}
\end{equation} 
which expresses the fact that $M_{\rm tot}$ can be viewed as a function of $A$,  $T_\infty$, $\sigma$, and $m$, instead of as a function of $M$, $m$, $\sigma$, and $x_0$. Because $dT_\infty$ and $dm$ are perturbative quantities, the thermodynamic length $\lambda$ and the redshift factor $z$ will remain unchanged from Eqs.~(\ref{lambda}) and (\ref{redshift}). We must then verify that the coefficient in front of $dA$ remains $\kappa/(8\pi)$, discover the identity of $E(M,x_0)$, and obtain an expression for $\omega$. 

For the moment we keep $\kappa$ unrelated to Eq.~(\ref{kappa}); we view it as an unknown coefficient in front of $dA$, and we write it as $(4M)^{-1}(1 + k \kappa_1 + \sigma \kappa_2)$, with $\kappa_1$ and $\kappa_2$ functions of $M$ and $x_0$. We write the candidate first law as 
\begin{equation} 
0 = d M_{\rm tot} - \frac{\kappa}{8\pi}\, dA  + \lambda\, dT_\infty - \omega\, d\sigma - z\, dm, 
\end{equation} 
and because each quantity is a function of $M$, $m$, $\sigma$, and $x_0$, the expression becomes 
\begin{equation} 
0 = {\cal P}_1\, dM + {\cal P}_2\, dm + {\cal P}_3\, d\sigma + {\cal P}_4\, dx_0
\end{equation} 
for some functions ${\cal P}_n$; ${\cal P}_1$ and ${\cal P}_4$ contain background terms as well as terms linear in both $m$ and $\sigma$; ${\cal P}_2$ and ${\cal P}_3$ have only background terms. Because $M$, $m$, $\sigma$, and $x_0$ are independent parameters, each ${\cal P}_n$ must vanish separately. The requirement that ${\cal P}_1 = 0$ implies that $\kappa_1 = -2/(x_0-1)$ and $\kappa_2 = 2\ln[2(x_0-1)] - \partial E/\partial M$; these expressions agree with $\kappa = (4M\beta)^{-1}$ provided that $E$ is independent of $M$. We find that ${\cal P}_2 = 0$ is automatically satisfied, and ${\cal P}_4 = 0$ implies that $\partial E/\partial x_0 = 0$. Finally, we find that ${\cal P}_3 = 0$ returns 
\begin{equation} 
\omega = M \Bigl[ \ln 2 + \frac{1}{2} (x_0+1)\ln(x_0+1) - \frac{1}{2}(x_0-1)\ln(x_0-1) \Bigr] 
- E(M). 
\end{equation} 
The contribution $\sigma E$ to the total mass must be independent of $x_0$, and this rules out the candidate $E = -M(x_0+1)$ delivered by the finite piece of the Komar mass. This analysis permits the existence of a contribution of the form $\sigma E(M)$ to the total mass, but it does not disclose its identity. 

The simplest assignment is $E(M) = 0$. With this choice, $\kappa$ re-acquires its meaning as the black hole's surface gravity, and $\omega$ becomes simply 
\begin{equation} 
\omega = M \Bigl[ \ln 2 + \frac{1}{2} (x_0+1)\ln(x_0+1) - \frac{1}{2}(x_0-1)\ln(x_0-1) \Bigr]. 
\end{equation} 
With this, and with Eqs.~(\ref{lambda}) and (\ref{redshift}) for $\lambda$ and $z$, we have established that Eq.~(\ref{firstlaw_massive}) is a valid formulation of the first law. It is puzzling that the total mass $M_{\rm tot} = M + kM$ identified by the law makes no reference to the massive string. In addition, the physical interpretation of the new thermodynamic length $\omega$ remains unclear. For these reasons, the first law of Eq.~(\ref{firstlaw_massive}) should be taken with a grain of salt. It is a valid mathematical identity implicating various quantities associated with the spacetime, but it is lacking in terms of a compelling physical interpretation.  

\section{Application: Particle and generic string in weak field} 
\label{sec:generic} 

In the previous sections we formulated string models that were particularly simple, so as to facilitate an exact integration of the perturbation equations. In this section we go beyond the simple and examine models of massive strings that might be more realistic. Unfortunately, this enhanced realism comes at the price of a lost ability to integrate the equations exactly. We shall have to resort to finding approximate solutions to the perturbation equations, taking the particle and string to lie in the weak-field region of the Schwarzschild spacetime.  

\subsection{String models} 

The intrinsic energy-momentum tensor $t^{ab}$ of a massive string was written down in Eq.~(\ref{tab_up}), and expressed in terms of an energy density $\mu$, tension $T$, and velocity field $u^a$. To these variables we add a rest-mass density $\rho$ and a density of internal energy $\epsilon$. The new and old variables are linked by $\mu = \rho + \epsilon$ and the first law of thermodynamics, $d(\epsilon/\rho) = T d(1/\rho)$. Assuming that the string is subjected to the continuity equation $D_a(\rho u^a) = 0$, with $D_a$ denoting the covariant-derivative operator compatible with the world-sheet metric $\gamma_{ab}$, the first law follows from the conservation equation $u_a D_b t^{ab} = 0$. We take the string to possess an equation of state of the form $T = T(\rho)$. 

We focus our attention on the upper string, placed at $\theta = 0$ and extending from particle to infinity. This shall now be a realistic string described by the variables introduced in the preceding paragraph --- we omit the label ``up'' on these variables. We continue to take the lower string, situated at $\theta = \pi$ and extending from black hole to infinity, to be a massless string with equal (and constant) energy density $\mu_{\rm dn}$ and tension $T_{\rm dn}$. To prevent the black hole from being accelerated in the perturbed spacetime, we continue to set $T_{\rm dn} = T_\infty := T(r=\infty)$.  

As was discussed in Sec.~\ref{sec:EMtensor}, the energy density $\mu$ and tension $T$ of the upper string are linked by the conservation equation [refer back to Eqs.~(\ref{conservation-up}) and (\ref{T0})]  
\begin{equation} 
\frac{dT}{dx} = \frac{\mu - T}{x^2-1}, \qquad 
T_0 := T(x=x_0) = \frac{k}{x_0^2-1}, \qquad 
k := \frac{m}{M} \sqrt{\frac{x_0-1}{x_0+1}}, 
\label{cons_eq}
\end{equation} 
where $x := r/M - 1$ and $x_0 = r_0/M - 1$. We recall that the particle has a mass $m$, and that it is situated at $r =  r_0$ (or $x = x_0$) on the upper axis of the Schwarzschild spacetime. 

As a first example of a realistic string model, we consider a polytropic string with equation of state 
\begin{equation} 
T = K \rho^{1 + 1/n}, 
\label{polytrope} 
\end{equation} 
where $K$ and $n$ are constants. The first law of thermodynamics produces $\epsilon = -nT$, and we have that $\mu = \rho - nT$. To integrate Eq.~(\ref{cons_eq}) we introduce the Lane-Emden variable $\vartheta$ defined by 
\begin{equation} 
\rho = \rho_0\, \vartheta^n, 
\end{equation} 
where $\rho_0$ is the density at $x = x_0$. The conservation equation becomes 
\begin{equation} 
(n+1) b\, \frac{d\vartheta}{dx} = \frac{1 - (n+1) b\, \vartheta}{x^2 - 1}, \qquad 
\vartheta(x=x_0) = 1, 
\end{equation}
where $b := T_0/\rho_0$. The solution is 
\begin{equation} 
\vartheta = \frac{\sqrt{x_0^2-1}}{x_0+1} \frac{x+1}{\sqrt{x^2-1}} \biggl[ 1 - \frac{1}{(n+1)b} \biggr]
+ \frac{1}{(n+1)b},  
\end{equation} 
and the profiles $\mu(x)$, $T(x)$ are determined. In the weak-field regime, in which $x_0 \gg 1$ and $x \gg 1$, this reduces to 
\begin{equation} 
\vartheta = \biggl[ 1 - \frac{1}{(n+1)b} \biggr] \biggl[ 1 + \biggl( \frac{1}{x} - \frac{1}{x_0} \biggr)
+ \frac{1}{2} \biggl( \frac{1}{x} - \frac{1}{x_0} \biggr)^2 + \cdots \biggr] + \frac{1}{(n+1)b}.   
\end{equation} 
It follows that in this regime, $\mu-T$ admits an expansion in powers of $(x^{-1} - x^{-1}_0)$.  

As a second example of a realistic string model, we consider the linear equation of state
\begin{equation} 
T = K\rho, 
\label{linear_string} 
\end{equation} 
where $K$ is a constant. In this case the first law returns $\epsilon = -K\rho\ln\rho$, and we have that $\mu = \rho(1 - K\ln\rho)$. To integrate Eq.~(\ref{cons_eq}) we write 
\begin{equation} 
\rho = \rho_0\, e^\psi, 
\end{equation} 
where $\rho_0$ is again the density at $x = x_0$. The equation becomes 
\begin{equation} 
K \frac{d\psi}{dx} = \frac{A - K\psi}{x^2 - 1}, \qquad 
\psi(x=x_0) = 0, 
\end{equation} 
where $A := 1 - K(1 + \ln\rho_0)$. The solution is 
\begin{equation} 
\psi = \frac{A}{K} \biggl( 1 - \frac{\sqrt{x_0^2-1}}{x_0+1} \frac{x+1}{\sqrt{x^2-1}} \biggr), 
\end{equation} 
and it becomes 
\begin{equation} 
\psi = -\frac{A}{K} \Biggl[ \biggl( \frac{1}{x} - \frac{1}{x_0} \biggr)
+ \frac{1}{2} \biggl( \frac{1}{x} - \frac{1}{x_0} \biggr)^2 + \cdots \Biggr] 
\end{equation} 
in the weak-field regime. Again we find that $\mu-T$ admits an expansion in powers of $(x^{-1} - x^{-1}_0)$.   

\subsection{Generic string} 

Summarizing our results from this (albeit limited) survey of string models, we take a generic string to have an energy density $\mu$ and tension $T$ related by 
\begin{equation} 
\mu - T = \sigma - C_1 \biggl( \frac{1}{x} - \frac{1}{x_0} \biggr)
+ C_2 \biggl( \frac{1}{x} - \frac{1}{x_0} \biggr)^2 + \cdots 
\label{muT_generic} 
\end{equation} 
in the weak-field region of the Schwarzschild spacetime. Here $\sigma$, $C_1$, and $C_2$ are constants determined by the string's equation of state. For this generic string, Eq.~(\ref{cons_eq}) implies 
\begin{equation} 
T = T_0 - \sigma \biggl[ \biggl( \frac{1}{x} - \frac{1}{x_0} \biggr) 
+ \frac{1}{3} \biggl( \frac{1}{x^3} - \frac{1}{x_0^3} \biggr) \biggr] 
+ \frac{1}{2} C_1 \biggl( \frac{1}{x} - \frac{1}{x_0} \biggr)^2  
- \frac{1}{3} C_2 \biggl( \frac{1}{x} - \frac{1}{x_0} \biggr)^3 + \cdots. 
\label{T_generic} 
\end{equation} 
The tension at $x = \infty$ is then
\begin{equation} 
T_\infty = T_0 +\sigma \biggl( \frac{1}{x_0} + \frac{1}{3x_0^3} \biggr) 
+ C_1\, \frac{1}{2 x_0^2} + C_2\, \frac{1}{3x_0^3} + \cdots. 
\label{Tinfty_generic} 
\end{equation} 
This, we recall, shall be matched to the tension in the lower, massless string.  

As we did in Sec.~\ref{subsec:difference}, we decompose the ($\mbox{particle} + \mbox{massive string}$) system into a superposition of ($\mbox{particle} + \mbox{massless string}$) and (difference string) systems. The ``difference'' string consists of our actual massive string from which we subtract a massless string with tension $T_0$, so that
\begin{subequations} 
\label{diffstring_generic}  
\begin{align} 
\mu^{\rm diff} - T^{\rm diff} &= \sigma - C_1 \biggl( \frac{1}{x} - \frac{1}{x_0} \biggr)
+ C_2 \biggl( \frac{1}{x} - \frac{1}{x_0} \biggr)^2 + \cdots, \\ 
T^{\rm diff} &= -\sigma \biggl[ \biggl( \frac{1}{x} - \frac{1}{x_0} \biggr) 
+ \frac{1}{3} \biggl( \frac{1}{x^3} - \frac{1}{x_0^3} \biggr) \biggr] 
+ \frac{1}{2} C_1 \biggl( \frac{1}{x} - \frac{1}{x_0} \biggr)^2  
- \frac{1}{3} C_2 \biggl( \frac{1}{x} - \frac{1}{x_0} \biggr)^3 + \cdots, \\ 
T^{\rm diff}_\infty &= \sigma \biggl( \frac{1}{x_0} + \frac{1}{3x_0^3} \biggr) 
+ C_1\, \frac{1}{2 x_0^2} + C_2\, \frac{1}{3x_0^3} + \cdots. 
\end{align} 
\end{subequations} 
The complete solution to the ($\mbox{particle} + \mbox{massive string}$) problem is then the solution to the ($\mbox{particle} + \mbox{massless string}$) problem, as worked out in Sec.~\ref{sec:particle-massless}, added to the solution to the (difference string) problem, to be obtained below. 

We extend the superposition principle even further, and decompose the (difference string) problem into the three separate problems of a $\sigma$-string, a $C_1$-string, and a $C_2$-string. The metric perturbation produced by a $\sigma$-string was already obtained in Sec.~\ref{sec:particle-massive}, and there is no need to duplicate this effort here. The previous work, however, was carried out in the exact Schwarzschild background, without the assumption that $x$ and $x_0$ are large. To adapt it to the current context, we shall have to specialize all expressions to the weak-field regime. 

\subsection{Integration of the perturbation equations} 

The equations to be integrated were presented in Sec.~\ref{sec:EFE}. The main two are 
\begin{subequations}
\begin{align}
(x^2-1) u''_\ell + 2 x u'_\ell - \ell(\ell+1) u_\ell
&= -(2\ell+1) W, \\
(x^2 - 1) g''_\ell - \ell(\ell+1) g_\ell &= 4u'_\ell, 
\end{align} 
\end{subequations}
where 
\begin{equation} 
W := \Biggl[ \sigma - C_1 \biggl( \frac{1}{x} - \frac{1}{x_0} \biggr)
+ C_2 \biggl( \frac{1}{x} - \frac{1}{x_0} \biggr)^2 \Biggr] \Theta(x - x_0).
\end{equation} 
As usual, these equations are accompanied by Eqs.~(\ref{EFEG}) and (\ref{Shat_def}), which provide constraints when $\ell = 0, 1$, and determine $G_\ell$ when $\ell \geq 2$. 

We saw back in Sec.~\ref{sec:EFE} that particular solutions to these equations are given by
\begin{subequations}
\begin{align} 
u^{\rm part}_\ell &= (2\ell+1) \biggl[ Q_\ell(x) \int_{x_1}^x P_\ell(x') W(x')\, dx' 
+ P_\ell(x) \int^\infty_x Q_\ell(x') W(x')\, dx' \biggr], \\ 
g^{\rm part}_\ell &= \frac{4 (x^2-1) }{\ell(\ell+1)} \biggl[
Q'_\ell(x) \int_{x_1}^x P'_\ell(x') u'_\ell(x')\, dx' 
+ P'_\ell(x) \int^\infty_x Q'_\ell(x') u'_\ell(x')\, dx' \biggr], 
\end{align} 
\end{subequations}
where $x_1$ is any constant. (The expression for $g^{\rm part}_\ell$ does not apply when $\ell = 0$.) Because we are interested in the weak-field regime, for which $x \gg 1$, we take $x_1$ to be the minimum value of $x$ at which the solution is to be evaluated; we have that $1 \ll x_1 < x_0$. We observe that in the integrals involving $P_\ell(x')$, the contributions from the boundary at $x = x_1$ give rise to terms in $u_\ell^{\rm part}$ and $g_\ell^{\rm part}$ that are  proportional to $Q_\ell(x)$ or its derivative, and that such terms would fail to be smooth at $x = 1$; we eliminate these boundary terms to obtain the correct physical solution to the perturbation equations. Regularity at $x = \infty$ is ensured by setting to infinity the upper bound of the integrals involving $Q_\ell(x')$. 

All integrals are defined in the large-$x'$ regime, and in these we may substitute the asymptotic behaviors\footnote{Refer to Sec.~14.8 of Ref.~\cite{NIST:10} for the numerical prefactors. The subleading terms are derived from Legendre's equation.} 
\begin{subequations} 
\begin{align} 
P_\ell(x) &= \frac{(2\ell-1)!!}{\ell!} x^\ell \biggl[ 1 
- \frac{1}{2} \frac{(\ell-1)\ell}{2\ell+1} \frac{1}{x^2} + \cdots \biggr], \\ 
Q_\ell(x) &= \frac{\ell!}{(2\ell+1)!!} \frac{1}{x^{\ell+1}} \biggl[ 1 
+ \frac{1}{2} \frac{(\ell+1)(\ell+2)}{2\ell+3} \frac{1}{x^2} + \cdots \biggr]
\end{align} 
\end{subequations} 
for the Legendre functions. All subsequent calculations are then straightforward. These computations, however, can be further simplified by exploiting the fact that the expression for $W$ is truncated beyond order $\varepsilon^2$, where $\varepsilon := O(x^{-1}, x_0^{-1})$. Consider, for example, the piece of the solution proportional to $C_1$. This piece originates from the $C_1$ term in $W$, which occurs at order $\varepsilon$. The leading contributions to $P_\ell$ and $Q_\ell$ give rise to terms that are also of order $\varepsilon$ in the solution, while the subleading contributions would produce terms of order $\varepsilon^3$; these are comparable to terms that would arise from neglected contributions of order $\varepsilon^3$ to $W$, and they may therefore be neglected as well. We conclude that the $C_1$-piece of the solution can be constructed solely from the leading terms in the Legendre functions. The same conclusion applies to the $C_2$-piece, which is of order $\varepsilon^2$. The conclusion, however, does not apply to the $\sigma$-piece, which leads at order $\varepsilon^0$ and comes with corrections of order $\varepsilon^2$; for this we do require the subleading terms in the Legendre functions. But as was pointed out previously, this piece of the solution was already obtained in Sec.~\ref{sec:particle-massive},  and it does not need to be calculated again. 

The upshot is that the computation of $u_\ell$ and $g_\ell$ for the $C_1$ and $C_2$ strings require only the leading contributions to the Legendre functions. This observation is equivalent to the statement that $x^2 - 1$ can be approximated by $x^2$ in the differential equations, which become 
\begin{subequations}
\label{weakfield_diffeqs} 
\begin{align}
x^2 u''_\ell + 2 x u'_\ell - \ell(\ell+1) u_\ell &= -(2\ell+1) W, \\
x^2 g''_\ell - \ell(\ell+1) g_\ell &= 4u'_\ell. 
\end{align} 
\end{subequations}
Integration is then a very simple matter. In this simplified setting, regularity at the event horizon ($x=1$) is replaced by regularity at $x = 0$; the solutions must still be well-behaved at $x = \infty$.  

\subsection{$C_1$-string} 

After imposing the boundary conditions at $x = 0$ and $x = \infty$, as well as continuity and differentiability at $x = x_0$, we find that the solutions to Eqs.~(\ref{weakfield_diffeqs}) for the $C_1$-string are 
\begin{subequations} 
\begin{align} 
u_\ell^< &= C_1\, \frac{1}{\ell(\ell+1)} \frac{x^\ell}{x_0^{\ell+1}}, \\ 
u_\ell^> &= C_1\, \biggl[ \frac{1}{\ell(\ell+1)} \frac{x_0^\ell}{x^{\ell+1}}
- \frac{2\ell+1}{\ell(\ell+1)} \biggl(\frac{1}{x} - \frac{1}{x_0} \biggr) \biggr] 
\end{align} 
\end{subequations} 
and 
\begin{subequations} 
\begin{align} 
g_\ell^< &= C_1 \biggl[ \frac{2}{(\ell+3)(2\ell+3)} \frac{x^{\ell+1}}{x_0^{\ell+3}} 
- \frac{2}{(\ell+1)(2\ell-1)} \frac{x^{\ell-1}}{x_0^{\ell+1}} \biggr], \\ 
g_\ell^> &= C_1 \biggl[ -\frac{2}{\ell(2\ell+3)} \frac{x_0^\ell}{x^{\ell+2}} 
+ \frac{2}{(\ell-2)(2\ell-1)} \frac{x_0^{\ell-2}}{x^\ell} 
- \frac{4(2\ell+1)}{(\ell-2)\ell(\ell+1)(\ell+3)} \frac{1}{x^2} \biggr]. 
\end{align} 
\end{subequations} 
We recall that $u^<_\ell$ and $g_\ell^<$ apply when $x < x_0$, while $u^>_\ell$ and $g_\ell^>$ apply when $x > x_0$. The special cases are 
\begin{subequations} 
\begin{align} 
u_0^< &=0, \\ 
u_0^> &= C_1 \biggl[ -2\biggl(\frac{1}{x} - \frac{1}{x_0} \biggr) 
- \biggl(\frac{1}{x} + \frac{1}{x_0} \biggr) \ln(x/x_0) \biggr], \\ 
g_0^< &= C_1 \biggl( \frac{2x}{9x_0^3} + \frac{1}{x_0^2} \biggr), \\ 
g_0^> &= C_1 \biggl[ \frac{11}{9x^2} - \frac{2}{x_0 x} + \frac{2}{x_0^2} 
+ \frac{2}{3 x^2} \ln(x/x_0) \biggr] 
\end{align} 
\end{subequations} 
and 
\begin{equation} 
g_2^> = C_1 \biggl[ -\frac{x_0^2}{7x^4} - \frac{1}{45 x^2} - \frac{2}{3x^2} \ln(x/x_0) \biggr]. 
\end{equation} 
The integration constant in $g_0$ is determined by imposing the constraints of Eqs.~(\ref{EFEG}) and (\ref{Shat_def}) when $\ell = 0$ and $\ell = 1$. The constraints also confirm that $T_{\rm dn} = T_\infty$, as given by the $C_1$-term in Eq.~(\ref{diffstring_generic}). Our results imply that the recursion relation of Eq.~(\ref{Shat_recursion}) is satisfied, together with the initial conditions of Eq.~(\ref{Shat_initial}). The perturbation created by the $C_1$-string therefore belongs to the Weyl class of Sec.~\ref{sec:Weyl-class}. 

The potentials 
\begin{equation} 
U = \sum_{\ell=0}^\infty u_\ell\, P_\ell(\cos\theta), \qquad 
\gamma = \sum_{\ell=0}^\infty g_\ell\, P_\ell(\cos\theta) 
\end{equation} 
are evaluated with the help of summation formulae developed in Appendix~\ref{sec:summation}. To obtain $U$ when $x < x_0$ we make use of Eq.~(\ref{WF1a}) with $y \equiv x_0$. When $x > x_0$ instead, we invoke Eq.~(\ref{WF1a}) again, but with $x$ and $y$ interchanged, and complete the task with Eq.~(\ref{SF0}). In either case we find that 
\begin{align} 
U[C_1] &= C_1 \biggl\{ \frac{1}{x} \bigl[ \ln x_0 + \ln(1-\cos\theta) - \ln(E + x - x_0\cos\theta) \bigr] 
\nonumber \\ & \quad \mbox{} 
+ \frac{1}{x_0} \bigl[ \ln 2 + \ln x_0 - \ln(E + x_0 - x\cos\theta) \bigr] \biggr\},  
\label{U_C1} 
\end{align} 
where 
\begin{equation} 
E := ( x^2 - 2 x_0 x \cos\theta + x_0^2 )^{1/2}
\label{Edef} 
\end{equation} 
is the Euclidean distance between a point at $(x, \theta)$ and the particle at $(x_0, 0)$. This expression for $U[C_1]$ reflects a choice of integration constant; this choice can always be altered by adding another constant $U_0$ to the potential.    

We proceed as follows to find $\gamma$. When $x < x_0$ we write 
\begin{equation} 
\frac{2}{(\ell+3)(2\ell+3)} = \frac{4}{3}\, \frac{1}{2\ell+3} - \frac{2}{3}\, \frac{1}{\ell+3}, \qquad 
\frac{2}{(\ell+1)(2\ell-1)} = \frac{4}{3}\, \frac{1}{2\ell-1} - \frac{2}{3}\, \frac{1}{\ell+1} 
\end{equation} 
and make use of Eqs.~(\ref{WF1b}) and (\ref{WF1c}). We obtain  
\begin{equation} 
\gamma[C_1] = C_1 \biggl\{ \frac{\sin^2\theta}{x^2} \bigl[ \ln(E + x - x_0\cos\theta) 
- \ln x_0 - \ln(1-\cos\theta) \bigr] 
+ \frac{x - x_0 \cos\theta}{x_0^2 x^2}\, E + \frac{1}{x_0^2} - \frac{2}{x_0 x} + \frac{\cos\theta}{x^2} \biggr\}. 
\label{gamma_C1}
\end{equation} 
When $x > x_0$ we write 
\begin{equation} 
\frac{2}{\ell(2\ell+3)} = -\frac{4}{3}\, \frac{1}{2\ell+3} + \frac{2}{3}\, \frac{1}{\ell}, \qquad 
\frac{2}{(\ell-2)(2\ell-1)} = -\frac{4}{3}\, \frac{1}{2\ell-1} + \frac{2}{3}\, \frac{1}{\ell-2} 
\end{equation} 
and invoke Eqs.~(\ref{SF0c}), (\ref{WF2a}), and (\ref{WF2b}); we obtain the same expression for $\gamma[C_1]$.  

\subsection{$C_2$-string} 

The solutions to Eqs.~(\ref{weakfield_diffeqs}) for the $C_2$-string are 
\begin{subequations} 
\begin{align} 
u_\ell^< &= C_2\, \frac{2}{\ell(\ell+1)(\ell+2)} \frac{x^\ell}{x_0^{\ell+2}}, \\ 
u_\ell^> &= C_2\, \biggl[ -\frac{2}{(\ell-1)\ell(\ell+1)} \frac{x_0^{\ell-1}}{x^{\ell+1}}
+ \frac{2\ell+1}{(\ell-1)(\ell+2)} \frac{1}{x^2} 
+ \frac{2\ell+1}{\ell(\ell+1)} \biggl( \frac{1}{x_0^2} - \frac{2}{x_0 x} \biggr) \biggr]  
\end{align} 
\end{subequations} 
and 
\begin{subequations} 
\begin{align} 
g_\ell^< &= C_2 \biggl[ \frac{4}{(\ell+3)(\ell+4))(2\ell+3)} \frac{x^{\ell+1}}{x_0^{\ell+4}} 
- \frac{4}{(\ell+1)(\ell+2)(2\ell-1)} \frac{x^{\ell-1}}{x_0^{\ell+2}} \biggr], \\ 
g_\ell^> &= C_2 \biggl[ \frac{4}{(\ell-1)\ell(2\ell+3)} \frac{x_0^{\ell-1}}{x^{\ell+2}} 
- \frac{4}{(\ell-3)(\ell-2)(2\ell-1)} \frac{x_0^{\ell-3}}{x^\ell} 
\nonumber \\ & \quad \mbox{} 
- \frac{8(2\ell+1)}{(\ell-2)\ell(\ell+1)(\ell+3)} \frac{1}{x_0 x^2} 
+ \frac{8(2\ell+1)}{(\ell-3)(\ell-1)(\ell+2)(\ell+4)} \frac{1}{x^3} \biggr]. 
\end{align} 
\end{subequations} 
The exceptional cases are 
\begin{subequations} 
\begin{align} 
u_0^< &=0, \\ 
u_0^> &= C_2 \biggl[ -\frac{1}{2x^2} - \frac{2}{x_0 x} + \frac{5}{2x_0^2} 
- \biggl( \frac{2}{x_0 x} + \frac{1}{x_0^2} \biggr) \ln(x/x_0), \\ 
g_0^< &= C_2 \biggl( \frac{x}{9x_0^4} + \frac{2}{3x_0^3} \biggr), \\ 
g_0^> &= C_2 \biggl[ \frac{1}{3x^3} + \frac{10}{9x_0 x^2} - \frac{2}{x_0^2 x} 
+ \frac{4}{3x_0^3} + \frac{4}{3 x_0 x^2} \ln(x/x_0) \biggr] 
\end{align} 
\end{subequations} 
and 
\begin{subequations} 
\begin{align} 
u^>_1 &= C_2 \biggl[ \frac{11}{6x^2} - \frac{3}{x_0 x} + \frac{3}{2x_0^2} + \frac{1}{x^2} \ln(x/x_0) \biggr], \\ 
g^>_1 &= C_2 \biggl[ -\frac{122}{75x^3} + \frac{3}{x_0 x^2} - \frac{2}{x_0^2 x} 
- \frac{4}{5x^3} \ln(x/x_0) \biggr], \\ 
g^>_2 &= C_2 \biggl[ \frac{2 x_0}{7x^4} - \frac{5}{3x^3} + \frac{58}{45x_0 x^2} 
- \frac{4}{3x_0 x^2} \ln(x/x_0) \biggr], \\
g^>_3 &= C_2 \biggl[ \frac{2x_0^2}{27x^5} + \frac{118}{175x^3} - \frac{7}{9x_0 x^2} 
+ \frac{4}{5x^3} \ln(x/x_0) \biggr]. 
\end{align} 
\end{subequations} 
Equations (\ref{EFEG}) and (\ref{Shat_def}), evaluated with $\ell = 0$ and $\ell = 1$, determine the integration constant in $g_0$ and confirm that $T_{\rm dn} = T_\infty$, as given by the $C_2$-term in Eq.~(\ref{diffstring_generic}). Our results imply that the recursion relation of Eq.~(\ref{Shat_recursion}) is satisfied, together with the initial conditions of Eq.~(\ref{Shat_initial}). The perturbation created by the $C_2$-string also belongs to the Weyl class. Because all pieces of the massive string produce a perturbation in the Weyl class, the superposition principle guarantees that the perturbation created by the entire system belongs to the Weyl class. The metric of the ($\mbox{particle} + \mbox{massive string}$) system can therefore be put in the form of Eq.~(\ref{weylclass_metric}). 

The potential $U$ is evaluated as follows. When $x < x_0$ we invoke Eq.~(\ref{WF1d}). When $x > x_0$ instead, we make use of Eqs.~(\ref{SF0}), (\ref{SF0b}), and (\ref{WF2c}). In both cases we obtain 
\begin{align} 
U[C_2] &= C_2 \biggl\{ \biggl( \frac{2}{x_0 x} - \frac{\cos\theta}{x^2} \biggr) 
\bigl[ \ln x_0 + \ln(1-\cos\theta) - \ln(E + x - x_0 \cos\theta) \bigr] 
\nonumber \\ & \quad \mbox{}
+ \frac{1}{x_0^2} \bigl[ \ln 2 + \ln x_0 - \ln(E + x_0 - x\cos\theta) \bigr]  
+ \frac{E}{x_0 x^2} - \frac{1}{x^2} + \frac{3}{2x_0^2} \biggr\},  
\label{U_C2} 
\end{align} 
where $E$ is still given by Eq.~(\ref{Edef}). Again this expression reflects a choice of integration constant, which can be altered at will by adding any constant $U_0$ to the potential.    

For $\gamma$ we proceed in the following way. When $x < x_0$ we write 
\begin{subequations} 
\begin{align} 
\frac{4}{(\ell+3)(\ell+4)(2\ell+3)} &= -\frac{4}{3}\, \frac{1}{\ell+3} + \frac{4}{5}\, \frac{1}{\ell+4} 
+ \frac{16}{15}\, \frac{1}{2\ell+3}, \\ 
\frac{4}{(\ell+1)(\ell+2)(2\ell-1)} &= -\frac{4}{3}\, \frac{1}{\ell+1} + \frac{4}{5}\, \frac{1}{\ell+2} 
+ \frac{16}{15}\, \frac{1}{2\ell-1}
\end{align} 
\end{subequations} 
and import Eqs.~(\ref{WF1b}), (\ref{WF1c}), and (\ref{WF1e}). When $x > x_0$ we write 
\begin{subequations} 
\begin{align} 
\frac{4}{(\ell-1)\ell (2\ell+3)} &= \frac{4}{5}\, \frac{1}{\ell-1} - \frac{4}{3}\, \frac{1}{\ell} 
+ \frac{16}{15}\, \frac{1}{2\ell+3}, \\
\frac{4}{(\ell-3)(\ell-2) (2\ell-1)} &= \frac{4}{5}\, \frac{1}{\ell-3} - \frac{4}{3}\, \frac{1}{\ell-2} 
+ \frac{16}{15}\, \frac{1}{2\ell-1}
\end{align} 
\end{subequations} 
and make use of Eqs.~(\ref{SF0c}), (\ref{SF0d}), (\ref{WF2a}), (\ref{WF2b}), and (\ref{WF2d}). The end result in either case is  
\begin{align} 
\gamma[C_2] &= C_2 \biggl\{ \frac{2(x-x_0\cos\theta)\sin^2\theta}{x_0 x^3} 
\bigl[ \ln(E + x - x_0 \cos\theta) - \ln(1-\cos\theta) - \ln(x_0) \bigr] 
\nonumber \\ & \quad \mbox{}
+ \biggl( -\frac{4}{3x_0 x^3} + \frac{2\cos^2\theta}{x_0 x^3} 
- \frac{4\cos\theta}{3x_0^2 x^2} + \frac{2}{3x_0^3 x} \biggr) E 
+ \frac{4}{3x^3} - \frac{2\cos^2\theta}{x^3} + \frac{2\cos\theta}{x_0 x^2} 
- \frac{2}{x_0^2 x} + \frac{2}{3 x_0^3} \biggr\}. 
\label{gamma_C2}
\end{align} 

\subsection{Other contributions to the potentials} 

To the potentials of the preceding subsections we must add those of the $\sigma$-string, which were obtained in Sec.~\ref{sec:particle-massive}. The potentials, however, must be expanded in powers of $x^{-1}$ and $x_0^{-1}$ to reflect the weak-field approximation exploited in this section. To the required order we obtain 
\begin{equation} 
U[\sigma] = \sigma \biggl[ -\ln(E + x_0 - x\cos\theta) 
+ \frac{\sin^2\theta}{2 E (E + x_0 -x\cos\theta)} \biggr] 
\label{U_Delta} 
\end{equation} 
from Eq.~(\ref{U_massive}), and 
\begin{equation} 
\gamma[\sigma] = \sigma \biggl[ \frac{2(1-\cos\theta)(E + x + x_0)}{x_0 (E + x_0 - x\cos\theta)} 
+ \frac{(1-\cos\theta)^2\, \Xi}{3 x_0^3 E (E + x_0 - x\cos\theta)^3} \biggr] 
\label{gamma_Delta} 
\end{equation} 
from Eq.~(\ref{gamma_massive1}), where 
\begin{align} 
\Xi &:= 2(1-\cos\theta) E^4 + 2(1-\cos\theta) \bigl[ x + x_0 (3 + 2\cos\theta) \bigr] E^3 
+ 3x_0 \sin^2\theta \bigl[ 2x + x_0 (2+\cos\theta) \bigr] E^2 
\nonumber \\ & \quad \mbox{} 
+ x_0^2 (1+\cos\theta) \bigl[ x (7 + \cos\theta - 8\cos^2\theta) 
+ 2 x_0 (1 + 2\cos\theta) \bigr] E 
+ 3 x_0^3(1+\cos\theta)^2 \bigl[ x (1-2\cos^2\theta) + x_0 \cos\theta \bigr]. 
\label{Xi_def} 
\end{align} 
 
The potentials obtained thus far make up the solution to the (difference string) problem. The final solution to the ($\mbox{particle} + \mbox{massive string}$) problem is then the sum of these with those of the 
($\mbox{particle} + \mbox{massless string}$) problem, which were obtained in Sec.~\ref{sec:particle-massless}. After an expansion in powers of $x^{-1}$ and $x_0^{-1}$, we find that 
\begin{equation} 
U[k] = k \biggl( \frac{1}{E} + \frac{\sin^2\theta}{2 E^3} \biggr) 
\label{U_k}
\end{equation} 
from Eq.~(\ref{U_particle}), and 
\begin{equation} 
\gamma[k] = k \biggl[ \frac{2}{x_0^2} + \frac{2(x-x_0\cos\theta)}{x_0^2 E} 
+ \frac{2}{x_0^4} + \frac{2(x-x_0\cos\theta)}{x_0^4 E} 
+ \frac{(x-x_0\cos\theta)\sin^2\theta}{x_0^2 E^3} \biggr] 
\label{gamma_k}
\end{equation} 
from Eq.~(\ref{gamma_particle}). 

\subsection{Complete potentials} 

The complete potentials for the ($\mbox{particle} + \mbox{massive string}$) problem are 
\begin{subequations}
\label{potentials_generic}
\begin{align} 
U &= U[k] + U[\sigma] + U[C_1] + U[C_2], \\ 
\gamma &= \gamma[k] + \gamma[\sigma] + \gamma[C_1] + \gamma[C_2], 
\end{align} 
\end{subequations} 
where $U[k]$ is given by Eq.~(\ref{U_k}), $U[\sigma]$ by Eq.~(\ref{U_Delta}), $U[C_1]$ by Eq.~(\ref{U_C1}), $U[C_2]$ by Eq.~(\ref{U_C2}), and where $\gamma[k]$ is given by Eq.~(\ref{gamma_k}), $\gamma[\sigma]$ by Eq.~(\ref{gamma_Delta}), $\gamma[C_1]$ by Eq.~(\ref{gamma_C1}), $\gamma[C_2]$ by Eq.~(\ref{gamma_C2}). 

It is instructive to examine the asymptotic behavior of $U$ when $x$ is much larger than $x_0$. The individual contributions are given by 
\begin{subequations} 
\begin{align} 
U[k] &= \frac{k}{x_0} \biggl\{ (1 + \tfrac{1}{2} x^{-2} \sin^2\theta)\uu + \cos\theta\, \uu^2 
- \frac{1}{2}(1-3\cos^2\theta)\, \uu^3 + O(\uu^4) \biggr\}, \\
U[\sigma] &= \sigma \biggl\{ \ln\uu - \ln\bigl[ \tfrac{1}{2}(1-\cos\theta)\bigr] 
- (1 + \tfrac{1}{2} x^{-2} \sin^2\theta) \uu 
- \frac{1}{2}\cos\theta\, \uu^2 + \frac{1}{6}(1 - 3\cos^2\theta)\, \uu^3 + O(\uu^4) \biggr\}, \\
U[C_1] &= \frac{C_1}{x_0} \biggl\{  (1 + \uu) \ln\uu 
- (1-\uu) \ln\bigl[ \tfrac{1}{2}(1-\cos\theta)\bigr] 
- \uu + \frac{1}{2} \cos\theta\, \uu^2 - \frac{1}{12}(1 - 3\cos^2\theta)\, \uu^3 
+ O(\uu^4) \biggr\}, \\ 
U[C_2] &= \frac{C_2}{x_0^2} \biggl\{ (1 + 2\uu - \cos\theta\, \uu^2) \ln\uu 
- (1 -2\uu + \cos\theta\, \uu^2) \ln\bigl[ \tfrac{1}{2}(1-\cos\theta)\bigr] 
\nonumber \\ & \quad \mbox{}
+ \frac{3}{2} -  \frac{1}{2} (2 - \cos\theta)\, \uu^2 
+ \frac{1}{6} (1 - 3\cos^2\theta)\, \uu^3 
+ O(\uu^4) \biggr\}, 
\end{align} 
\end{subequations} 
where $\uu := x_0/x$. We see that as expected, the particle contributes to $U$ a term that decays as $1/x$, while the massive string contributes terms that diverge logarithmically, both at infinity and on the upper portion of the axis (at $\theta = 0$). 

It is also interesting to evaluate $\gamma$ on the axis. When $\theta = 0$ and $x > x_0$ (upper axis, above the particle) we find that 
\begin{equation} 
\gamma(\theta = 0, x > x_0) = \frac{4k}{x_0^2} \biggl(1 + \frac{1}{x_0^2} \biggr) 
- 4 \sigma \biggl[ \biggl( \frac{1}{x} - \frac{1}{x_0} \biggr) 
+ \frac{1}{3} \biggl( \frac{1}{x^3} - \frac{1}{x_0^3} \biggr) \biggr] 
+ 2 C_1 \biggl( \frac{1}{x} - \frac{1}{x_0} \biggr)^2 
- \frac{4}{3} C_2 \bigg( \frac{1}{x} - \frac{1}{x_0} \biggr)^3.
\end{equation} 
When $\theta = 0$ and $x < x_0$ (upper axis, below the particle) we find instead that
\begin{equation} 
\gamma(\theta = 0, x < x_0) = 0. 
\end{equation} 
And when $\theta = \pi$ (lower axis), we have that 
\begin{equation} 
\gamma(\theta = \pi) = \frac{4k}{x_0^2} \biggl(1 + \frac{1}{x_0^2} \biggr) 
+ 4 \sigma \biggl( \frac{1}{x_0} + \frac{1}{3x_0^3} \biggr)  
+ 2 C_1\, \frac{1}{x_0^2}  + \frac{4}{3} C_2\, \frac{1}{x_0^3}. 
\end{equation} 
All these results are summarized in the statement 
\begin{equation} 
\gamma^{\rm axis} = \left\{ 
\begin{array}{ll}
4T_{\rm up}(x) & \quad \mbox{upper axis},\ x > x_0 \\ 
0 & \quad \mbox{upper axis},\ x < x_0 \\ 
4T_{\rm dn} & \quad \mbox{lower axis} 
\end{array} \right., 
\end{equation} 
where the varying upper tension $T_{\rm up}(x)$ is given by Eq.~(\ref{T_generic}), with $T_0$ found in Eq.~(\ref{cons_eq}) and expanded in powers of $x_0^{-1}$, and where the constant lower tension $T_{\rm dn}$ is given by $T_\infty$, as written in Eq.~(\ref{Tinfty_generic}). This result was previously given a simplified expression in Eq.~(\ref{gamma_vs_T}).   

\begin{acknowledgments} 
This work was supported by the Natural Sciences and Engineering Research Council of Canada.  
\end{acknowledgments} 

\appendix

\section{Sum over tensional harmonics}
\label{sec:summation1} 

We provide a derivation of Eq.~(\ref{weylclass_G}), which features an infinite sum over the tensorial harmonics $P^\ell_{AB}$ defined by Eq.~(\ref{tensor_harmonics}). 

We multiply Eq.~(\ref{weylclass_Gcond}) by $\Omega^{AC} \Omega^{BD} P^{\ell'}_{CD} \sin\theta$, and integrate with respect to $\theta$, making use of the orthogonality relations of Eq.~(\ref{ortho_P}). We obtain
\begin{equation}
\frac{(\ell-1)\ell(\ell+1)(\ell+2)}{2\ell+1} G_\ell
= 2 \sum_{\ell'=0}^\infty J(\ell,\ell')\, g_{\ell'}
\label{Geq} 
\end{equation}
with 
\begin{equation}
J(\ell,\ell') := \int_0^\pi \Omega^{AC} \Omega^{BD} e_\stf{AB} P^\ell_{CD}\, P^{\ell'}\, \sin\theta\, d\theta.
\end{equation}
It is understood that $\ell \geq 2$. Working out the integrand and making the change of variable $u = \cos\theta$, we have that
\begin{equation}
J(\ell,\ell') = \int_{-1}^1 u P_\ell' P_{\ell'}\, du
- \frac{1}{2} \ell(\ell+1) \int_{-1}^1 P_\ell P_{\ell'}\, du
= \int_{-1}^1 u P_\ell' P_{\ell'}\, du - \frac{\ell(\ell+1)}{2\ell+1} \delta_{\ell\ell'},
\end{equation}
where a prime on $P_\ell(x)$ indicates differentiation with respect to $u$. If we apply to this the recursion relation $u P'_\ell = P_{\ell+1}' - (\ell+1) P_\ell$, we have that
\begin{equation}
J(\ell,\ell') = K(\ell+1, \ell') - \frac{(\ell+1)(\ell+2)}{2\ell+1} \delta_{\ell\ell'}, 
\end{equation} 
where
\begin{equation}
K(\ell, \ell') := \int_{-1}^1 P'_\ell P_{\ell'}\, du. 
\end{equation}
If we apply $u P'_\ell = P_{\ell-1}' + \ell P_\ell$ instead, we obtain
\begin{equation}
J(\ell,\ell') = K(\ell-1, \ell') - \frac{(\ell-1)\ell}{2\ell+1} \delta_{\ell\ell'}.  
\end{equation} 
The two equivalent expressions for $J(\ell,\ell')$ imply that $K(\ell,\ell')$ satisfies the recursion relation $K(\ell+1,\ell') = K(\ell-1,\ell') + 2 \delta_{\ell \ell'}$. Initial values can be computed from the definition. We have that $K(1,\ell') = 2\delta_{0\ell'}$ and $K(2,\ell') = 2\delta_{1\ell'}$, and the recursion relation gives
\begin{equation}
K(\ell+1,\ell') = 2 \left\{
\begin{array}{ll}
  \delta_{0\ell'} + \delta_{2\ell'} + \cdots + \delta_{\ell\ell'} & \qquad \mbox{$\ell$ even} \\ 
  \delta_{1\ell'} + \delta_{3\ell'} + \cdots + \delta_{\ell\ell'} & \qquad \mbox{$\ell$ odd}
\end{array} \right. . 
\end{equation}
Making the substitution in $J(\ell,\ell')$, we find that Eq.~(\ref{Geq}) gives
\begin{equation}
\frac{(\ell-1)\ell(\ell+1)(\ell+2)}{2\ell+1} G_\ell
= -\frac{2(\ell+1)(\ell+2)}{2\ell+1} g_\ell
+ 4 \left\{
\begin{array}{ll}
  g_0 + g_2 + \cdots + g_\ell & \qquad \mbox{$\ell$ even} \\ 
  g_1 + g_3 + \cdots + g_\ell & \qquad \mbox{$\ell$ odd} 
\end{array} \right. . 
\end{equation}
A slight rearrangement turns this into Eq.~(\ref{weylclass_G}).

\section{Regularity of the metric perturbation at $r=2M$}
\label{sec:regularity} 

We identify the conditions that ensure that a metric perturbation presented in the Weyl gauge is regular (as a tensor field) at $r=2M$.

According to Eqs.~(\ref{perturbation}), (\ref{Ug_decomp}), and (\ref{Wgauge}), the temporal and radial components of the metric tensor are given by
\begin{equation}
g_{tt} = -(1 - 2U) f, \qquad g_{rr} = (1 + 2U + 2\gamma) f^{-1}
\end{equation}
in the Weyl gauge. The angular components are $g_{AB} = r^2 \Omega_{AB} + p_{AB}$, and these are regular provided that $U$ and $\gamma$ both are, and provided also that the sum over all terms implicating $G_\ell$ is  regular. 

We decompose $U$ and $\gamma$ according to
\begin{equation}
U = u_0(r) + \bar{U}(r,\theta), \qquad
\gamma = g_0(r) + \bar{\gamma}(r,\theta),
\end{equation}
where $\bar{U}$ and $\bar{g}$ are the sums of Eq.~(\ref{Ug_decomp}) with the $\ell = 0$ terms omitted. The purpose of this is to isolate the spherically-symmetric piece of each perturbation variable. With the understanding that the perturbations are small, we have that
\begin{equation}
g_{tt} = -(1-2u_0)(1 - 2\bar{U}) f, \qquad
g_{rr} = (1 + 2u_0 + 2g_0)(1 + 2\bar{U} + 2\bar{\gamma}) f^{-1}.
\end{equation}

The regularity of the metric perturbation must be ascertained in a coordinate system that is itself regular at $r=2M$. For this purpose we make use of a variant of the Eddington-Finkelstein coordinates $(v,r)$, with the advanced-time coordinate $v$ defined by
\begin{equation}
dv = dt + (1 + 2u_0 + g_0) f^{-1}\, dr.
\end{equation}
A quick computation reveals that the transformation brings the line element to the new form
\begin{equation}
ds^2 = -(1 - 2u_0)(1 - 2 \bar{U}) f\, dv^2 + 2(1 + g_0)(1 - 2 \bar{U})\, dvdr
+ 2(2\bar{U} + \bar{\gamma}) f^{-1}\, dr^2 + \cdots,
\end{equation}
where the ellipsis represents the angular piece of the line element. Regularity at $r=2M$ requires that $u_0$ and $g_0$ be bounded there, that $\bar{U}$ be bounded, and that $2 \bar{U} + \bar{\gamma} = 0$, in order to compensate for the factor of $f^{-1}$ in front of $dr^2$. These are the conditions specified at the beginning of Sec.~\ref{sec:weyl-gauge}. 

\section{Hypergeometric and Legendre functions}
\label{sec:FvsL} 

We establish useful relations between hypergeometric functions and Legendre functions. A main resource for this material is the {\it NIST Handbook of Mathematical Functions} \cite{NIST:10}, hereafter refereed to as ``NIST''.  We let $z := 2M/r$ and $x := r/M-1$, so that $x = 2/z-1$ and $z = 2/(x+1)$. For $r \geq 2M$ we have that $z \leq 1$ and $x \geq 1$. It is also useful to note that $1-z = (x-1)/(x+1)$.

It is easy to verify that the functions
\begin{equation}
h_1 :=z^{-\ell} F(-\ell,-\ell;-2\ell;z), \qquad h_2 := z^{\ell+1} F(\ell+1, \ell+1; 2\ell+2; z)
\end{equation}
and
\begin{equation}
h_3 := P_\ell(x), \qquad h_4 := Q_\ell(x)
\end{equation}
all satisfy the differential equation
\begin{equation}
r^2 f \frac{d^2 h}{dr^2} + 2(r-M) \frac{dh}{dr} - \ell(\ell+1) h = 0.
\end{equation}
Because $h_1$ is a terminating polynomial in $z$, and $h_3$ is a terminating polynomial in $x$, these functions must be proportional to each other. In principle, $h_2$ could be a linear superposition of $h_3$ and $h_4$; an examination of many special cases reveals instead that $h_2$ is simply proportional to $h_4$.

To identify the ratio $h_1/h_3$ we examine the $z \to 0$ behavior of $h_1$, which corresponds to the $x \to \infty$ behavior of $h_3$. The leading-order term in $h_1$ is $z^{-\ell}$, and according to [NIST (14.8.12)], the leading-order term in $h_3$ is $[(2\ell-1)!!/\ell!] x^\ell$. Because $z \sim 2/x$ in this regime, we conclude that
\begin{equation}
z^{-\ell} F(-\ell,-\ell;-2\ell;z) = \frac{(\ell!)^2}{(2\ell)!}\, P_\ell(x).
\label{id1}
\end{equation}
To find the ratio $h_2/h_4$ we examine the $z \to 1$ behavior of $h_2$, which must match the $x \to 1$ behavior of $h_4$. According to [NIST (14.7.7)], $h_4 \sim -1/2 \ln(x-1)$, and from [NIST (15.8.10)] we infer that $h_2 \sim -[(2\ell+1)!/\ell!^2] \ln(1-z)$. With $\ln(1-z) \sim \ln(x-1)$, we have that
\begin{equation}
z^{\ell+1} F(\ell+1,\ell+1;2\ell+2;z) = \frac{2(2\ell+1)!}{(\ell!)^2}\, Q_\ell(x).
\label{id2}
\end{equation}

The identity [NIST (15.5.3)]
\begin{equation}
z \frac{d}{dz} \bigl[ z^a F(a,b;c;z) \bigr] = a z^a F(a+1, b; c; z)
\end{equation}
allows us to derive other relations between hypergeometric and Legendre functions; we note that $z d/dz = -(x+1) d/dx$. With $a=-\ell$, $b=-\ell$, and $c=-2\ell$ we get
\begin{equation}
z^{-\ell} F(-\ell+1,-\ell;-2\ell;z) = \frac{(\ell-1)!^2}{2(2\ell-1)!} (x+1) P'_\ell(x),
\label{id3}
\end{equation}
in which a prime indicates differentiation with respect to $x$. With $a = \ell+1$, $b=\ell+1$, $c=2\ell+2$ we get instead
\begin{equation}
z^{\ell+1} F(\ell+2,\ell+1;2\ell+2;z) = -\frac {2(2\ell+1)!}{\ell!\, (\ell+1)!} (x+1) Q'_\ell(x). 
\label{id4}
\end{equation}
With $a=-\ell+1$, $b=-\ell$, and $c=-2\ell$ we obtain
\begin{equation}
z^{-\ell} F(-\ell+2,-\ell;-2\ell;z) = \frac{(\ell-2)!\, (\ell-1)!}{2(2\ell-1)!} (x+1)^2 P''_\ell(x). 
\label{id5}
\end{equation}
And with $a=\ell+2$, $b=\ell+1$, $c=2\ell+2$ we arrive at
\begin{equation}
z^{\ell+1} F(\ell+3,\ell+1;2\ell+2;z) = \frac {2(2\ell+1)!}{\ell!\, (\ell+2)!} (x+1)^2 Q''_\ell(x). 
\label{id6}
\end{equation}

Another identity [NIST (15.5.5)],
\begin{equation}
z \frac{d}{dz} \bigl[ z^{c-a} (1-z)^{a+b-c} F(a,b;c;z) \bigr]
= (c-a) z^{c-a} (1-z)^{a+b-c-1} F(a-1,b;c;z),
\end{equation}
applied with $a=-\ell$, $b=-\ell$, $c=-2\ell$, allows us to deduce that
\begin{equation}
z^{-\ell} F(-\ell-1,-\ell;-2\ell;z) = \frac{(\ell-1)!^2}{2(2\ell-1)!} (x-1) P'_\ell(x).
\label{id7}
\end{equation}

\section{Integrals of products of Legendre functions} 
\label{sec:integrals}

We evaluate indefinite integrals of the form
\begin{equation}
J[A_\ell, B_\ell] := \int A'_\ell(x) B'_\ell(x)\, dx, 
\end{equation}
in which $A_\ell$ and $B_\ell$ are any solution to Legendre's equation (with the same value of $\ell$). We note that the integral is symmetric under an exchange of $A_\ell$ and $B_\ell$.

The first step is to integrate by parts, so that the derivative acting on $A_\ell$ is moved to $B'_\ell$. We then recall that $B_\ell^2 := (x^2-1) B''_\ell$ is an associated Legendre function, and write
\begin{equation}
J  = A_\ell B'_\ell - \int \frac{A_\ell\, B^2_\ell}{x^2-1}\, dx.
\end{equation}
In the second step we invoke the Legendre equation for $A_\ell$, the associated Legendre equation for $B_\ell^2$, and deduce the identity
\begin{equation}
\frac{A_\ell\, B^2_\ell}{x^2-1} = \frac{1}{4} \frac{d}{dx} \Bigl[ (x^2-1) \bigl( A_\ell B^{2\prime}_\ell
- B_\ell^2 A'_\ell \bigr) \Bigr].
\end{equation}
Integration is now immediate. In the third step we relate $B^2_\ell$ and $B^{2\prime}_\ell$ to $B'_\ell$ and $B_\ell$ by making repeated use of Legendre's equation for $B_\ell$. After some simplifying algebra we obtain
\begin{equation}
J =\frac{1}{2} \ell(\ell+1) x A_\ell B_\ell - \frac{1}{2} (x^2-1) A_\ell B'_\ell
- \frac{1}{4} \ell(\ell+1) (x^2 - 1) {\cal W} 
- \frac{1}{2} x (x^2-1) A'_\ell B'_\ell + \mbox{constant},
\end{equation}
where ${\cal W} := A_\ell B'_\ell - A'_\ell B_\ell$ is the Wronskian of the two solutions to Legendre's equation. This, of course, vanishes when $A_\ell$ and $B_\ell$ are linearly dependent. 

We notice that the symmetry with respect to $A_\ell$ and $B_\ell$ appears to be lost. In the fourth step we restore it by writing $A_\ell B'_\ell = \frac{1}{2} (A_\ell B'_\ell + A'_\ell B_\ell) + \frac{1}{2} {\cal W}$, and by recalling that $(x^2 - 1) {\cal W}$ is a constant. The final result is
\begin{equation}
J[A_\ell, B_\ell] = \frac{1}{2} \ell(\ell+1) x A_\ell B_\ell
- \frac{1}{4} (x^2-1) \bigl( A_\ell B'_\ell + A'_\ell B_\ell \bigr) 
- \frac{1}{2} x (x^2-1) A'_\ell B'_\ell + \mbox{constant}.
\label{JAB} 
\end{equation}

\section{Summation formulae} 
\label{sec:summation} 

We establish a number of summation identities that are required in the main text. 

\subsection{Functions of $\cos\theta$} 

The first set of identities involves functions of $\cos\theta$ only. They are 
\begin{subequations} 
\begin{align} 
\sum_{\ell=1}^\infty \frac{2\ell+1}{\ell(\ell+1)}\, P_\ell(\cos\theta) 
&= -\ln(1-\cos\theta) + \ln 2 - 1, 
\label{SF0} \\ 
\sum_{\ell=2}^\infty \frac{2\ell+1}{(\ell-1)(\ell+2)}\, P_\ell(\cos\theta) 
&= -\cos\theta \bigl[ \ln(1-\cos\theta) - \ln 2 \bigr] - \frac{4}{3} \cos\theta - \frac{1}{2}, 
\label{SF0b} \\ 
\sum_{\ell=3}^\infty \frac{2\ell+1}{(\ell-2)\ell(\ell+1)(\ell+3)}\, P_\ell(\cos\theta) 
&= \frac{1}{4} \sin^2\theta \bigl[ \ln(1-\cos\theta) - \ln 2 \bigr] 
- \frac{2}{15} \cos^2\theta + \frac{1}{8} \cos\theta + \frac{11}{60}, 
\label{SF0c}  \\
\sum_{\ell=4}^\infty \frac{2\ell+1}{(\ell-3)(\ell-1)(\ell+2)(\ell+4)}\, P_\ell(\cos\theta) 
&= \frac{1}{4} \cos\theta \sin^2\theta \bigl[ \ln(1-\cos\theta) - \ln 2 \bigr] 
\nonumber \\ & \quad \mbox{} 
- \frac{71}{420} \cos^3\theta + \frac{1}{16} \cos^2\theta + \frac{43}{210} \cos\theta + \frac{1}{48}. 
\label{SF0d} 
\end{align} 
\end{subequations} 
If we change the sign in front of $\cos\theta$ in Eq.~(\ref{SF0}) and exploit the even/odd nature of the Legendre polynomials, we find that it becomes 
\begin{equation} 
\sum_{\ell=1}^\infty \frac{2\ell+1}{\ell(\ell+1)}\, (-1)^\ell P_\ell(\cos\theta)
= -\ln(1+\cos\theta) + \ln 2 - 1. 
\label{SF00} 
\end{equation} 

The derivation of these results is a straightforward application of Legendre series, in which a function $C(u)$ is decomposed as 
\begin{equation} 
C(u) = \sum_{\ell=0}^\infty c_\ell\, P_\ell(u), 
\end{equation} 
with coefficients given by 
\begin{equation} 
c_\ell = \frac{1}{2} (2\ell+1) \int_{-1}^1 C(u)\, P_\ell(u)\, du.  
\label{cell} 
\end{equation} 
We use the notation $u := \cos\theta$. 

We begin with Eq.~(\ref{SF0}), and set $C(u) = \ln(1-u)$. For $\ell = 0$ the integration in Eq.~(\ref{cell}) is immediate, and we obtain $c_0 = \ln 2 - 1$. For $\ell \geq 1$ we make use of Legendre's equation, 
\begin{equation}
\frac{d}{du} \Bigl[ (1-u^2) P_\ell'(u) \Bigr] + \ell(\ell+1) P_\ell(u) = 0, 
\label{Legendre1} 
\end{equation} 
to replace $P_\ell(u)$ within the integral; a prime indicates differentiation with respect to $u$. After integrating by parts and setting the boundary terms to zero, we find that 
\begin{equation} 
c_\ell = -\frac{2\ell+1}{2\ell(\ell+1)} \int_{-1}^1 (1 + u) P'_\ell(u)\, du. 
\end{equation} 
Another integration by parts returns 
\begin{equation} 
c_\ell = -\frac{2\ell+1}{2\ell(\ell+1)} \biggl[ (1 + u) P_\ell(u) \biggr|^1_{-1} 
- \int_{-1}^1 P_\ell(u)\, du \biggr].  
\end{equation} 
The boundary terms evaluate to $2$, and the remaining integral vanishes when $\ell \geq 1$. We arrive at 
\begin{equation} 
c_\ell = -\frac{2\ell+1}{\ell(\ell+1)}, 
\end{equation} 
in agreement with Eq.~(\ref{SF0}). 

We follow the same steps to establish Eq.~(\ref{SF0b}). In this case we set $C(u) = u\ln(1-u)$, and for $\ell = \{0, 1\}$ the corresponding coefficients are $c_0 = -1/2$ and $c_1 = \ln 2 - 4/3$. For $\ell \geq 2$ we obtain 
\begin{equation} 
c_\ell = \frac{2\ell+1}{2\ell(\ell+1)} \int_{-1}^1 \bigl[ (1-u^2) \ln(1-u) - u(1+u) \bigr] P_\ell'(u)\, du 
\end{equation} 
after the first integration by parts. The second one produces 
\begin{equation} 
c_\ell = \frac{2\ell+1}{2\ell(\ell+1)} \biggl\{ 
\bigl[ (1-u^2) \ln(1-u) - u(1+u) \bigr] P_\ell(u) \biggr|^1_{-1} 
+ 2 \int_{-1}^1 u \ln(1-u)\, P_\ell(u)\, du 
+ \int_{-1}^1 (1 + 2u)\, P_\ell(u)\, du \biggr\}. 
\end{equation} 
The boundary terms evaluate to $-2$, the last integral vanishes when $\ell \geq 2$, and the first integral is proportional to $c_\ell$. We arrive at $c_\ell = -(2\ell+1)/[(\ell-1)(\ell+2)]$, in agreement with Eq.~(\ref{SF0b}). 

To derive Eq.~(\ref{SF0c}) we first set $C(u) = u^2\ln(1-u)$ and go through the preceding steps to obtain 
\begin{equation} 
c_\ell = -\frac{(2\ell+1)(\ell-1)(\ell+2)}{(\ell-2)\ell(\ell+1)(\ell+3)} 
\end{equation} 
when $\ell \geq 3$. We next combine this with Eq.~(\ref{SF0}) and get the Legendre series for $(1-u^2)\ln(1-u)$. We arrive at Eq.~(\ref{SF0c}) after calculating the coefficients for the special cases $\ell = \{0, 1, 2\}$. 

For Eq.~(\ref{SF0d}) we begin with $C(u) = u^3\ln(1-u)$, for which we get 
\begin{equation} 
c_\ell = -\frac{(2\ell+1)(\ell^2+\ell-8)}{(\ell-3)(\ell-1)(\ell+2)(\ell+4)}
\end{equation} 
when $\ell \geq 4$. Then we combine this with Eq.~(\ref{SF0b}) to obtain the Legendre series for $u(1-u^2)\ln(1-u)$. The final result is Eq.~(\ref{SF0d}).  

\subsection{Strong-field formulae} 

We present a derivation of the identities \cite{bini-geralico-ruffini:07}
\begin{subequations} 
\begin{align} 
\frac{1}{D} &= \sum_{\ell=0}^\infty (2\ell+1) 
\left\{ \begin{array}{c}
Q_\ell(y) P_\ell(x) \\ 
P_\ell(y) Q_\ell(x) 
\end{array} \right\} P_\ell(\cos\theta), 
\label{SF1} \\ 
\frac{xy-\cos\theta}{D} &= 
\left\{ \begin{array}{c}
x \\
y 
\end{array} \right\} 
- (x^2-1)(y^2-1) \sum_{\ell=1}^\infty \frac{2\ell+1}{\ell(\ell+1)}  
\left\{ \begin{array}{c}
Q'_\ell(y) P'_\ell(x) \\ 
P'_\ell(y) Q'_\ell(x) 
\end{array}
\right\} P_\ell(\cos\theta), 
\label{SF2} 
\end{align} 
\end{subequations} 
where 
\begin{equation} 
D := (x^2 - 2xy \cos\theta + y^2 - \sin^2\theta)^{1/2}. 
\label{D_def} 
\end{equation} 
In these equations, the upper row refers to the case $x < y$, while the lower row refers to $x > y$. We also establish that   
\begin{subequations} 
\begin{align} 
\sum_{\ell=1}^\infty \frac{2\ell+1}{\ell(\ell+1)} (y^2-1) Q'_\ell(y)\, P_\ell(x)\, P_\ell(\cos\theta) 
&= \ln(D + y - x\cos\theta) + \frac{1}{2} (y-1) \ln(y-1) - \frac{1}{2} (y+1)\ln(y+1)
\nonumber \\ & \quad \mbox{}  
+ 1 - \ln 2 \qquad (x < y), 
\label{SF3a} \\ 
\sum_{\ell=1}^\infty \frac{2\ell+1}{\ell(\ell+1)} (y^2-1) P'_\ell(y)\, Q_\ell(x)\, P_\ell(\cos\theta) 
&= \ln(D + y - x\cos\theta) + \frac{1}{2} (y-1) \ln(x-1) - \frac{1}{2} (y+1)\ln(x+1)
\nonumber \\ & \quad \mbox{}  
- \ln(1-\cos\theta) \qquad (x > y),
\label{SF3b} \\ 
\sum_{\ell=1}^\infty \frac{2\ell+1}{[\ell(\ell+1)]^2} (y^2-1) Q'_\ell(y)\, (x^2-1) P'_\ell(x)\, P_\ell(\cos\theta) 
&= \frac{1}{2} (xy - x - y) \ln(y-1) - \frac{1}{2} (xy + x + y) \ln(y+1) 
\nonumber \\ & \quad \mbox{}  
- y\ln(1-\cos\theta) - x\ln 2 + x\ln(D + y - x\cos\theta) 
\nonumber \\ & \quad \mbox{}  
+ y\ln(D + x - y\cos\theta) 
+ \frac{1}{2} \ln \frac{\Phi_+}{\Phi_-}  \qquad (x < y), 
\label{SF4} 
\end{align}
\end{subequations} 
where 
\begin{equation} 
\Phi_\pm := (y \pm \cos\theta)(D + y \pm \cos\theta) - (y\cos\theta \pm 1)(x \pm 1). 
\end{equation} 

To obtain Eq.~(\ref{SF1}) we rely on the fact that ${\cal G}(\bm{r},\bm{r'}) = |\bm{r}-\bm{r'}|^{-1}$ is a Green's function for Poisson's equation, so that 
\begin{equation} 
\nabla^2 {\cal G}(\bm{r},\bm{r'}) = -4\pi \delta(\bm{r}-\bm{r'}). 
\end{equation} 
Here, the position vectors $\bm{r}$ and $\bm{r'}$ are defined in a flat, three-dimensional space, and $|\bm{r}-\bm{r'}|$ is the Euclidean distance between points at $\bm{r}$ and $\bm{r'}$. The strategy is to express Green's equation in elliptical coordinates $(s,\theta,\phi)$, defined in terms of Cartesian coordinates $(X,Y,Z)$ by 
\begin{equation} 
X = \sqrt{s^2-1}\, \sin\theta\cos\phi, \qquad 
Y = \sqrt{s^2-1}\, \sin\theta\sin\phi, \qquad 
Z = s\cos\theta, 
\end{equation} 
and to represent ${\cal G}$ as a sum over Legendre polynomials. The sum will then be identified with $|\bm{r}-\bm{r'}|^{-1}$, also expressed in elliptical coordinates. 

The metric of flat space is given by 
\begin{equation} 
ds^2 = \frac{s^2-\cos^2\theta}{s^2-1}\, ds^2 + (s^2-\cos^2\theta)\, d\theta^2 + (s^2-1)\sin^2\theta\, d\phi^2 
\end{equation} 
in elliptical coordinates. We have that $\sqrt{g} = (s^2-\cos^2\theta)\sin\theta$, and for any function $\psi(s,\theta)$, the action of the Laplacian operator is given by 
\begin{equation} 
\nabla^2 \psi = \frac{1}{s^2-\cos^2\theta} \biggl\{ \partial_s \bigr[ (s^2-1) \partial_s \psi \bigr] 
+ \frac{1}{\sin\theta} \partial_\theta \bigl[ \sin\theta\, \partial_\theta \psi \bigr] \biggr\}. 
\end{equation} 
The point at $\bm{r}$ is given the coordinates $(s,\theta,\phi)$, and the point at $\bm{r'}$ is placed on the polar axis, so that $s' = s_0$ and $\theta' = 0$. The $\phi$-average of $\delta(\bm{r}-\bm{r'})$ is given by 
\begin{equation} 
\langle \delta(\bm{r}-\bm{r'}) \rangle = \frac{1}{2\pi(s^2-\cos^2\theta)}\delta(s-s_0) \delta(\cos\theta-1),  
\end{equation} 
and Green's equation becomes 
\begin{equation} 
\partial_s \bigr[ (s^2-1) \partial_s {\cal G} \bigr] 
+ \frac{1}{\sin\theta} \partial_\theta \bigl[ \sin\theta\, \partial_\theta {\cal G} \bigr] 
= -2 \delta(s-s_0) \delta(\cos\theta-1). 
\end{equation} 

To integrate this equation we expand ${\cal G}$ and $\delta(\cos\theta-1)$ in Legendre polynomials, 
\begin{equation} 
{\cal G}(s,\theta) = \sum_{\ell=0}^\infty {\cal G}_\ell(s) P_\ell(\cos\theta), \qquad 
\delta(\cos\theta-1) = \frac{1}{2} \sum_{\ell=0}^\infty (2\ell+1) P_\ell(\cos\theta). 
\end{equation} 
Making the substitution, we obtain 
\begin{equation} 
(s^2-1) {\cal G}_\ell'' + 2s\, {\cal G}_\ell' - \ell(\ell+1) {\cal G}_\ell = -(2\ell+1) \delta(s-s_0). 
\end{equation} 
The solution is 
\begin{equation} 
{\cal G}_\ell = (2\ell+1) \left\{ \begin{array}{ll} 
Q_\ell(s_0) P_\ell(s) & \quad s < s_0 \\ 
P_\ell(s_0) Q_\ell(s) & \quad s > s_0 
\end{array} \right. , 
\end{equation} 
and the Green's function can therefore be expressed as 
\begin{equation} 
{\cal G} = \sum_{\ell = 0}^\infty (2\ell+1) \left\{ \begin{array}{c} 
Q_\ell(s_0) P_\ell(s) \\ 
P_\ell(s_0) Q_\ell(s)  
\end{array} \right\} P_\ell(\cos\theta). 
\end{equation} 
With $s = x$ and $s_0 = y$, this is the same as the right-hand side of Eq.~(\ref{SF1}). 

On the other hand, simple algebra reveals that 
\begin{equation} 
|\bm{r} - \bm{r'}| = \bigr( s^2 - 2s_0 s\cos\theta + s_0^2 - \sin^2\theta\bigr)^{1/2}, 
\end{equation} 
and since ${\cal G} = |\bm{r} - \bm{r'}|^{-1}$, we also have the left-hand side of Eq.~(\ref{SF1}). The summation formula is therefore established. 

Next we turn to  Eq.~(\ref{SF2}). We begin with Eq.~(\ref{SF1}), which we differentiate with respect to $x$ and $y$. We have 
\begin{align} 
\partial_{xy} D^{-1} &= \sum_{\ell=1}^\infty (2\ell+1) \left\{ \begin{array}{c} 
Q_\ell'(y) P'_\ell(x) \\ 
P_\ell'(y) Q'_\ell(x) 
\end{array} \right\} P_\ell(\cos\theta)
\nonumber \\ 
&=-\sum_{\ell=1}^\infty \frac{2\ell+1}{\ell(\ell+1)} \left\{ \begin{array}{c} 
Q_\ell'(y) P'_\ell(x) \\ 
P_\ell'(y) Q'_\ell(x) 
\end{array} \right\} 
\frac{1}{\sin\theta} \frac{d}{d\theta} \biggl( \sin\theta \frac{dP_\ell}{d\theta} \biggr).
\end{align} 
We used Legendre's equation in the second step, and a prime indicates differentiation with respect to the argument. We multiply both sides by $\sin\theta$ and integrate with respect to $\theta$, to obtain 
\begin{equation} 
\frac{\sin^2\theta}{D^3} = -\sum_{\ell=1}^\infty \frac{2\ell+1}{\ell(\ell+1)} \left\{ \begin{array}{c} 
Q_\ell'(y) P'_\ell(x) \\ 
P_\ell'(y) Q'_\ell(x) 
\end{array} \right\} \sin\theta \frac{dP_\ell}{d\theta}. 
\end{equation} 
The constant of integration, a function of $x$ and $y$, is set to zero by evaluating both sides of the equation at $\theta = 0$. Next we divide by $\sin\theta$ and integrate again, to find that 
\begin{equation} 
\frac{xy-\cos\theta}{D} = h(x,y) - (x^2-1)(y^2-1) 
\sum_{\ell=1}^\infty \frac{2\ell+1}{\ell(\ell+1)} \left\{ \begin{array}{c} 
Q_\ell'(y) P'_\ell(x) \\ 
P_\ell'(y) Q'_\ell(x) 
\end{array} \right\} P_\ell(\cos\theta). 
\label{tmp1} 
\end{equation} 
We have obtained the left-hand side of Eq.~(\ref{SF2}) and most of the right-hand side, but we have yet to identify the function $h(x,y)$. 

To determine $h(x,y)$ we first assume that $x < y$, and set $\theta = 0$ in Eq.~(\ref{tmp1}), so that   
\begin{equation} 
\frac{xy-1}{y-x} - h(x,y) = -(x^2-1)(y^2-1) 
\sum_{\ell=1}^\infty \frac{2\ell+1}{\ell(\ell+1)} Q_\ell'(y) P'_\ell(x).
\label{tmp2} 
\end{equation} 
We make the same assumption and substitution in Eq.~(\ref{SF1}), and get 
\begin{equation} 
\frac{1}{y-x} = \sum_{\ell=0}^\infty (2\ell+1) Q_\ell(y) P_\ell(x) 
 = Q_0(y) +\sum_{\ell=1}^\infty \frac{2\ell+1}{\ell(\ell+1)} Q_\ell(y) 
\frac{d}{dx} \biggl[ (x^2-1) \frac{d P_\ell}{dx} \biggr], 
\end{equation} 
where we used Eq.~(\ref{Legendre1}) in the second step. We integrate with respect to $x$, 
\begin{equation} 
-\ln(y-x) + h(y) - xQ_0(y) =\sum_{\ell=1}^\infty \frac{2\ell+1}{\ell(\ell+1)} Q_\ell(y) 
(x^2-1) P'_\ell(x), 
\end{equation} 
and to determine the constant of integration, we set $x = 1$ to find that $h(y) = \frac{1}{2} \ln(y^2-1)$. Inserting this within the preceding equation and differentiating with respect to $y$, we arrive at 
\begin{equation} 
\frac{x^2-1}{y-x} = -(x^2-1)(y^2-1) \sum_{\ell=1}^\infty \frac{2\ell+1}{\ell(\ell+1)} Q'_\ell(y) P'_\ell(x). 
\end{equation} 
We have the right-hand side of Eq.~(\ref{tmp2}), and comparing the left-hand sides, we conclude that $h(x,y) = x$ when $x < y$. 

The case $x > y$ can be handled in a similar fashion, but it is simpler to observe that since the left-hand side of Eq.~(\ref{tmp1})  is symmetric under an exchange of $x$ and $y$, the same must be true of $h(x,y)$. So if $h$ is equal to $x$ when $x < y$, it must be equal to $y$ when $x > y$. We therefore have 
\begin{equation} 
h(x,y) = \left\{ \begin{array}{c} 
x \\ 
y \end{array} \right\}, 
\end{equation} 
and inserting this within Eq.~(\ref{tmp1}), we obtain Eq.~(\ref{SF2}).  

To establish Eq.~(\ref{SF3a}) we begin with Eq.~(\ref{SF1}) for $x < y$, re-expressed as 
\begin{equation} 
\sum_{\ell=1}^\infty \frac{2\ell+1}{\ell(\ell+1)} \frac{d}{dy} \Bigl[ (y^2-1) Q'_\ell(y) \Bigr]\, P_\ell(x)\, P_\ell(\cos\theta) = \frac{1}{D} + \frac{1}{2} \ln(y-1) - \frac{1}{2} \ln(y+1). 
\end{equation} 
We integrate with respect to $y$ and obtain 
\begin{equation} 
\sum_{\ell=1}^\infty \frac{2\ell+1}{\ell(\ell+1)} (y^2-1) Q'_\ell(y)\, P_\ell(x)\, P_\ell(\cos\theta) 
= \ln(D + y - x\cos\theta) + \frac{1}{2} (y-1) \ln(y-1) - \frac{1}{2} (y+1)\ln(y+1) + 1 + h(x,\theta), 
\end{equation} 
where $h(x,\theta)$ is a constant of integration. In the limit $y \to \infty$ the sum evaluates to zero, thanks to the decaying property of $Q'_\ell(y)$, and we find that $h(x,\theta) = -\ln 2$. We have arrived at Eq.~(\ref{SF3a}). 

We follow a very similar strategy to derive Eq.~(\ref{SF3b}). Equation (\ref{SF1}) with $x > y$ can be expressed as 
\begin{equation} 
\sum_{\ell=1}^\infty \frac{2\ell+1}{\ell(\ell+1)} \frac{d}{dy} \Bigl[ (y^2-1) P'_\ell(y) \Bigr]\, Q_\ell(x)\, P_\ell(\cos\theta) = \frac{1}{D} + \frac{1}{2} \ln(x-1) - \frac{1}{2} \ln(x+1),  
\end{equation} 
and integration with respect to $y$ yields 
\begin{equation} 
\sum_{\ell=1}^\infty \frac{2\ell+1}{\ell(\ell+1)} (y^2-1) P'_\ell(y)\, Q_\ell(x)\, P_\ell(\cos\theta) 
= \ln(D + y - \cos\theta) + \frac{1}{2} y\ln(x-1) - \frac{1}{2} y \ln(x+1) + h(x,\theta), 
\end{equation} 
where $h(x,\theta)$ is a new constant of integration. We evaluate this equation at $y = 1$, where the sum vanishes, and where $D = x - \cos\theta$. We find that $h = -\frac{1}{2} \ln(x-1) - \frac{1}{2} \ln(x+1) - \ln(1-\cos\theta)$, and this gives us Eq.~(\ref{SF3b}). 

For Eq.~(\ref{SF4}) we begin with Eq.~(\ref{SF3a}), which we rewrite as 
\begin{align} 
\sum_{\ell=1}^\infty \frac{2\ell+1}{[\ell(\ell+1)]^2} (y^2-1) Q'_\ell(y)\, 
\frac{d}{dx} \Bigl[ (x^2-1) P'_\ell(x) \Bigr]\, P_\ell(\cos\theta) 
&= \ln(D + y - x\cos\theta) +\frac{1}{2} (y-1) \ln(y-1) 
\nonumber \\ & \quad \mbox{} 
- \frac{1}{2} (y+1)\ln(y+1) + 1 - \ln 2. 
\end{align} 
We integrate with respect to $x$ and find 
\begin{align} 
\sum_{\ell=1}^\infty \frac{2\ell+1}{[\ell(\ell+1)]^2} (y^2-1) Q'_\ell(y)\, 
(x^2-1) P'_\ell(x)\, P_\ell(\cos\theta) 
&= \frac{1}{2} x(y-1) \ln(y-1) - \frac{1}{2} x(y+1)\ln(y+1) - x \ln 2 
\nonumber \\ & \quad \mbox{} 
+ x\ln(D + y - x\cos\theta) + y\ln(D + x - y\cos\theta) 
\nonumber \\ & \quad \mbox{} 
+ \frac{1}{2} \ln\frac{\Phi_+}{\Phi_-} + h(y,\theta), 
\end{align} 
where $h(y,\theta)$ is yet another constant of integration. To determine it we evaluate the preceding equation at $x=1$, noting that $D = y -\cos\theta$, $\Phi_+ = 2(y^2-1)$, $\Phi_- = 2(y-\cos\theta)^2$, and that the sum vanishes. This gives us $h = -\frac{1}{2}y\ln(y-1) - \frac{1}{2}y\ln(y+1) - y\ln(1-\cos\theta)$, and we arrive at Eq.~(\ref{SF4}).  

\subsection{Weak-field formulae} 

We conclude with two more sets of summation identities. For the first set we let $x < y$. We have 
\begin{subequations} 
\label{WF1} 
\begin{align} 
\sum_{\ell=1}^\infty \frac{1}{\ell(\ell+1)} \frac{x^\ell}{y^{\ell+1}}\, P_\ell(\cos\theta)
&= \frac{1}{x} \bigl[ \ln y + \ln(1-\cos\theta) - \ln(E + x - y\cos\theta) \bigr] 
\nonumber \\ & \quad \mbox{} 
+ \frac{1}{y} \bigl[ 1 + \ln 2 + \ln y - \ln(E + y - x\cos\theta) \bigr], 
\label{WF1a} \\ 
\sum_{\ell=1}^\infty \biggl( \frac{1}{2\ell+3} \frac{x^{\ell+1}}{y^{\ell+3}} 
- \frac{1}{2\ell-1} \frac{x^{\ell-1}}{y^{\ell+1}} \biggr)\, P_\ell(\cos\theta) 
&= \frac{E}{xy^2} - \frac{x}{3y^3} - \frac{1}{xy}, 
\label{WF1b} \\   
\sum_{\ell=1}^\infty \biggl( \frac{1}{\ell+3} \frac{x^{\ell+1}}{y^{\ell+3}}
- \frac{1}{\ell+1} \frac{x^{\ell-1}}{y^{\ell+1}} \biggr)\, P_\ell(\cos\theta)
&= \frac{3\sin^2\theta}{2x^2} \bigl[ \ln y + \ln(1-\cos\theta) - \ln(E + x - y\cos\theta) \bigr] 
\nonumber \\ & \quad \mbox{} 
+ \frac{x + 3y\cos\theta}{2 x^2 y^2}\, E - \frac{3\cos\theta}{2x^2} + \frac{1}{xy} - \frac{x}{3y^3}, 
\label{WF1c} \\ 
\sum_{\ell=1}^\infty \frac{2}{\ell(\ell+1)(\ell+2)} \frac{x^\ell}{y^{\ell+2}}\, P_\ell(\cos\theta) 
&= \biggl( \frac{\cos\theta}{x^2} - \frac{2}{xy} \biggr) \bigl[ \ln(E + x - y\cos\theta) 
- \ln(1-\cos\theta) - \ln y \bigr] 
\nonumber \\ & \quad \mbox{} 
- \frac{1}{y^2} \bigl[ \ln(E + y - x\cos\theta) - \ln y - \ln 2 \bigr] 
+ \frac{E}{x^2 y} - \frac{1}{x^2} + \frac{3}{2y^2}, 
\label{WF1d} \\
\sum_{\ell=1}^\infty \biggl( \frac{1}{\ell+4} \frac{x^{\ell+1}}{y^{\ell+4}} 
- \frac{1}{\ell+2} \frac{x^{\ell-1}}{y^{\ell+2}} \biggr)\, P_\ell(\cos\theta) 
&= \frac{5 \cos\theta \sin^2\theta}{2 x^3} \bigl[ \ln y + \ln(1-\cos\theta) - \ln(E + x - y\cos\theta) \bigr] 
\nonumber \\ & \quad \mbox{} 
+ \biggl( \frac{1}{3xy^3} - \frac{5}{3x^3y} + \frac{5\cos\theta}{6x^2 y^2} 
+ \frac{5\cos^2\theta}{2x^3 y} \biggr) E 
\nonumber \\ & \quad \mbox{} 
+ \frac{5}{6x^3} (2-3\cos^2\theta) + \frac{1}{2x y^2} - \frac{x}{4 y^4}, 
\label{WF1e}
\end{align} 
\end{subequations}
where 
\begin{equation} 
E := (x^2 - 2xy \cos\theta + y^2)^{1/2}. 
\end{equation} 
For the second set we let $x > y$. We have 
\begin{subequations} 
\label{WF2} 
\begin{align} 
\sum_{\ell=3}^\infty \biggl( \frac{1}{\ell} \frac{y^\ell}{x^{\ell+2}} 
- \frac{1}{\ell-2} \frac{y^{\ell-2}}{x^\ell} \biggr)\, P_\ell(\cos\theta) 
&= \frac{3\sin^2\theta}{2x^2} \bigl[ \ln 2 + \ln x - \ln(E + x - y\cos\theta) \bigr] 
+ \frac{x + 3y\cos\theta}{2 x^2 y^2} E 
\nonumber \\ & \quad \mbox{} 
- \frac{1}{2y^2} - \frac{\cos\theta}{xy} 
+ \frac{7\cos^2\theta - 1}{4 x^2} - \frac{y \cos\theta}{x^3} 
- \frac{y^2}{4 x^4} (3\cos^2\theta - 1), 
\label{WF2a} \\ 
\sum_{\ell=3}^\infty \biggl( \frac{1}{2\ell+3} \frac{y^\ell}{x^{\ell+2}} 
- \frac{1}{2\ell-1} \frac{y^{\ell-2}}{x^\ell} \biggr)\, P_\ell(\cos\theta) 
&= \frac{E}{x y^2} - \frac{1}{3x^2} - \frac{1}{y^2} 
- \biggl( \frac{y}{5x^3} - \frac{1}{xy} \biggr) \cos\theta
- \biggl( \frac{y^2}{14 x^4} - \frac{1}{6 x^2} \biggr) (3\cos^2\theta - 1), 
\label{WF2b} \\ 
\sum_{\ell=2}^\infty \frac{2}{(\ell-1)\ell(\ell+1)} \frac{y^{\ell-1}}{x^{\ell+1}}\, P_\ell(\cos\theta) 
&= -\biggl( \frac{\cos\theta}{x^2} - \frac{2}{xy} \biggr) \bigl[ \ln(E + x - y\cos\theta) 
- \ln x - \ln 2 \bigr] 
\nonumber \\ & \quad \mbox{} 
+ \frac{1}{y^2} \bigl[ \ln(E + y - x\cos\theta) - \ln x - \ln(1-\cos\theta) \bigr] 
- \frac{E}{x^2 y} + \frac{\cos\theta}{2x^2}, 
\label{WF2c} \\ 
\sum_{\ell=4}^\infty \biggl( \frac{1}{\ell-1} \frac{y^{\ell-1}}{x^{\ell+2}} 
- \frac{1}{\ell - 3} \frac{y^{\ell-3}}{x^\ell} \biggr)\, P_\ell(\cos\theta) 
&= \frac{5\cos\theta \sin^2\theta}{2 x^3} \bigl[ \ln 2 + \ln x - \ln(E + x - y\cos\theta) \bigr] 
\nonumber \\ & \quad \mbox{} 
+ \biggl( \frac{1}{3x y^3} - \frac{5}{3 x^3 y} + \frac{5\cos\theta}{6 x^2 y^2} 
+ \frac{5 \cos^2\theta}{2 x^3 y} \biggr) E 
- \biggl( \frac{5y^2}{4 x^5} - \frac{37}{12 x^3} \biggr) \cos^3\theta 
\nonumber \\ & \quad \mbox{} 
- \frac{3}{2} \biggl( \frac{1}{x^2 y} + \frac{y}{x^4} \biggr) \cos^2\theta 
- \biggl( \frac{1}{2 x y^2} - \frac{3y^2}{4x^5} + \frac{9}{4x^3} \biggr) \cos\theta 
\nonumber \\ & \quad \mbox{} 
- \frac{1}{3y^3} + \frac{3}{2x^2 y} + \frac{y}{2x^4}. 
\label{WF2d} 
\end{align} 
\end{subequations} 
 
The function $E^{-1}$ is intimately tied to the generating function for Legendre polynomials, and each identity in the listing of Eqs.~(\ref{WF1}) originates from
\begin{equation} 
\frac{1}{E} = \sum_{\ell=0}^\infty \frac{x^\ell}{y^{\ell+1}}\, P_\ell(\cos\theta). 
\end{equation} 
To establish Eq.~(\ref{WF1a}) we write the left-hand side as 
\begin{equation} 
\sum_{\ell=1}^\infty \biggl( \frac{1}{\ell} - \frac{1}{\ell+1} \biggr) \frac{x^\ell}{y^{\ell+1}}\, P_\ell(\cos\theta) 
= \int \frac{1}{x} \biggl( \frac{1}{E} - \frac{1}{y} \biggr)\, dx 
+ \int \frac{1}{y} \biggl( \frac{1}{E} - \frac{1}{y} \biggr)\, dy, 
\end{equation} 
and evaluate the integrals. In the first instance the constant of integration is determined by taking the limit $x \to 0$, and demanding that it vanishes. For the second integral we take the limit $y \to \infty$, and also ensure that it evaluates to zero. The end result is the right-hand side of Eq.~(\ref{WF1a}). Very similar steps produce Eqs.~(\ref{WF1c}) and (\ref{WF1d}). 

For Eq.~(\ref{WF1b}) we proceed slightly differently. We introduce the new variable $t := (x/y)^{1/2}$ and write the left-hand side as 
\begin{equation} 
\frac{t}{xy} \sum_{\ell=1}^\infty P_\ell(\cos\theta) \int (t^2 + t^{-2}) t^{2\ell}\, dt  
= \frac{t}{xy} \int (t^2 + t^{-2}) \bigl[ (1 - 2t^2 \cos\theta + t^4)^{-1/2} - 1 \bigr]\, dt, 
\end{equation} 
where we again made use of the generating function. Evaluating the integral returns the right-hand side of Eq.~(\ref{WF1b}); the constant of integration is set to zero to eliminate odd powers of $t$ in the final result. 

For the listing of Eqs.~(\ref{WF2}) we begin instead with 
\begin{equation} 
\frac{1}{E} = \sum_{\ell=0}^\infty \frac{y^\ell}{x^{\ell+1}}\, P_\ell(\cos\theta). 
\end{equation} 
To derive Eq.~(\ref{WF2a}) we write the left-hand side as 
\begin{equation} 
\sum_{\ell=3}^\infty P_\ell(\cos\theta) \int \biggl( \frac{1}{xy} - \frac{x}{y^3} \biggr) \frac{y^\ell}{x^{\ell+1}}\, dy 
= \int \biggl( \frac{1}{xy} - \frac{x}{y^3} \biggr) \biggl[ \frac{1}{E} - \frac{1}{x} - \frac{y}{x^2} \cos\theta 
- \frac{y^2}{2x^3} (3\cos^2\theta - 1) \biggr]\, dy, 
\end{equation} 
and evaluate the integral. The requirement that the result vanish in the limit $y \to 0$ determines the constant of integration, and we arrive at the right-hand side of Eq.~(\ref{WF2a}). We proceed in the same way for Eqs.~(\ref{WF2c}) and (\ref{WF2d}); in the first instance we express $2/[(\ell-1)\ell(\ell+1)]$ as $1/(\ell-1)  - 2/\ell + 1/(\ell+1)$ and deal with each sum separately. For Eq.~(\ref{WF2b}) we can re-introduce $t := (x/y)^{1/2}$ as before, but it is simpler to notice that the equation is a version of Eq.~(\ref{WF1b}) with $x$ and $y$ interchanged.   

\bibliography{/Users/poisson/writing/papers/tex/bib/master}
\end{document}